\documentclass[10pt]{article}
\usepackage[letterpaper,textwidth=5.5in,textheight=9in,top=1in]{geometry}
\usepackage[T1]{fontenc}
\usepackage{newtxtext,newtxmath}
\usepackage{microtype}
\usepackage{amsmath}
\usepackage{booktabs}
\usepackage{graphicx}
\usepackage{xcolor}
\usepackage{caption}
\usepackage{titlesec}
\usepackage{enumitem}
\usepackage{listings}
\usepackage[numbers,sort&compress]{natbib}
\definecolor{linkblue}{HTML}{0F4D92}
\usepackage[colorlinks=true,allcolors=linkblue]{hyperref}

\titleformat{\section}{\large\bfseries}{\thesection}{0.8em}{}
\titleformat{\subsection}{\normalsize\bfseries}{\thesubsection}{0.8em}{}
\titleformat{\paragraph}[runin]{\normalsize\bfseries}{}{0pt}{}[.]
\titlespacing*{\section}{0pt}{1.6ex plus .4ex}{0.8ex}
\titlespacing*{\subsection}{0pt}{1.2ex plus .3ex}{0.5ex}
\titlespacing*{\paragraph}{0pt}{0.8ex plus .2ex}{0.6em}
\setlist{nosep,leftmargin=1.4em}
\graphicspath{{figures/}}

\newcommand{\papertitle}{Weight-Space Programming for Materials Science}

\begin{document}

\begin{center}
\rule{\linewidth}{2.5pt}\\[10pt]
{\LARGE\bfseries \papertitle\par}
\vspace{6pt}
\rule{\linewidth}{0.8pt}\\[14pt]
{\large \textbf{Sk Md Ahnaf Akif Alvi} \qquad \textbf{Raymundo Arr\'oyave}}\\[5pt]
{\normalsize Department of Materials Science and Engineering, Texas A\&M University\\ College Station, TX 77843, USA}
\end{center}
\vspace{6pt}

\begin{center}{\large\bfseries Abstract}\end{center}
\vspace{-4pt}
\begin{quote}
A materials discovery campaign often begins with almost no measurements of the property it is searching for, and the models that rank candidates or screen the output of a generator then have to be built from those first few labels. Following recent work on weight-space programming, in which a language model writes the weights of a small network directly as text, we show that a general-purpose model, given the task in words, the names of the inputs and a network of 16 hidden units with every weight set to zero, can provide this starting point by filling in all of the weights from what it has learned about materials. With no labels, the networks of the strongest model rank experimental alloys at Spearman $\rho = 0.64$ to $0.79$, where the best of five simple descriptors reaches $0.47$ to $0.62$. They rank the bulk modulus of crystals added during generative discovery loops at $0.95$, where the inverse volume per atom reaches $0.89$, and 35 phase-equilibrium targets computed with Thermo-Calc at a mean of $0.67$, close to the $0.70$ of an extrapolation from calculated binary and ternary data. As the prior mean of a Gaussian process they raise the ranking at five labels from $0.35$ to $0.69$ on alloys and from $0.15$ to $0.95$ on crystals, and they lower the regret of Bayesian optimization significantly on 4 of 5 experimental tasks and on 13 to 16 of 33 computed targets, depending on how multiple tests are corrected, without raising it significantly on any. The same networks also serve as the mean of a Bayesian neural network and as a screen for the samples of two crystal generators, where they do as well as standard models trained on 400 labeled structures. The gain is confined to the first tens of labels and to targets that no rule or sub-system extrapolation already ranks.

\end{quote}

\section{Introduction}

A materials discovery campaign often begins with almost no data on the property it is searching for. The surrogate that decides which candidate to make or simulate next is a Gaussian process or a Bayesian neural network fitted to a handful of labels, and with a constant prior mean it carries no information about the target until those labels arrive. The first tens of experiments are therefore spent on trends that a metallurgist could have stated in advance, such as the hardening of titanium by aluminum and oxygen or the rise of the solidus with refractory content. The knowledge can be supplied in advance by a physics-informed mean, a multi-fidelity model or transfer from an earlier campaign, but each of these needs a resource that is often missing at the start of a new campaign.

\begin{figure}[t]
\centering
\includegraphics[width=\linewidth]{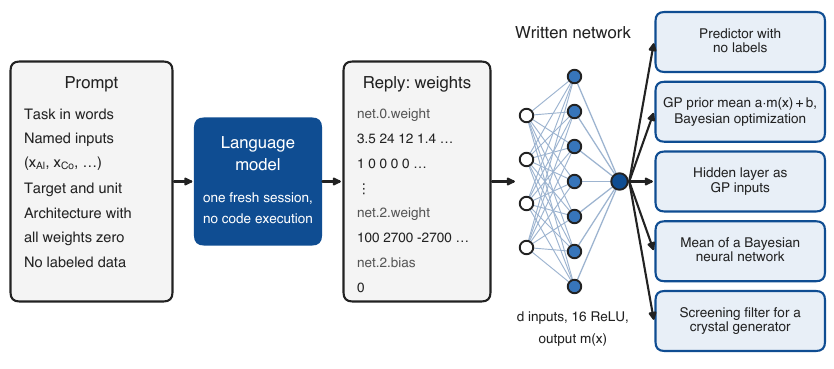}
\caption{A language model writes every weight of a small network from a description of the task. The written network $m(x)$ is then used as it is: as a predictor, as the mean of a probabilistic surrogate, as a feature map, or as a filter.}
\label{fig:method}
\end{figure}

One source of such knowledge that needs none of those resources is a language model. Takehi et al.\ recently showed that a frontier language model can produce the weights of a small working network directly, as text, with no optimizer \cite{takehi2026sparks}. They call this weight-space programming, and a number of studies have since examined its limits and applied it in other fields \cite{hu2026exploring,ahmed2026memorization,nusrat2026auditing,li2026weightloop,butkeraites2026write}, which we review in Section~\ref{sec:related}. For materials this suggests a direct route from what a language model has read to a function a surrogate can use. We ask the model to write a property predictor from a statement of what the inputs are and what the target is, with no labeled example of either, and it returns every weight of a network with 16 hidden units (Figure~\ref{fig:method}).

The result is an ordinary network, and the rest of this paper measures how much of use such a network is at each point in a discovery campaign where prior knowledge is lacking. We treat it first as a property-prediction model in its own right and measure how well it ranks candidates with no labels. We then use it in the places where a discovery loop lacks knowledge, which are the prior mean of a Gaussian process, including inside Bayesian optimization, the mean of a Bayesian neural network, a set of input features, and a filter on the samples of a crystal generator before an oracle call. The tasks cover experimental alloy data, phase equilibria that we computed ourselves, so that the labels are unlikely to have been seen as a table, and crystals proposed by generative models. Every written network is compared with the cheapest predictor that also uses no labels of the task, since a result that a one-line rule also reaches is not evidence for the language model. On the computed targets that reference is strong, since it is an extrapolation from calculated binary and ternary values, the standard way to estimate a multicomponent alloy, and it uses information the model is not given unless stated.

With no labels, the written networks of the strongest model rank experimental alloys at $\rho = 0.64$ to $0.79$, where the best of five simple descriptors reaches $0.47$ to $0.62$, and rank the 35 computed phase-equilibrium targets at a mean of $0.67$, close to the $0.70$ of an extrapolation from calculated binary and ternary values. Their main use is as a prior mean when labels are few, where they lower the regret of Bayesian optimization on 4 of 5 experimental tasks and on 13 to 16 of 33 computed targets, depending on the correction for multiple tests, without raising it significantly on any. When binary and ternary values are placed in the prompt, the networks come within $0.02$ of the extrapolation on the smooth targets and are level with it overall. The same network serves as the mean of a Bayesian neural network, as a feature map and as a screen for generated crystals, where it does as well as standard models trained on 400 labeled crystals. The gain is confined to the first tens of labels, and a network that ranks a pool well can still mislead an optimizer on its best members.

We also tested the other modes of weight-space programming, reading and editing weights that were trained elsewhere, on surrogates and interatomic potentials. Those experiments were less successful and are reported in the supplementary material (Section~\ref{sec:other}).

\section{Related work}
\label{sec:related}

\paragraph{Weight-space programming} Takehi et al.\ showed that a frontier language model, given a small network as text, can write, explain and edit its weights with no optimizer, for networks of up to roughly $10^5$ weights \cite{takehi2026sparks}. Later studies find editing outcomes uneven across runs \cite{hu2026exploring}, ask how much of the effect is recall of familiar datasets \cite{ahmed2026memorization}, apply weight reading to audits of clinical networks \cite{nusrat2026auditing}, edit a readout on a frozen encoder after each new assay measurement \cite{li2026weightloop}, and write small scoring networks for optimization solvers \cite{butkeraites2026write}. We take the one mode that needs no trained network to start from, writing, and ask what a written network is worth in materials discovery, where the downstream model must rank candidates from a handful of labels and report calibrated uncertainty. Our tests of the other modes on materials models (reading trained surrogates and editing interatomic potentials) and of weight arithmetic on crystal generators \cite{ilharco2023task,verma2026gffmerge} are reported in the supplementary material.

\paragraph{Prior knowledge in Bayesian optimization} Knowledge of the target can be built into a Gaussian process before the first label through a physics-informed mean function \cite{ziatdinov2022physics}, a multi-fidelity model \cite{kennedy2000predicting,khatamsaz2021efficiently} or transfer from a related campaign \cite{swersky2013multitask}, and each of these approaches depends on a resource, a simulator, a cheaper data source or an earlier campaign, that the present paper assumes is absent. Thermodynamic databases play that role in alloy design, where values extrapolated from binary and ternary assessments guide the search of multicomponent space \cite{andersson2002thermocalc}. A written network is a mean function whose only source of information is what the language model absorbed in training, so it requires none of these resources. We therefore compare it with rules that are equally free of task labels, including an extrapolation from computed sub-systems.

\paragraph{Models under test} The surrogates are Gaussian processes with expected improvement \cite{jones1998ego,balandat2020botorch} and three Bayesian neural networks: Monte Carlo dropout \cite{gal2016dropout}, deep ensembles \cite{lakshminarayanan2017ensembles} and Hamiltonian Monte Carlo as studied for Bayesian optimization by Li et al.\ \cite{li2024bnnbo}. The generators are MatterGen \cite{zeni2025mattergen} and CrystalFlow \cite{luo2025crystalflow}, with the reinforcement-learning fine-tunes of a closed discovery loop \cite{chen2025matinvent}.

\section{Method}

\subsection{Writing a network}
The network is a multilayer perceptron with one hidden layer of 16 ReLU units, $m(x) = w_2^\top \mathrm{ReLU}(W_1 x + b_1) + b_2$. For $d$ inputs it has $16d + 33$ weights, which for the tasks here is 113 to 417 numbers. These networks are small beside those of the original work, which wrote game-playing policies of about 20{,}500 weights in one reply and edited networks of about 172{,}000 \cite{takehi2026sparks}. The size was chosen because the inputs are 5 to 24 named quantities and the output is one number, and because the network has to be small enough for a surrogate to use as a mean that can be inspected. On four tasks we also asked for larger networks, with one hidden layer of 64 units or two hidden layers of 32 units, which have 577 to 1{,}825 weights. These ranked no better than the 16-unit network (supplementary Table~\ref{tab:size}). The \emph{write} prompt states the task in two or three sentences, lists the inputs by name in order (for an alloy, the atomic fraction of each element), names the target and its unit, and gives the architecture with every weight set to zero. It asks for every weight, such that the output is the model's best estimate of the target. No labeled example is shown and no range of the target. The prompt does name the data source (for measured sets the published dataset and its reference), the number of candidates, the composition range of a computed pool and, for crystals, typical ranges of the inputs. The reply is a block of decimal numbers under one header per tensor (Figure~\ref{fig:method}).

Each run is a fresh session of one model. It reads one prompt file, thinks, and writes one reply file and a short account of how it arrived at the reply. It is instructed to use no code execution, calculator or web access. The transcript of the session that wrote each reply is audited for any tool call beyond those file operations, and all 358 runs behind the main text passed. Seven runs were started a second time after a usage limit interrupted the first session. In three of them both sessions wrote a reply, and the file written last is the one kept. The first sessions (the Opus alloy runs and two Fable replies) followed an earlier form of the protocol with fewer file operations, and those two Fable replies have no self-report. A reply is loaded only if it parses into the requested tensors with the requested shapes. A reply that fails this check is counted as a failure and is not repaired. We use Claude Opus~5.5 and Fable~5.2 on every task and Sonnet~5.5 and Haiku~5.5 on the experimental alloy and crystal tasks, with two or three replies per model and task, except one reply each from Sonnet and Haiku on crystal heat capacity. All results come from one vendor's models. The instructions given to every session, and the prompt, raw reply, self-report and a diagram of the written network for every task, are in the supplementary material.

\subsection{Tasks}
\paragraph{Experimental alloys} Five regression tasks take atomic fractions as inputs (15 to 24 inputs: 14 to 23 elements and the summed fraction of the rest): Young's modulus (237 alloys), yield strength (271) and hardness (152) of titanium alloys \cite{salvador2022tialloys}, and hardness (530) and yield strength (722) of multi-principal-element alloys \cite{borg2020mpea}.

\paragraph{Computed phase equilibria} To test targets that the models are unlikely to have seen as a table, we computed nine pools of 200 to 400 compositions with Thermo-Calc 2025b \cite{andersson2002thermocalc}: five high-entropy families sampled on a 5~at.\% grid (Al--Co--Cr--Fe--Mn--Ni, Mo--Nb--Ta--Ti--V--W--Zr, Hf--Mo--Nb--Ta--Ti--Zr, Co--Cr--Cu--Fe--Ni--Ti and Al--Cr--Mo--Nb--Ti, with the TCHEA8 database) and four families with a balance element (nickel superalloys with TCNI12, titanium alloys with TCTI6, steels with TCFE10, aluminum alloys with TCAL9). Each family has three to five targets: the equilibrium solidus, the freezing range, and mole fractions of named phases at a fixed temperature (for example the $\gamma'$ fraction of a nickel alloy at 1173~K or the austenite fraction of a steel). A target is dropped when more than 80\% of a pool shares one value, which leaves 35 targets. Solidus and liquidus are located to within 4~K, and compositions for which a quantity failed to compute are dropped. Each target has a direction for optimization (a high solidus, a narrow freezing range, a low fraction of intermetallic phases).

\paragraph{Crystals behind a structure embedding} The surrogates of a generative discovery loop \cite{chen2025matinvent} see a crystal as the first 50 principal components of a pretrained graph-network embedding (ORB-v3 \cite{rhodes2025orbv3}). Those coordinates have no names, and a model cannot write a network on inputs it cannot interpret. The network is therefore written on 15 named descriptors of the same crystal, which are the volume per atom, the mean atomic mass, the mean electronegativity, the mean elemental melting point and eleven composition-group fractions. We use two targets from that loop, bulk modulus (6 runs) and heat capacity per gram at 300~K (5 runs). All runs of a target start from the same set, which is 500 Materials Project crystals with bulk moduli from density functional theory \cite{dejong2015elastic}, and 445 crystals labeled in an earlier loop for heat capacity. The crystals added during the runs, 479 for bulk modulus and 212 for heat capacity, were proposed by a generator and labeled with a machine-learned potential (eSEN-30M-OAM \cite{fu2025esen}), except 63 and 35 that the loop drew from a seed pool. In each run the training pool is the starting set and the test set is the crystals added in that run. For bulk modulus the surrogate is therefore trained on DFT labels of known crystals and tested on potential labels of generated ones, as in the loop itself, so a gap between a trained surrogate and a label-free predictor on this task includes that shift. Scores of a predictor with no labels are computed on the added crystals, each counted once. Because the runs share their starting set they are not independent, so tests on these tasks take the run as the unit, and with six runs the smallest attainable two-sided $p$ is $0.031$.

\subsection{References with no language model}
Every written network is compared with the cheapest predictor that also uses no labels of the task. For experimental alloys this is a rule of mixtures over tabulated elemental properties (Young's modulus for the modulus target, Vickers hardness for strength and hardness). That rule turned out to be a weak reference, so after the first analysis we scored four more label-free descriptors with signs fixed from metallurgy, namely rules of mixtures of melting point and shear modulus, the valence electron concentration, and the atomic size mismatch $\delta = \sqrt{\sum_i c_i (1 - r_i/\bar{r})^2}$ \cite{zhang2008solidsolution}. The size mismatch is the best of them on four of the five tasks and we report it beside the rule of mixtures, although it was chosen with the targets in view, which favors it. Its radii are metallic radii where tabulated and empirical atomic radii for boron, carbon and silicon, and it has no stated relation to an elastic modulus, where it is scored only for completeness. For crystals it is the inverse of the volume per atom for bulk modulus and the Dulong--Petit value $3R/\bar{M}$ for heat capacity per gram. For the Thermo-Calc targets it is an extrapolation from sub-systems: the target is computed for the pure elements, at three points of every binary and at the equimolar point of every ternary (for balance-element families, at two contents of each solute and two of each solute pair), and a polynomial of Redlich--Kister--Muggianu form through exactly those points is evaluated at the multicomponent composition. This is how an alloy designer would estimate a multicomponent value from binary and ternary diagrams, and it is a strong reference because it uses computed data from the same database as the targets, excluding only the multicomponent pool itself.

\subsection{Phase-diagram context}
The same sub-system values can be given to the language model. In the \emph{write with facts} arm the prompt is unchanged except for an inserted list of the binary and ternary values in words (for example, the solidus of Ni with 10~at.\% Al, or the phases present in equimolar Co--Cr--Ni at the fixed temperature). The multicomponent target itself never appears. The arm without facts measures what the model carries from training, and this arm measures whether it can combine that with thermodynamic context supplied in the prompt.

\subsection{Uses of a written network}
\paragraph{As a predictor} With no labels, we report the Spearman correlation $\rho$ between $m(x)$ and the target over the whole pool, the share of the best tenth of the pool that is in the network's best tenth, and the coefficient of determination $R^2$, as written and after the best straight-line rescaling.

\paragraph{As the prior mean of a Gaussian process} The network enters a Gaussian process as the mean $a\,m(x)+b$. We compare three fits of $a$ and $b$: least squares on the labels followed by a constant-mean process on the residuals (two-stage); a linear mean on the network's output whose $a$ and $b$ are learned with the kernel by marginal likelihood; and a mean equal to the network's output with a linear kernel term on it, which integrates Gaussian $a$ and $b$ out. Models are fitted with BoTorch \cite{balandat2020botorch} on 5, 10, 20 and 50 labels and scored on held-out candidates by $\rho$, coverage of the 95\% interval and negative log likelihood.

\paragraph{In Bayesian optimization} A loop starts from 5 random labels and takes 25 steps of expected improvement \cite{jones1998ego} over a fixed pool, 20 seeds per arm, with the two-stage fit of the mean unless stated. All targets are maximized except the Young's modulus of titanium alloys, where a low value is the goal. An arm is defined by the replies of one model. At each step it uses the reply whose output correlates best with the labels so far, and it falls back to no prior if that correlation is below 0.3. This rule was fixed before the 20-seed runs, after a pilot on the same tasks in which an unselected prior raised regret on some of them. We report the area under the regret curve and a paired Wilcoxon test against the plain Gaussian process, with Holm correction over the arms of a target. Counts of significant targets are also given with the correction applied across targets. Comparisons between two prior means on the experimental tasks are corrected over the five tasks.

\paragraph{As inputs of a Gaussian process} The 16 hidden activations of a written network are a feature map obtained without training. On the crystal tasks we fit Gaussian processes on these activations, alone or appended to the embedding, with and without the network's output as the mean.

\paragraph{In a Bayesian neural network} The network is used as the prior mean (the Bayesian network predicts the residual), as the starting weights, and as the center of a Gaussian prior over weights, for Monte Carlo dropout \cite{gal2016dropout}, deep ensembles \cite{lakshminarayanan2017ensembles} and Hamiltonian Monte Carlo \cite{li2024bnnbo}.

\paragraph{As a screening filter} Crystals sampled from two generative models, MatterGen \cite{zeni2025mattergen} and CrystalFlow \cite{luo2025crystalflow} (128 from each base model and from ten fine-tuned variants of each), are scored by the written network from composition and as-generated volume, and only the top quarter would be sent to the oracle. References are the textbook rule and three models trained on the same descriptors of 400 relaxed Materials Project crystals with DFT bulk moduli (the two descriptors that are zero in all 400 crystals are left out of the network and the ridge regression), so that their training labels differ from the oracle's \cite{jain2013materials,dejong2015elastic}: a network of the same size as the written one, a ridge regression on the logarithm of the bulk modulus, and a random forest. The last two were added after the first analysis, so that the comparison does not rest on one trained model.

\section{Results}

\subsection{A written network is a usable predictor before any label}
\label{sec:predictor}
\begin{figure}[t]
\centering
\includegraphics[width=\linewidth]{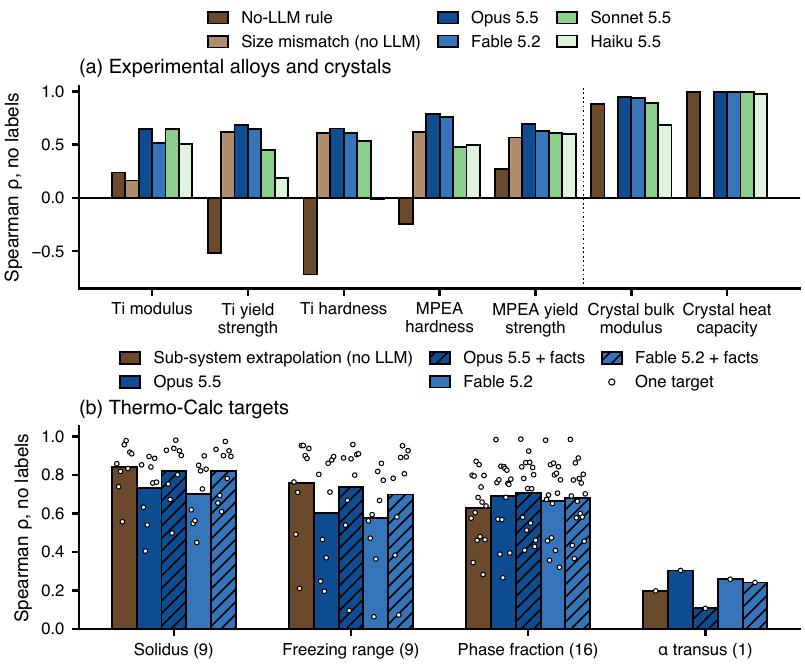}
\caption{Rank correlation between a written network and the target, with no labels used. (a) Experimental alloys and crystals; bars are means over the replies of one model. The rule is a rule of mixtures over elemental properties for alloys, the inverse volume per atom for bulk modulus and the Dulong--Petit value for heat capacity; the atomic size mismatch is a second label-free reference for alloys. Crystal scores use the crystals added during the loops, each counted once. (b) The 35 Thermo-Calc targets by kind; bars are means over targets and dots are single targets. ``Facts'' are sub-system values added to the prompt; the sub-system extrapolation uses the same values with no language model.}
\label{fig:predictor}
\end{figure}

Figure~\ref{fig:predictor} shows how well a written network ranks candidates before it has seen a label. On the five experimental alloy tasks Opus~5.5 reaches $\rho = 0.64$ to $0.79$ and Fable~5.2 $0.51$ to $0.76$. The rule of mixtures gives $-0.72$ to $0.27$. On both hardness tasks and on titanium yield strength it ranks alloys backwards, because the hardness of the pure elements carries little information about what hardens an alloy. Sonnet~5.5 and Haiku~5.5 are weaker and less even ($-0.01$ to $0.65$). The rule of mixtures is a weak reference. The atomic size mismatch ranks the four strength and hardness tasks at $0.57$ to $0.62$, and a rule of mixtures of the shear modulus ranks titanium-alloy modulus at $0.47$ (supplementary Table~\ref{tab:rules}), and the Opus networks are ahead of the best of these descriptors on every task, by $0.04$ to $0.17$. On the crystals added during the loops, pooled over runs, the two strongest models rank bulk modulus at $\rho = 0.95$ and $0.94$ against $0.89$ for the inverse volume per atom, so most of what the networks know about this target is the volume rule. For heat capacity per gram the Dulong--Petit value alone gives $\rho = 1.00$ and the networks reproduce it. That task is solved by the rule, and we keep it as a check that a written network does no harm where a rule suffices.

The Thermo-Calc targets test the same thing against a much stronger reference (Figure~\ref{fig:predictor}b). Over all 35 targets the Opus networks rank the pool at a mean $\rho = 0.67$ and the Fable networks at $0.64$, given the names of the elements, the composition range of the pool and the definition of the target. The extrapolation from computed binary and ternary values reaches $0.70$, and which of the two is ahead depends on the kind of target. For the solidus and the freezing range, which vary smoothly with composition, the extrapolation is ahead ($0.84$ and $0.76$ against $0.73$ and $0.60$ for Opus). For the 16 phase-fraction targets, which depend on whether a phase appears at all, the written networks have a higher mean, at $0.69$ for Opus and $0.67$ for Fable against $0.63$ for the extrapolation. Per target Opus is ahead on 7 of the 16 and Fable on 9, three targets carry most of the difference, and it is not significant. Paired over all 35 targets the Opus networks are not significantly behind the extrapolation and the Fable networks are (supplementary Section~\ref{sec:robustness}). Given the sub-system values in its prompt, Opus comes within $0.02$ of the extrapolation on solidus and freezing range ($0.82$ and $0.74$) and has a higher mean on phase fractions ($0.71$) and overall ($0.73$ against $0.70$), although neither difference is significant over targets. Fable with the same facts is $0.02$ behind the extrapolation on solidus and $0.06$ behind on freezing range. Per target the networks with facts trail the extrapolation by more than $0.05$ on 9 (Opus) and 12 (Fable) of the 35. The models use supplied thermodynamic context but do not reproduce everything an explicit extrapolation gets from it. Every predictor does poorly on two targets: the freezing range of the Co--Cr--Cu--Fe--Ni--Ti family ($\rho \le 0.21$) and the $\alpha$ transus of titanium alloys ($\rho \le 0.30$).

Of 60 replies on the experimental alloy tasks, 3 were rejected as malformed, and none of the 280 Thermo-Calc replies or the 18 crystal replies was rejected. Replies of one model agree with each other more than models agree with one another (on the alloy tasks the mean pairwise rank correlation of predictions is $0.87$ to $0.94$ within a model and $0.60$ to $0.78$ across models), so the choice of model matters more than the choice of reply. The models write sparse networks, in which a typical reply uses 5 or 6 of the 16 hidden units on the experimental alloys and the Thermo-Calc targets and 10 on crystals (medians) and leaves the rest at zero. Each unit that is used has a stated meaning in the model's own account, for example a solid-solution term, a threshold for a second phase or a saturating contribution of one element (supplementary Section~\ref{supp:written}).

\begin{table}[t]
\centering
\caption{Written networks as regression models with no labels. $\rho$: Spearman rank correlation with the target. $R^2$: coefficient of determination of the network's output as written, and after the best straight-line rescaling of that output (the squared linear correlation, an upper bound on what a slope and an offset can recover). Experimental alloys and crystals: mean over the replies of a model; crystals are those added during the loops, each counted once. Thermo-Calc: mean of $\rho$ and median of $R^2$ over targets; the single $\alpha$-transus target is counted only in the last rows. The rule is a rule of mixtures for alloys, the inverse volume per atom and the Dulong--Petit value for crystals, and the sub-system extrapolation for Thermo-Calc targets. MPEA: multi-principal-element alloys; TC: Thermo-Calc.}
\label{tab:regression}
\footnotesize
\setlength{\tabcolsep}{2.4pt}
\begin{tabular}{lrrrrrrrrr}
\toprule
 & \multicolumn{3}{c}{Rule, no language model} & \multicolumn{3}{c}{Opus 5.5} & \multicolumn{3}{c}{Fable 5.2} \\
\cmidrule(lr){2-4}\cmidrule(lr){5-7}\cmidrule(lr){8-10}
Task & $\rho$ & $R^2$ & $R^2$ resc. & $\rho$ & $R^2$ & $R^2$ resc. & $\rho$ & $R^2$ & $R^2$ resc. \\
\midrule
Ti alloy Young's modulus & 0.24 & $-$2.64 & 0.03 & 0.64 & 0.19 & 0.34 & 0.51 & 0.02 & 0.25 \\
Ti alloy yield strength & $-$0.52 & $-$0.79 & 0.23 & 0.69 & 0.42 & 0.46 & 0.64 & 0.34 & 0.37 \\
Ti alloy hardness & $-$0.72 & $<-10$ & 0.44 & 0.65 & 0.40 & 0.43 & 0.61 & 0.29 & 0.39 \\
MPEA hardness & $-$0.24 & $<-10$ & 0.01 & 0.79 & 0.45 & 0.55 & 0.76 & 0.47 & 0.55 \\
MPEA yield strength & 0.27 & 0.06 & 0.10 & 0.69 & 0.33 & 0.44 & 0.63 & 0.21 & 0.35 \\
\addlinespace
Crystal bulk modulus & 0.89 & $-$2.13 & 0.58 & 0.95 & 0.80 & 0.83 & 0.94 & 0.77 & 0.83 \\
Crystal heat capacity & 1.00 & 0.93 & 0.97 & 1.00 & 0.97 & 0.98 & 1.00 & 0.96 & 0.98 \\
\midrule
TC solidus (9) & 0.84 & 0.45 & 0.80 & 0.73 & $-$0.12 & 0.60 & 0.70 & 0.03 & 0.60 \\
\quad with facts & & & & 0.82 & 0.70 & 0.83 & 0.82 & 0.58 & 0.82 \\
TC freezing range (9) & 0.76 & 0.29 & 0.65 & 0.60 & $-$0.55 & 0.42 & 0.57 & $-$0.26 & 0.37 \\
\quad with facts & & & & 0.74 & 0.50 & 0.68 & 0.70 & 0.18 & 0.63 \\
TC phase fractions (16) & 0.63 & 0.20 & 0.41 & 0.69 & 0.42 & 0.60 & 0.67 & 0.46 & 0.59 \\
\quad with facts & & & & 0.71 & 0.43 & 0.58 & 0.68 & 0.36 & 0.56 \\
TC, all 35 targets & 0.70 & 0.29 & 0.60 & 0.67 & 0.32 & 0.57 & 0.64 & 0.16 & 0.49 \\
\quad with facts & & & & 0.73 & 0.46 & 0.65 & 0.71 & 0.35 & 0.59 \\
\bottomrule
\end{tabular}
\end{table}

\paragraph{As a regression model} Table~\ref{tab:regression} scores the same networks as regression models, by the coefficient of determination as well as by rank. A written network gets the order of candidates right more often than their values. On the experimental alloy tasks the Opus networks reach $R^2 = 0.19$ to $0.45$ as written, where the rule of mixtures is below zero on four of the five tasks because its scale is wrong, and on crystal bulk modulus $0.80$. A straight-line rescaling of the output raises these only to $0.34$--$0.55$ and $0.83$, so on these tasks the scale was roughly right and what remains is error in the shape of the function. On the Thermo-Calc targets without facts, the median $R^2$ as written is below zero ($-0.12$ and $-0.55$ for Opus) and rises to $0.60$ and $0.42$ after rescaling, which means that the networks have the trend of a solidus or a freezing range but not its absolute level. With the sub-system values in the prompt the level is right as written ($0.70$ for the solidus) and after rescaling the networks match the extrapolation ($0.83$ against $0.80$). On phase fractions the written networks are ahead of the extrapolation on average, as written and rescaled ($0.42$ and $0.60$ against $0.20$ and $0.41$), but not consistently, since the Opus networks rank better on 7 of the 16 targets and the difference in $\rho$ is not significant (paired Wilcoxon $p = 0.50$, and for Fable 9 of 16 with $p = 0.60$). These are modest values for a regression model, and on the alloy tasks a Gaussian process trained on a few hundred labels does better. The written network is a zero-label predictor of trends, and the downstream models below are given labels to correct its scale and shape.

\subsection{As the prior mean of a Gaussian process}
\label{sec:fewlabels}
\begin{figure}[t]
\centering
\includegraphics[width=\linewidth]{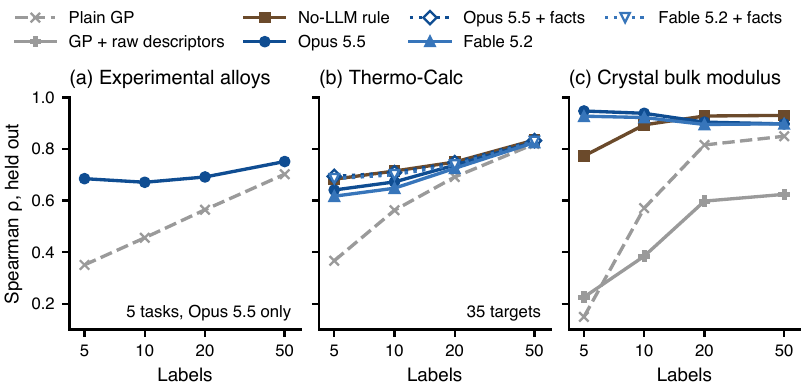}
\caption{Ranking of held-out candidates against the number of labels, for a plain Gaussian process and for the same process with a data-free predictor as its mean. (a) Experimental alloys, Opus networks, mean of five tasks, slope and offset integrated out. (b) Thermo-Calc targets, mean over targets, slope and offset integrated out. (c) Crystal bulk modulus, mean of six runs; the process sees a 50-dimensional embedding, the written network sees 15 named descriptors; slope and offset learned with the kernel (two-stage for the rule).}
\label{fig:fewlabels}
\end{figure}

With five labels a Gaussian process with a constant mean ranks held-out candidates at $\rho = 0.35$ on the experimental alloys, $0.37$ on the Thermo-Calc targets and $0.15$ on crystal bulk modulus (Figure~\ref{fig:fewlabels}). The same process with a written network as its mean starts at the ranking of the network alone, which is $0.69$ on the alloys with the Opus networks, $0.64$ on the Thermo-Calc targets and $0.95$ on crystals. From there it changes little up to 20 labels on the alloys ($0.69$) and falls slightly on crystals ($0.90$), and it rises to $0.83$ at 50 labels on the Thermo-Calc targets. The plain process needs about 50 labels to reach the same ranking on the alloys and about 20 on the Thermo-Calc targets, where the sub-system extrapolation and the networks written with facts start a little higher still ($0.68$ and $0.69$). On crystal bulk modulus it reaches $0.88$ with all 500 labels of a run, below the $0.95$ of the written network with none and level with the volume rule with none ($0.88$, where these two values are means over runs and differ slightly from the pooled values above). Those 500 labels are DFT values of known crystals and the test crystals are generated ones labeled by a potential, so this comparison measures the shift between the two as much as the value of the labels. All six runs improve with the Opus and Fable means at 5, 10 and 20 labels ($p = 0.031$, the smallest value six runs can give).

In the crystal case the surrogate works on a structure embedding and the written network reads 15 descriptors that the embedding process does not see, so the gain could come from the extra inputs alone. A plain process given the embedding and the same 15 descriptors is worse than the embedding alone from 10 labels on ($\rho = 0.60$ against $0.82$ at 20 labels), so the extra inputs on their own do not explain the gain. The descriptors help once they are combined into one number, and the combination need not come from a language model, since the volume rule, which uses one of the 15, gives most of the gain as a two-stage mean ($0.77$, $0.89$ and $0.93$ at 5, 10 and 20 labels).

The advantage is confined to the regime of few labels. With 20 labels on crystals the inverse-volume rule as a two-stage mean ($0.93$) is ahead of the written networks on the same fit ($0.91$ and $0.89$). With the joint fits the rule stays near the plain process at few labels, because its raw scale is far from the target's and the slope starts at one. With the full pool no mean differs from the plain process. The Haiku networks ($\rho = 0.66$ alone, mean over runs) are not significantly better than the plain process at 10 labels and are no better at 20, so a written mean from a weaker model does not help.

\paragraph{Calibration} How the slope $a$ and offset $b$ are fitted decides whether the error bars can be trusted. The two-stage fit estimates them from the same few labels and passes on none of their uncertainty, so at 5 labels its 95\% intervals cover only $0.63$ of held-out alloys, $0.74$ of held-out Thermo-Calc compositions and $0.69$--$0.73$ of the added crystals. Integrating $a$ and $b$ out raises coverage to $0.79$, $0.88$ and $0.98$ (plain process: $0.76$, $0.86$ and $0.86$; nominal $0.95$) with the same or better ranking (Table~\ref{tab:joint}), and learning them with the kernel is nearly as good. We therefore recommend a joint fit as the default and use the two-stage fit only where stated.

\begin{table}[t]
\centering
\caption{Three ways to fit the mean $a\,m(x)+b$ of a Gaussian process, on five experimental alloy tasks with the Opus and Fable networks (120 cells per row, pooled: 75 Opus and 45 Fable, the latter from three of the tasks). NLL: median negative log likelihood of held-out alloys.}
\label{tab:joint}
\small
\begin{tabular}{llrrr}
\toprule
Labels & Fit & $\rho$ & 95\% coverage & NLL \\
\midrule
5 & Plain GP, no prior & 0.34 & 0.76 & 7.46 \\
  & Two-stage & 0.64 & 0.63 & 8.51 \\
  & Joint, $a$ and $b$ learned & 0.66 & 0.75 & 7.63 \\
  & Joint, $a$ and $b$ integrated out & \textbf{0.68} & \textbf{0.79} & \textbf{7.39} \\
\midrule
20 & Plain GP, no prior & 0.58 & 0.68 & 7.96 \\
  & Two-stage & 0.70 & 0.66 & 8.28 \\
  & Joint, $a$ and $b$ learned & 0.69 & 0.67 & 8.42 \\
  & Joint, $a$ and $b$ integrated out & 0.70 & 0.69 & \textbf{7.87} \\
\bottomrule
\end{tabular}
\end{table}

\subsection{In Bayesian optimization}
\label{sec:bo}
\begin{figure}[t]
\centering
\includegraphics[width=\linewidth]{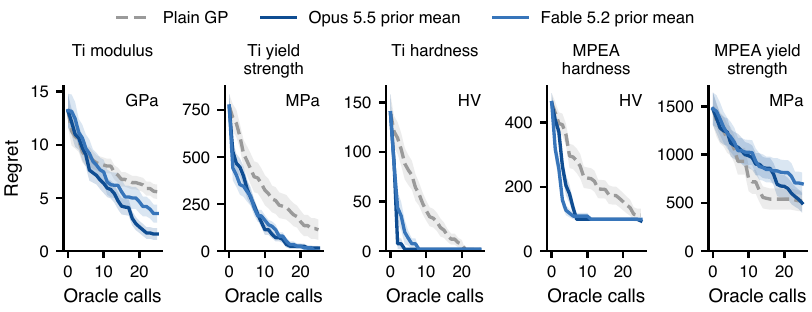}
\caption{Bayesian optimization on the experimental alloy tasks with a written prior mean. Regret is the gap between the best alloy in the pool and the best found; lines are means over 20 seeds and bands are standard errors.}
\label{fig:bo}
\end{figure}

\begin{table}[t]
\centering
\caption{Area under the regret curve over 25 steps on the experimental alloy tasks, mean of 20 seeds, two-stage fit (lower is better; Young's modulus is minimized). Bold: below the plain Gaussian process with Holm-corrected $p<0.05$, paired Wilcoxon test.}
\label{tab:bo}
\small
\begin{tabular}{lrrrrr}
\toprule
Task & GP & Opus 5.5 & Fable 5.2 & Sonnet 5.5 & Haiku 5.5 \\
\midrule
Ti alloy Young's modulus (GPa) & 7.68 & \textbf{5.55} & 6.92 & 7.34 & 7.61 \\
Ti alloy yield strength (MPa) & 314 & \textbf{153} & \textbf{148} & \textbf{216} & 255 \\
Ti alloy hardness (HV) & 40.1 & \textbf{5.95} & \textbf{10.0} & \textbf{11.3} & 27.2 \\
MPEA hardness (HV) & 226 & \textbf{137} & \textbf{120} & 247 & 259 \\
MPEA yield strength (MPa) & 772 & 900 & 966 & 735 & 641 \\
\bottomrule
\end{tabular}
\end{table}

Inside an optimization loop the written mean pays off in the first ten oracle calls (Figure~\ref{fig:bo}, Table~\ref{tab:bo}). With the Opus networks the regret on titanium-alloy hardness falls from 140 to 2~HV within four steps (no seed then finds the single best alloy), where the plain process needs about twenty steps to get as low. Four of the five experimental tasks improve significantly with Opus and three with Fable. The area under the regret curve relative to the plain process is $0.69$ for the two-stage fit and $0.68$ with $a$ and $b$ integrated out (means of per-task ratios), and no task differs at $p<0.05$, so the choice between the two fits makes no measurable difference here. A rule-of-mixtures mean, run as a further arm, is never significantly better than the plain process. On the yield strength of multi-principal-element alloys it does have a lower regret area than the plain process (676 against 772, not significant), where both written means have a higher one, a case taken up in Section~\ref{sec:failures}. The atomic size mismatch is a much better mean than the rule of mixtures on one task and no better on the rest. On titanium-alloy hardness it lowers the regret area to $5.7$~HV and is significantly better than both written means (Opus $5.95$, better in 3 of 20 seeds, and Fable $10.0$, better in 1). On the other four tasks it is not significantly better than the plain process. With Holm correction over the five tasks, the Opus mean is significantly better than the size mismatch on the hardness of multi-principal-element alloys ($p = 0.003$) and not on titanium-alloy yield strength or modulus ($p = 0.053$ and $0.097$, uncorrected $0.018$ and $0.049$), and the Fable mean is significantly better on that hardness task and on titanium-alloy yield strength (supplementary Table~\ref{tab:borule}).

\begin{figure}[p]
\centering
\includegraphics[width=\linewidth]{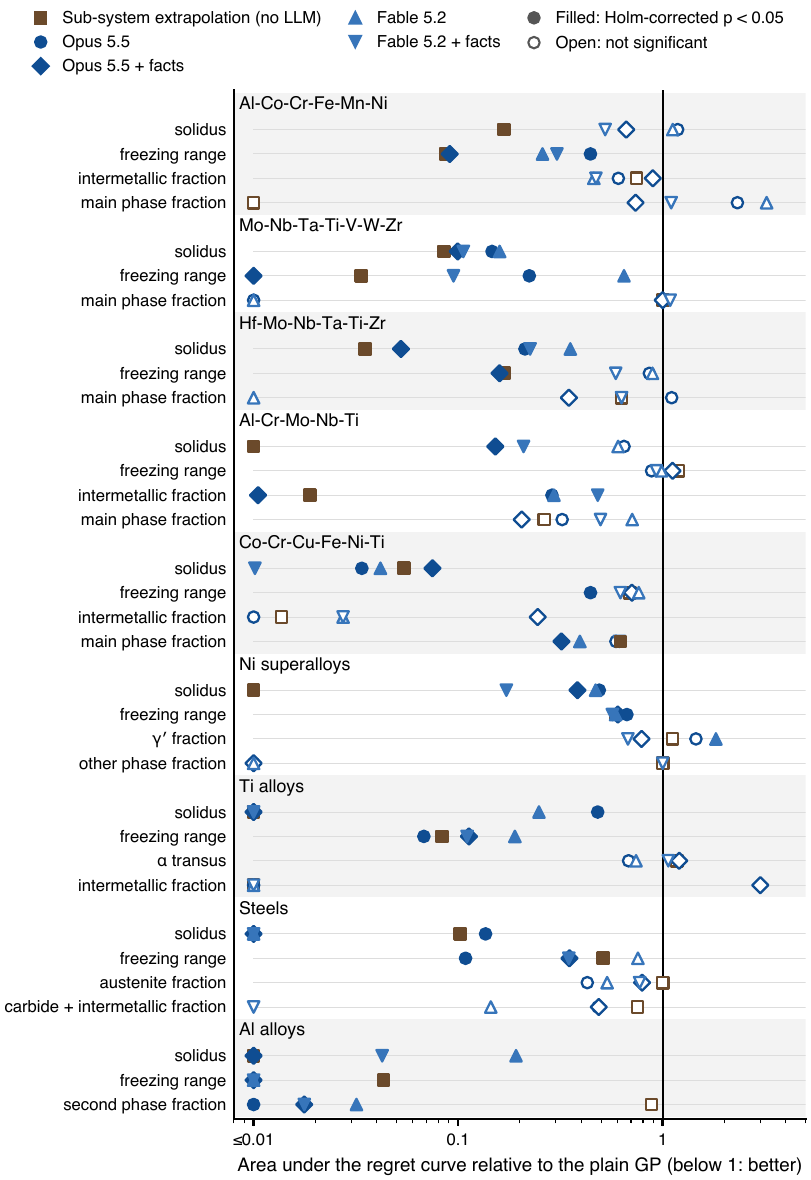}
\caption{Bayesian optimization on the Thermo-Calc targets: area under the regret curve relative to the plain Gaussian process (left of the line is better), 20 seeds, two-stage fit. Filled markers: Holm-corrected $p<0.05$. Two targets on which the plain process has zero regret are left out; on several phase-fraction targets its regret is close to zero and the ratios carry little information.}
\label{fig:bocalphad}
\end{figure}

The Thermo-Calc targets show the same on a larger and harder set (Figure~\ref{fig:bocalphad}, Table~\ref{tab:bocalphad}). On the 33 targets where the plain process does not already reach zero regret, the Opus networks lower the regret significantly on 16 and raise it on none. The Fable networks lower it on 15 and raise it on one. The median area under the regret curve is $0.43$ and $0.35$ of the plain process's. The sub-system extrapolation, which uses computed thermodynamic data, wins on 18 targets with a median of $0.17$, and the networks written with the same data in the prompt win on 18 (Opus) and 16 (Fable), with medians of $0.24$ and $0.35$. On the 18 solidus and freezing-range targets, where the plain process has the most regret, the Opus networks win 14 without facts and 15 with them, against 16 for the extrapolation. On the phase-fraction targets few differences are significant for any prior, largely because the plain process already finds a single-phase or precipitate-free composition within a few steps.

\begin{table}[t]
\centering
\caption{Bayesian optimization on 33 Thermo-Calc targets: number of targets on which a prior mean is significantly better or worse than the plain Gaussian process (Holm-corrected $p<0.05$), and the median area under the regret curve relative to the plain process. Holm correction over the arms of a target. Two-stage fit; the joint fits give the same counts within three targets.}
\label{tab:bocalphad}
\small
\begin{tabular}{lrrrrr}
\toprule
 & \multicolumn{2}{c}{All 33 targets} & \multicolumn{2}{c}{Solidus, freezing range (18)} & Median \\
Prior mean & better & worse & better & worse & relative area \\
\midrule
Sub-system extrapolation (no LLM) & 18 & 0 & 16 & 0 & 0.17 \\
Opus 5.5 & 16 & 0 & 14 & 0 & 0.43 \\
Opus 5.5 + facts & 18 & 0 & 15 & 0 & 0.24 \\
Fable 5.2 & 15 & 1 & 12 & 0 & 0.35 \\
Fable 5.2 + facts & 16 & 0 & 14 & 0 & 0.35 \\
\bottomrule
\end{tabular}
\end{table}

The counts in Table~\ref{tab:bocalphad} depend on how multiple tests are treated. With the Holm correction applied across the 33 targets of an arm, the Opus networks are better on 15 targets and the Fable networks on 13; across all tests of the five arms, on 13 and 11 (supplementary Table~\ref{tab:holm}). The extrapolation keeps 18 and 17.

\subsection{The hidden layer as inputs}
\label{sec:features}
\begin{figure}[t]
\centering
\includegraphics[width=3.3in]{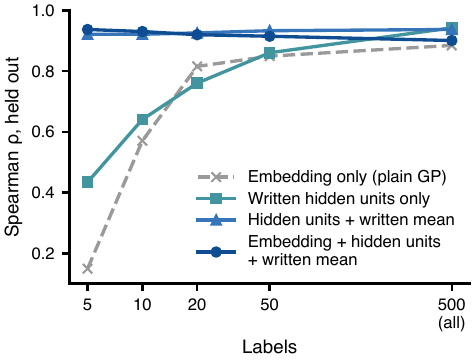}
\caption{Crystal bulk modulus: Gaussian processes on different inputs, mean over the Opus and Fable networks and six runs. The written hidden layer is 16 numbers computed from 15 named descriptors; the embedding is 50 principal components of a pretrained graph network.}
\label{fig:features}
\end{figure}

A written network also defines 16 features, its hidden activations, and these were obtained without a training run. Used as inputs alone they need labels like any other features, and a process on them ranks at only $\rho = 0.43$ with 5 labels (Figure~\ref{fig:features}). With enough labels they are the better representation. On the full pool of 500 crystals a process on the 16 written features reaches $\rho = 0.94$, against $0.88$ for the process on the 50 embedding coordinates and $0.80$ for the process on embedding plus raw descriptors. Using the hidden layer as inputs and the output as the mean gives $0.92$ to $0.94$ at every label count from 5 to 500, the most even behavior of any arm we tested. Appending the embedding to the written features does not help and costs a little at 500 labels ($0.90$).

\subsection{In a Bayesian neural network}
\label{sec:bnn}
The gain from a written mean is not specific to Gaussian processes. Used as the mean of a Bayesian neural network, so that the network learns the residual, it raises the ranking of Monte Carlo dropout from $\rho = 0.54$ to $0.65$, averaged over six tasks at 10 and 20 labels, which compares with a rise from $0.53$ to $0.69$ for the Gaussian process. Each pair is the mean over 24 cells (six tasks, two models, and 10 or 20 labels) without and with the written mean, and the ranking improves in 22 of the 24 cells for dropout and in 23 for the Gaussian process (supplementary Figure~\ref{fig:bnn}). The two other ways of using it are not recommended. As starting weights it helps a deep ensemble a little ($0.50$ to $0.58$), does nothing for the ranking of dropout and slightly lowers that of Hamiltonian Monte Carlo ($0.45$ to $0.40$). As the center of a Gaussian prior over weights it harms Hamiltonian Monte Carlo ($0.45$ to $0.21$), because a written network uses a few hidden units with large weights and a unit-variance prior around it pins the rest of the network at zero.

\subsection{As a screening filter for crystal generators}
\label{sec:filter}
\begin{figure}[t]
\centering
\includegraphics[width=\linewidth]{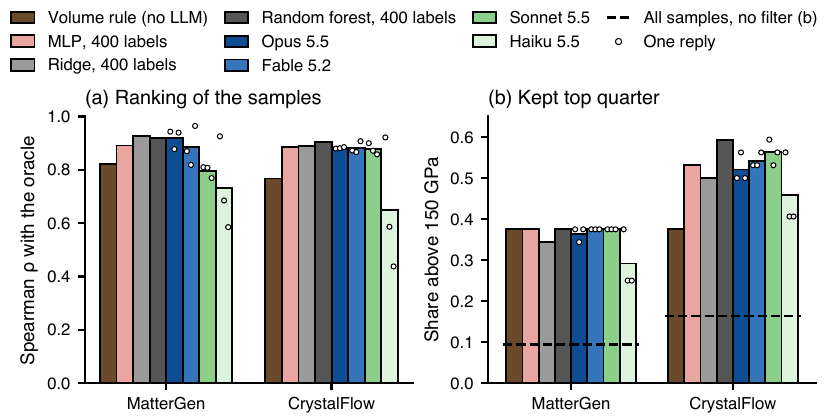}
\caption{Scoring the samples of two crystal generators with a predictor that used no data from either; bulk modulus, 128 samples of each base model. The three trained models saw 400 Materials Project crystals with DFT labels. (a) Rank correlation with the oracle. (b) Share of crystals above 150~GPa among the top quarter kept by each scorer; dashed lines mark the share among all samples.}
\label{fig:filter}
\end{figure}

The last use involves no surrogate. A generative model proposes crystals, most of which are not worth an oracle call, and a written network can be used to choose which of them to send. On MatterGen samples the Opus networks rank bulk modulus at $\rho = 0.92$ with the oracle and on CrystalFlow samples at $0.88$, against $0.82$ and $0.77$ for the rule that a small volume per atom means a stiff crystal (Figure~\ref{fig:filter}). Three models trained on 400 Materials Project crystals do as well. A network of the same size as the written one ranks at $0.89$ on MatterGen samples and $0.88$ on CrystalFlow samples, a ridge regression on the logarithm of the bulk modulus at $0.93$ and $0.89$, and a random forest at $0.92$ and $0.90$. The written filter therefore matches what 400 labeled crystals give a standard model and does not exceed it, and its advantage is that it needs none. Keeping the top quarter of CrystalFlow samples by the written score raises the share of crystals above 150~GPa from $0.16$ to $0.52$--$0.56$ for the Opus, Fable and Sonnet networks (Haiku $0.46$; volume rule $0.38$; trained network $0.53$; ridge $0.50$; random forest $0.59$), and half of all such crystals are found in 17 or 18 oracle calls instead of 59 (volume rule: 28). On MatterGen the top quarter by the Opus score has a mean bulk modulus of $144$~GPa against $72$ for all samples and $147$ for the best possible quarter, though the share above 150~GPa in that quarter is no higher than with the volume rule ($0.36$ against $0.38$), because few MatterGen samples exceed the threshold at all ($0.09$). For heat capacity the Dulong--Petit rule ranks the samples of both generators perfectly and the networks match it.

\subsection{Failures}
\label{sec:failures}
 On the yield strength of multi-principal-element alloys the Opus and Fable networks rank the pool well ($\rho = 0.69$ and $0.63$) and still make the optimization worse than with no prior, in 14 of 20 seeds (Table~\ref{tab:bo}; $p = 0.03$ and $0.02$ before Holm correction, not significant after). The networks are weakest on the best tenth of that pool, of which they place only $0.31$ and $0.17$ in their own top tenth. Inside that tenth their predictions carry no ranking (rank correlation $0.02$ and $0.00$) and are low by 1.3 to 1.5 standard deviations of the pool. A high rank correlation over the pool is therefore not sufficient for a prior mean to help the optimizer, because the optimizer has to reach the top tenth of the pool and the correlation says little about the ordering there.

On the $\gamma'$ fraction of nickel superalloys the Fable networks written without facts raise the regret to $1.82$ times that of the plain process ($p<0.05$), the one significant loss among 165 prior--target pairs on the Thermo-Calc set. The same model given the binary and ternary values does not lose ($0.68$). Among targets on which the plain process has real regret, no prior of any kind helps on the $\alpha$ transus of titanium alloys, which no predictor ranks (relative areas $0.68$ to $1.20$, none significant), or on the freezing range of the Al--Cr--Mo--Nb--Ti family ($0.88$ to $1.19$).

On heat capacity per gram, and on the solidus and freezing range of most alloy families, a rule, or the extrapolation from computed sub-system data, is as good as or better than a written network without using any label of the task. The network is worth its cost where no such rule exists.

\subsection{Cost}
On the experimental alloy tasks a reply from Opus~5.5 took a median of about 4 minutes and 73{,}000 output tokens, almost all of it reasoning. Fable~5.2 took under a minute and about 9{,}400 tokens, and the two smaller models about 20 seconds. A network is written once per task, and its evaluation afterwards is that of any network of its size.

\section{Discussion}

A written network holds no information that was not in the language model or in its prompt, and the prompt contains no label. What it adds to a discovery loop is prior knowledge in the form of a function of the right inputs, and that function carries trends, thresholds and interactions that a constant mean cannot. The accounts the models give of their own networks support this reading. They describe hidden units as named effects (solid-solution strengthening per element, a threshold above which a second phase forms, the depression of a solidus by a eutectic former), and the replies of one model agree with each other. In the comparison with the sub-system extrapolation, the written networks are behind on the solidus and the freezing range and ahead on average on the phase fractions, although that lead is not significant over targets. The comparison therefore suggests, without establishing it, that what the model contributes is a picture of which phases form in the multicomponent system rather than the level of a quantity that varies smoothly with composition.

A written network is safer in function space than in weight space. Used as a mean, a written network shifts the predictions of the surrogate and does nothing else, and when it is uninformative the fitted slope reduces its contribution to zero. This is why, across 38 optimization targets and the two strongest models, it led to only one significant loss and one consistent loss that did not reach significance. In weight space it constrains the surrogate to a parameterization that a language model chose for legibility, a few large units, which is a poor center for a prior and a poor start for a sampler. Using the hidden layer as inputs carries the model's choice of features into the surrogate without committing it to the output weights, and the surrogate then needs labels to make use of them.

Because the network is written from the names of the inputs and the target, it can only be used where the inputs are named quantities, and because it is a prior, its contribution lasts for as long as the surrogate has less information than it does, which in these campaigns was the first tens of labels. The cost is one model call per task, which for the strongest model is minutes of reasoning, and on a target that a rule already ranks, such as heat capacity per gram with the Dulong--Petit value, the call has nothing to add. In the arm with phase-diagram facts the supplied binary and ternary values appear in the written networks, and the level of the solidus becomes right as written, so calculated or measured data can be given to the model together with the task.

All language-model results use the models of one vendor, and the two smaller models of that vendor wrote networks that are weaker on average and less even. A written network may partly reflect recall of published data, since the experimental alloy sets are public and the prompts name them, and for this reason we added targets computed for this study, although the thermodynamic assessments behind them are built from literature the models were trained on. The self-reports of the networks also show calibration against remembered values of alloys that are in the pools, so the experimental results measure recall as well as reasoning. Tool use was excluded by instruction and by an audit of every transcript rather than by a technical block, so a tool call that left no trace in the transcript would not have been detected. The Thermo-Calc targets are equilibrium quantities from one software version and its databases, so they test agreement with those assessments, and several phase-fraction targets leave an optimizer little to gain. The optimization pools are fixed and hold 152 to 722 candidates. The references matter as much as the networks. The rule of mixtures we first chose for alloys was weak, and against a better descriptor the margin of the written networks shrinks, while against models trained on a few hundred labeled crystals the written filter is level and no better. The crystals added during the loops are labeled with a machine-learned potential, as in the loop they come from, while the starting set for bulk modulus carries DFT labels, and because the runs of a crystal task share that starting set they give six (bulk modulus) or five (heat capacity) paired comparisons and no independent replicates. Significance in the optimization tables is over 20 seeds with the written replies held fixed, and says nothing about the variation between replies. The main study uses networks of one size and one prompt wording. Larger networks were tried on four tasks with one model only, and variants of the prompt were not studied.

\section{Conclusion}

In this work we have examined whether weight-space programming, in which a language model writes the weights of a network directly as text, applies to the tasks of materials discovery, where the inputs are compositions and descriptors with physical names and the labels that would train a model are the very measurements the campaign has yet to make. We asked four general-purpose models to write a network of 16 hidden units for five experimental alloy properties, 35 phase-equilibrium targets computed with Thermo-Calc and two crystal properties, and we measured the written networks as predictors with no labels, as the mean of a Gaussian process and of a Bayesian neural network, as a prior for Bayesian optimization and as a screen for the samples of two crystal generators. The findings suggest that a strong model holds enough of the qualitative knowledge of a metallurgist, trends, thresholds and the appearance of phases, to supply the prior that a campaign lacks at its start, and that this knowledge is most valuable in the first tens of labels, on targets that no rule already ranks. They also suggest that the right place for such a prior is the mean function, where a fitted slope removes it when it is wrong, and that its value should be judged on the top of the candidate pool, since a network that ranked a pool well still misled the optimizer on its best members. Whether the method extends to models from other vendors, to inputs without physical names and to the processing and microstructure variables of a real campaign remains to be established.

\section*{Code and data availability}
The code, every prompt, every raw reply with the model's own account, the scores and the commands that produce each number will be released under the MIT license at \url{https://github.com/sheikhahnaf/weightspace-programming-for-material-science}. The experimental alloy data are public \cite{borg2020mpea,salvador2022tialloys}. The compositions and phase-equilibrium values computed with Thermo-Calc, and the networks written from sub-system values, are not redistributed under the terms of its license. The release contains the script that regenerates them with Thermo-Calc 2025b and the databases named below, together with the aggregate results. Networks written without sub-system values are included for all 35 targets.

\section*{Acknowledgments}
The authors acknowledge support from the U.S.\ Department of Energy ARPA-E CHADWICK Program through Project DE-AR0001988 and from the U.S.\ National Science Foundation under Award DMR-2534344 (Autonomous Robotic Metallurgist: An NSF Materials Innovation Platform, ARM-MIP). Portions of this research were conducted with the advanced computing resources provided by Texas A\&M High Performance Research Computing.

\section*{Generative AI Statement}
The manuscript text, figures and analysis code were drafted with an AI assistant under the authors' direction, and the text was then extensively edited by the authors. The authors reviewed all content and take responsibility for it.

\bibliographystyle{unsrtnat}
\bibliography{refs}

\clearpage
\appendix
\setcounter{section}{0}\setcounter{figure}{0}\setcounter{table}{0}
\renewcommand{\thesection}{S\arabic{section}}
\renewcommand{\thefigure}{S\arabic{figure}}
\renewcommand{\thetable}{S\arabic{table}}
\makeatletter
\@ifundefined{theHsection}{}{\renewcommand{\theHsection}{S\arabic{section}}}
\@ifundefined{theHfigure}{}{\renewcommand{\theHfigure}{S\arabic{figure}}}
\@ifundefined{theHtable}{}{\renewcommand{\theHtable}{S\arabic{table}}}
\makeatother
\providecommand{\suppneed}[1]{\par\penalty0\relax\dimen0=\pagegoal \advance\dimen0 by -\pagetotal \ifdim\dimen0<#1\relax\newpage\fi}
\providecommand{\suppentry}{\suppneed{1.5in}}
\lstset{basicstyle=\ttfamily\scriptsize, breaklines=true, breakatwhitespace=false, columns=fullflexible, keepspaces=true,
        frame=single, framesep=3pt, rulecolor=\color{black!30}, backgroundcolor=\color{black!3}, xleftmargin=2pt, xrightmargin=2pt,
        aboveskip=4pt, belowskip=4pt, breakindent=0pt}
\begin{center}{\Large\bfseries Supplementary material}\end{center}
\addcontentsline{toc}{section}{Supplementary material}

Every language-model run in this paper is a folder in the code repository (the replies written from Thermo-Calc sub-system values are kept by the authors and are not part of the public release) holding the prompt (\texttt{prompt.txt}), the raw reply (\texttt{reply.txt}), the model's own account of how it arrived at the reply (\texttt{self\_report.txt}), the model identifier, token counts and duration read from the session transcript (\texttt{meta.json}), and the score (\texttt{score.json}). Text reproduced here is verbatim. The changes made for typesetting are marked where they occur: long blocks of numbers are cut in the middle, long files are cut at the end, and characters outside the basic Latin set are written as ASCII.

\section{Prompts, replies and written networks for every task}\label{supp:written}
This section covers the 42 tasks on which a model writes every weight of a network with one hidden layer of 16 ReLU units: 5 experimental alloy tasks, 35 Thermo-Calc targets in nine alloy systems and 2 crystal tasks. It gives the instructions every session received (Section~\ref{supp:wrapper}), one complete prompt per task family (Section~\ref{supp:prompts}), and then one entry per task (Sections~\ref{supp:alloys} to~\ref{supp:crystals}). Each entry shows the first valid reply of Opus~5.5 that has a self-report, which is replicate~1 unless the entry says otherwise. An entry has four parts, in this order: the part of the prompt that is specific to the task (the first box), with the sentences shared by every prompt of its family replaced by \texttt{[...]} and blank lines removed; a diagram of the written network; the raw reply with the middle rows of the first weight matrix omitted; and the model's own account. The task name is the part of the run-folder name between the arm prefix and the model.

\paragraph{How to read the diagrams} Every line is one weight parsed from the reply file. Blue lines are positive weights and red lines negative. Line width grows with the absolute weight divided by the largest absolute weight in the same layer, and weights below $10^{-6}$ of that largest value are not drawn. A filled hidden unit has a nonzero weight to the output and is counted as used; an open one has a zero output weight, and first-layer weights into it are drawn in gray because they cannot change the output. Hidden-unit biases are not drawn; the output bias is printed. Inputs are not standardized, so a thin line from an input with a large numerical range can still carry a large contribution.

\subsection{Instructions given to every model session}\label{supp:wrapper}
Each session receives one line naming an instruction file and a run name (the earliest sessions, the Opus alloy runs and two Fable replies, received an earlier form of these instructions). The file is the same for every experiment apart from the folder path it names. \texttt{experiments/\allowbreak e7\_\allowbreak calphad\_\allowbreak prior/\allowbreak WRAPPER.md}:
\begin{lstlisting}
# Instructions for the model under test

You are the subject of an experiment on whether a language model can read, write or edit neural-network weights by reasoning alone. The rules are strict and your tool use is audited afterwards.

You were given a run name. Call it RUN. Its folder is
`<repository>/experiments/e7_calphad_prior/runs/RUN/`.

1. Read `prompt.txt` in that folder with the Read tool, once.
2. Do the task in it purely by thinking. No code, no Bash, no calculator, no web, and no other file or directory than the two you write below.
3. Write `reply.txt` in that folder. It must contain ONLY what the prompt's "Reply format" section asks for: no commentary, no code fences.
4. Write `self_report.txt` in that folder: two to four sentences on how you arrived at your reply, and, if the task asked you to find something, how confident you are as a percentage.
5. Answer with the single word `done`.

Including the read of this file, that is exactly four tool calls.
\end{lstlisting}

\subsection{One complete prompt per task family}\label{supp:prompts}
Every weight in the weights block of a writing prompt is zero; the middle rows of the first matrix are omitted here and the omission is marked.
\subsubsection{Experimental alloy task: titanium-alloy hardness}
{\raggedright File \texttt{experiments/\allowbreak e1\_\allowbreak surrogate\_\allowbreak head/\allowbreak runs/\allowbreak write\_\allowbreak \_\allowbreak ti\_\allowbreak alloy\_\allowbreak hardness\_\allowbreak \_\allowbreak opus\_\allowbreak \_\allowbreak rep1/\allowbreak prompt.txt}.\par}
\begin{lstlisting}
You are given a small neural network as its raw weights. You cannot run it and you cannot train it.

# Task
Write the weights of this network from what you know about materials. No training data is given. Inputs are atomic fractions of Al, Cr, Cu, Fe, Hf, Mn, Mo, Nb, Si, Sn, Ta, Ti, V, Zr, and the summed fraction of all other elements; each row sums to 1. The target is Vickers hardness in HV. Measured titanium alloys (Salvador et al., Sci. Data 9:188, 2022). 152 alloys.

Input order: x[0] = x_Al, x[1] = x_Cr, x[2] = x_Cu, x[3] = x_Fe, x[4] = x_Hf, x[5] = x_Mn, x[6] = x_Mo, x[7] = x_Nb, x[8] = x_Si, x[9] = x_Sn, x[10] = x_Ta, x[11] = x_Ti, x[12] = x_V, x[13] = x_Zr, x[14] = x_other.

The output y must be your best estimate of the Vickers hardness in HV. The network will be used as the prior mean of a Gaussian process that is then given a few measured alloys, so getting the trends and the rough scale right matters more than exact values.

# Architecture
Input x has 15 entries. h0 = x.
h1 = relu(net.0.weight @ h0 + net.0.bias)   (net.0.weight is 16 x 15, net.0.bias has 16 entries)
y = net.2.weight @ h1 + net.2.bias   (net.2.weight is 1 x 16, net.2.bias has 1 entries)

# Weights
## net.0.weight shape=(16, 15)
0 0 0 0 0 0 0 0 0 0 0 0 0 0 0
0 0 0 0 0 0 0 0 0 0 0 0 0 0 0
0 0 0 0 0 0 0 0 0 0 0 0 0 0 0
0 0 0 0 0 0 0 0 0 0 0 0 0 0 0
0 0 0 0 0 0 0 0 0 0 0 0 0 0 0
[... 9 lines of numbers omitted here; the full file is in the repository ...]
0 0 0 0 0 0 0 0 0 0 0 0 0 0 0
0 0 0 0 0 0 0 0 0 0 0 0 0 0 0
## net.0.bias shape=(16,)
0 0 0 0 0 0 0 0 0 0 0 0 0 0 0 0
## net.2.weight shape=(1, 16)
0 0 0 0 0 0 0 0 0 0 0 0 0 0 0 0
## net.2.bias shape=(1,)
0

# Reply format
Reply with the complete weights in exactly the format shown above: one '## name shape=(...)' header per tensor, then its rows of numbers. No other text.
\end{lstlisting}

\subsubsection{Thermo-Calc task without phase-diagram facts: nickel-base superalloy solidus}
{\raggedright File \texttt{experiments/\allowbreak e7\_\allowbreak calphad\_\allowbreak prior/\allowbreak runs/\allowbreak write\_\allowbreak \_\allowbreak ni\_\allowbreak superalloy\_\allowbreak solidus\_\allowbreak \_\allowbreak opus\_\allowbreak \_\allowbreak rep1/\allowbreak prompt.txt}.\par}
\begin{lstlisting}
You are given a small neural network as its raw weights. You cannot run it and you cannot train it.

# Task
Write the weights of this network from what you know about materials. No training data is given. Inputs are atomic fractions of Ni, Al, Co, Cr, Mo, Ti, W in nickel-base superalloys Ni-Al-Co-Cr-Mo-Ti-W; each row sums to 1. Composition range: Ni is the balance and at least 50 at.%; solutes in at.%: Al 0 to 12 (step 2), Co 0 to 20 (step 5), Cr 0 to 20 (step 5), Mo 0 to 6 (step 2), Ti 0 to 6 (step 2), W 0 to 6 (step 2); at least 4 elements present. The target is the equilibrium solidus temperature in K, calculated at 1 atm with the CALPHAD method (Thermo-Calc, TCNI12 database). 250 alloys.

Input order: x[0] = x_Ni, x[1] = x_Al, x[2] = x_Co, x[3] = x_Cr, x[4] = x_Mo, x[5] = x_Ti, x[6] = x_W.

The output y must be your best estimate of the equilibrium solidus temperature in K. The network will be used as the prior mean of a Gaussian process that is then given a few measured alloys, so getting the trends and the rough scale right matters more than exact values.

# Architecture
Input x has 7 entries. h0 = x.
h1 = relu(net.0.weight @ h0 + net.0.bias)   (net.0.weight is 16 x 7, net.0.bias has 16 entries)
y = net.2.weight @ h1 + net.2.bias   (net.2.weight is 1 x 16, net.2.bias has 1 entries)

# Weights
## net.0.weight shape=(16, 7)
0 0 0 0 0 0 0
0 0 0 0 0 0 0
0 0 0 0 0 0 0
0 0 0 0 0 0 0
0 0 0 0 0 0 0
[... 9 lines of numbers omitted here; the full file is in the repository ...]
0 0 0 0 0 0 0
0 0 0 0 0 0 0
## net.0.bias shape=(16,)
0 0 0 0 0 0 0 0 0 0 0 0 0 0 0 0
## net.2.weight shape=(1, 16)
0 0 0 0 0 0 0 0 0 0 0 0 0 0 0 0
## net.2.bias shape=(1,)
0

# Reply format
Reply with the complete weights in exactly the format shown above: one '## name shape=(...)' header per tensor, then its rows of numbers. No other text.
\end{lstlisting}

\subsubsection{The same Thermo-Calc task with phase-diagram facts}
{\raggedright File \texttt{experiments/\allowbreak e7\_\allowbreak calphad\_\allowbreak prior/\allowbreak runs/\allowbreak writepd\_\allowbreak \_\allowbreak ni\_\allowbreak superalloy\_\allowbreak solidus\_\allowbreak \_\allowbreak opus\_\allowbreak \_\allowbreak rep1/\allowbreak prompt.txt}.\par}
The lines of computed sub-system values are not reproduced here; their format is described in the prompt itself.
\begin{lstlisting}
You are given a small neural network as its raw weights. You cannot run it and you cannot train it.

# Task
Write the weights of this network from what you know about materials. No training data is given. Inputs are atomic fractions of Ni, Al, Co, Cr, Mo, Ti, W in nickel-base superalloys Ni-Al-Co-Cr-Mo-Ti-W; each row sums to 1. Composition range: Ni is the balance and at least 50 at.%; solutes in at.%: Al 0 to 12 (step 2), Co 0 to 20 (step 5), Cr 0 to 20 (step 5), Mo 0 to 6 (step 2), Ti 0 to 6 (step 2), W 0 to 6 (step 2); at least 4 elements present. The target is the equilibrium solidus temperature in K, calculated at 1 atm with the CALPHAD method (Thermo-Calc, TCNI12 database). 250 alloys.

Input order: x[0] = x_Ni, x[1] = x_Al, x[2] = x_Co, x[3] = x_Cr, x[4] = x_Mo, x[5] = x_Ti, x[6] = x_W.

The output y must be your best estimate of the equilibrium solidus temperature in K. The network will be used as the prior mean of a Gaussian process that is then given a few measured alloys, so getting the trends and the rough scale right matters more than exact values.

# Phase-diagram facts for sub-systems (computed with CALPHAD)
These facts cover only sub-systems with at most three elements: pure Ni, Ni with one solute at half and at the whole of that solute's range, and Ni with two solutes, both at half and both at the whole of their ranges. Nothing here was calculated for a composition of the kind you are asked about (four or more elements). Source: Thermo-Calc, TCNI12 database, 1 atm, the same database as the target.

Each line: composition in at.%; s/l = solidus/liquidus in K; then the stable phases at 1173 K with their mole fractions, where bcc, fcc and hcp are the disordered solid solutions, B2 and L12 are ordered, and the same phase listed twice is a miscibility gap.

[43 lines of sub-system values computed with Thermo-Calc, one composition per line, omitted]

# Architecture
Input x has 7 entries. h0 = x.
h1 = relu(net.0.weight @ h0 + net.0.bias)   (net.0.weight is 16 x 7, net.0.bias has 16 entries)
y = net.2.weight @ h1 + net.2.bias   (net.2.weight is 1 x 16, net.2.bias has 1 entries)

# Weights
## net.0.weight shape=(16, 7)
0 0 0 0 0 0 0
0 0 0 0 0 0 0
0 0 0 0 0 0 0
0 0 0 0 0 0 0
0 0 0 0 0 0 0
[... 9 lines of numbers omitted here; the full file is in the repository ...]
0 0 0 0 0 0 0
0 0 0 0 0 0 0
## net.0.bias shape=(16,)
0 0 0 0 0 0 0 0 0 0 0 0 0 0 0 0
## net.2.weight shape=(1, 16)
0 0 0 0 0 0 0 0 0 0 0 0 0 0 0 0
## net.2.bias shape=(1,)
0

# Reply format
Reply with the complete weights in exactly the format shown above: one '## name shape=(...)' header per tensor, then its rows of numbers. No other text.
\end{lstlisting}

\subsubsection{Crystal task: bulk modulus}
{\raggedright File \texttt{experiments/\allowbreak e3\_\allowbreak crystal\_\allowbreak filter/\allowbreak runs/\allowbreak filter\_\allowbreak K\_\allowbreak \_\allowbreak opus\_\allowbreak \_\allowbreak rep1/\allowbreak prompt.txt}.\par}
\begin{lstlisting}
You are given a small neural network as its raw weights. You cannot run it and you cannot train it.

# Task
Write the weights of this network from what you know about materials. No training data is given. The network scores crystals proposed by a generative model (inorganic, up to 20 atoms per cell, any elements) so that only the most promising are sent to an expensive calculation. The output y must be your best estimate of the crystal's bulk modulus in GPa; what matters most is that crystals with a high true value get a high score.

Inputs, not standardised:
x[0] = volume_per_atom_A3
x[1] = mean_atomic_mass_u
x[2] = mean_pauling_electronegativity
x[3] = mean_elemental_melting_point_K
x[4] = frac_alkali
x[5] = frac_alkaline_earth
x[6] = frac_3d_transition
x[7] = frac_4d_5d_transition
x[8] = frac_lanthanide_actinide
x[9] = frac_p_block_metal
x[10] = frac_metalloid
x[11] = frac_C_N_O
x[12] = frac_S_Se_P
x[13] = frac_halogen
x[14] = frac_H

volume_per_atom_A3 is the cell volume divided by the number of atoms, as generated (not relaxed). The means are over the atoms of the cell. The fractions are:
frac_alkali: atomic fraction of Li Na K Rb Cs
frac_alkaline_earth: atomic fraction of Be Mg Ca Sr Ba
frac_3d_transition: atomic fraction of Sc Ti V Cr Mn Fe Co Ni Cu Zn
frac_4d_5d_transition: atomic fraction of Y Zr Nb Mo Tc Ru Rh Pd Ag Cd Hf Ta W Re Os Ir Pt Au Hg
frac_lanthanide_actinide: atomic fraction of La Ce Pr Nd Pm Sm Eu Gd Tb Dy Ho Er Tm Yb Lu Ac Th Pa U Np Pu
frac_p_block_metal: atomic fraction of Al Ga In Sn Tl Pb Bi
frac_metalloid: atomic fraction of B Si Ge As Sb Te
frac_C_N_O: atomic fraction of C N O
frac_S_Se_P: atomic fraction of S Se P
frac_halogen: atomic fraction of F Cl Br I
frac_H: atomic fraction of H
Typical ranges: volume per atom 8 to 60, mean atomic mass 7 to 200, mean electronegativity 0.8 to 3.5, mean melting point 300 to 3700.

# Architecture
Input x has 15 entries. h0 = x.
h1 = relu(net.0.weight @ h0 + net.0.bias)   (net.0.weight is 16 x 15, net.0.bias has 16 entries)
y = net.2.weight @ h1 + net.2.bias   (net.2.weight is 1 x 16, net.2.bias has 1 entries)

# Weights
## net.0.weight shape=(16, 15)
0 0 0 0 0 0 0 0 0 0 0 0 0 0 0
0 0 0 0 0 0 0 0 0 0 0 0 0 0 0
0 0 0 0 0 0 0 0 0 0 0 0 0 0 0
0 0 0 0 0 0 0 0 0 0 0 0 0 0 0
0 0 0 0 0 0 0 0 0 0 0 0 0 0 0
[... 9 lines of numbers omitted here; the full file is in the repository ...]
0 0 0 0 0 0 0 0 0 0 0 0 0 0 0
0 0 0 0 0 0 0 0 0 0 0 0 0 0 0
## net.0.bias shape=(16,)
0 0 0 0 0 0 0 0 0 0 0 0 0 0 0 0
## net.2.weight shape=(1, 16)
0 0 0 0 0 0 0 0 0 0 0 0 0 0 0 0
## net.2.bias shape=(1,)
0

# Reply format
Reply with the complete weights in exactly the format shown above: one '## name shape=(...)' header per tensor, then its rows of numbers. No other text.
\end{lstlisting}

\subsection{Experimental alloy tasks}\label{supp:alloys}
Run folders are under \texttt{experiments/\allowbreak e1\_\allowbreak surrogate\_\allowbreak head/\allowbreak runs}.
\suppneed{2.2in}\subsubsection{Titanium alloys}
\suppentry{\raggedright\paragraph{Young's modulus (GPa)} Run folder \texttt{write\_\allowbreak \_\allowbreak ti\_\allowbreak alloy\_\allowbreak youngs\_\allowbreak modulus\_\allowbreak \_\allowbreak opus\_\allowbreak \_\allowbreak rep1}.\par}
\begin{lstlisting}
[...] Inputs are atomic fractions of Al, Cr, Cu, Fe, Hf, Mn, Mo, Nb, Si, Sn, Ta, Ti, V, Zr, and the summed fraction of all other elements; each row sums to 1. The target is Young's modulus in GPa. Measured titanium alloys (Salvador et al., Sci. Data 9:188, 2022). 237 alloys.
Input order: x[0] = x_Al, x[1] = x_Cr, x[2] = x_Cu, x[3] = x_Fe, x[4] = x_Hf, x[5] = x_Mn, x[6] = x_Mo, x[7] = x_Nb, x[8] = x_Si, x[9] = x_Sn, x[10] = x_Ta, x[11] = x_Ti, x[12] = x_V, x[13] = x_Zr, x[14] = x_other.
The output y must be your best estimate of the Young's modulus in GPa. [...]
\end{lstlisting}

{\centering\resizebox{\ifdim\width>\linewidth\linewidth\else\width\fi}{!}{\includegraphics{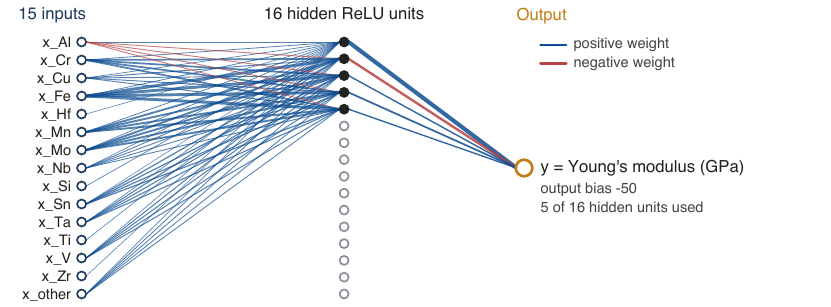}}\par}
\emph{Raw reply, shortened}
\begin{lstlisting}
## net.0.weight shape=(16, 15)
2.16 4.06 2.36 4.06 1.06 4.06 3.06 1.46 2.56 0.36 2.06 1.56 2.06 0.96 2.56
-5.6 17.3 6.0 34.0 0 19.5 20.0 5.4 0 10.0 8.3 0 7.1 0 5.0
[... 13 rows omitted ...]
0 0 0 0 0 0 0 0 0 0 0 0 0 0 0
## net.0.bias shape=(16,)
0 0 -0.5 -1.2 -1.6 0 0 0 0 0 0 0 0 0 0 0
## net.2.weight shape=(1, 16)
100 -50 20 30 20 0 0 0 0 0 0 0 0 0 0 0
## net.2.bias shape=(1,)
-50
\end{lstlisting}

\emph{The model's own account}
\begin{lstlisting}
Hidden unit 0 is an always-positive rule-of-mixtures term: pure Ti is 106 GPa, and each element gets a slope in GPa per at% (Al +0.6; Mo +1.5; Fe, Cr, Mn +2.5; Ta and V +0.5; Nb -0.1; Zr -0.6; Hf -0.5; Sn -1.2). A shift of +0.5 is added to every weight and removed again through the output bias, which keeps the unit active. Units 1-4 are ReLUs that share one beta-stability index, an at%-based molybdenum equivalent (Mo 2.0, Fe 3.4, Cr 1.73, Mn 1.95, Ta 0.83, V 0.71, Nb 0.54, Sn 1.0, Cu 0.6, Al -0.56), with hinges at index values of 0, 5, 12 and 16. Together they make a piecewise-linear drop: about 5 GPa per unit through the alpha-prime martensite range and 3 per unit approaching the metastable-beta minimum near 60 GPa, then a plateau, then a rise of +2 per unit for strongly stabilized beta. I checked the result by hand against remembered moduli for CP-Ti, Ti-6Al-4V, Ti-6Al-7Nb, Ti-13Nb-13Zr, TNZT, TNTZ, Ti-15Mo, TMZF and Ti-Ta/Ti-Nb binaries; the predictions came within roughly 10 GPa of those values.
\end{lstlisting}

\suppentry{\raggedright\paragraph{Yield strength (MPa)} Run folder \texttt{write\_\allowbreak \_\allowbreak ti\_\allowbreak alloy\_\allowbreak yield\_\allowbreak strength\_\allowbreak \_\allowbreak opus\_\allowbreak \_\allowbreak rep1}.\par}
\begin{lstlisting}
[...] Inputs are atomic fractions of Al, Cr, Cu, Fe, Hf, Mn, Mo, Nb, Si, Sn, Ta, Ti, V, Zr, and the summed fraction of all other elements; each row sums to 1. The target is yield strength in MPa. Measured titanium alloys (Salvador et al., Sci. Data 9:188, 2022). 271 alloys.
Input order: x[0] = x_Al, x[1] = x_Cr, x[2] = x_Cu, x[3] = x_Fe, x[4] = x_Hf, x[5] = x_Mn, x[6] = x_Mo, x[7] = x_Nb, x[8] = x_Si, x[9] = x_Sn, x[10] = x_Ta, x[11] = x_Ti, x[12] = x_V, x[13] = x_Zr, x[14] = x_other.
The output y must be your best estimate of the yield strength in MPa. [...]
\end{lstlisting}

{\centering\resizebox{\ifdim\width>\linewidth\linewidth\else\width\fi}{!}{\includegraphics{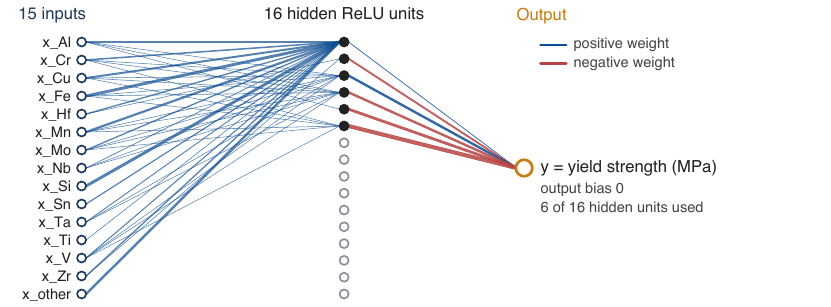}}\par}
\emph{Raw reply, shortened}
\begin{lstlisting}
## net.0.weight shape=(16, 15)
38.5 43.5 63.5 103.5 33.5 73.5 33.5 8.5 103.5 28.5 18.5 3.5 23.5 33.5 83.5
0 0 0 0 1 0 0 0 0 0 0 0 0 1 0
[... 13 rows omitted ...]
0 0 0 0 0 0 0 0 0 0 0 0 0 0 0
## net.0.bias shape=(16,)
0 -0.15 -0.12 -0.17 -0.13 -0.05 0 0 0 0 0 0 0 0 0 0
## net.2.weight shape=(1, 16)
100 -2700 4000 -4000 -5000 -8000 0 0 0 0 0 0 0 0 0 0
## net.2.bias shape=(1,)
0
\end{lstlisting}

\emph{The model's own account}
\begin{lstlisting}
Hidden unit 0 is an always-active additive solid-solution model: pure Ti contributes about 350 MPa (CP Ti), and each element adds a per-atomic-fraction increment estimated from known alloys (strong for Fe, Si, Mn, Cu, interstitial-laden "other"; moderate for Al, Cr, Mo, V, Zr, Hf, Sn; weak for Nb and Ta, which gives low-strength biomedical beta alloys). Unit 1 saturates Zr+Hf strengthening above about 15 at%, and unit 4 caps Al above about 13 at% so that Ti3Al and TiAl compositions do not explode. Units 2 and 3 form a bounded (at most 200 MPa) bonus for alpha+beta and near-beta alloys that combine Al with a Mo-equivalent of beta stabilisers (Ti-6246, Ti-5553, Beta C), and unit 5 reduces the slope once beta-stabiliser strengthening exceeds about 500 MPa. I checked the scale by hand against Ti-6Al-4V (~900), Ti-5553 (~1160), Ti-15Mo (~590), TNZT (~640), Ti-24Nb-4Zr-8Sn (~600) and IMI 834 (~920).
\end{lstlisting}

\suppentry{\raggedright\paragraph{Vickers hardness (HV)} Run folder \texttt{write\_\allowbreak \_\allowbreak ti\_\allowbreak alloy\_\allowbreak hardness\_\allowbreak \_\allowbreak opus\_\allowbreak \_\allowbreak rep1}.\par}
\begin{lstlisting}
[...] Inputs are atomic fractions of Al, Cr, Cu, Fe, Hf, Mn, Mo, Nb, Si, Sn, Ta, Ti, V, Zr, and the summed fraction of all other elements; each row sums to 1. The target is Vickers hardness in HV. Measured titanium alloys (Salvador et al., Sci. Data 9:188, 2022). 152 alloys.
Input order: x[0] = x_Al, x[1] = x_Cr, x[2] = x_Cu, x[3] = x_Fe, x[4] = x_Hf, x[5] = x_Mn, x[6] = x_Mo, x[7] = x_Nb, x[8] = x_Si, x[9] = x_Sn, x[10] = x_Ta, x[11] = x_Ti, x[12] = x_V, x[13] = x_Zr, x[14] = x_other.
The output y must be your best estimate of the Vickers hardness in HV. [...]
\end{lstlisting}

{\centering\resizebox{\ifdim\width>\linewidth\linewidth\else\width\fi}{!}{\includegraphics{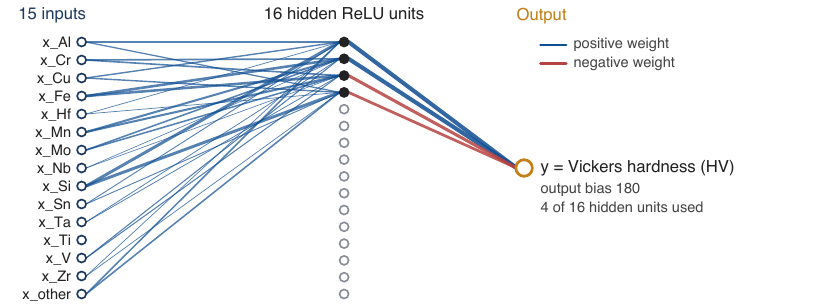}}\par}
\emph{Raw reply, shortened}
\begin{lstlisting}
## net.0.weight shape=(16, 15)
1100 0 1300 0 300 0 0 0 3500 600 0 0 0 300 1500
0 1600 0 3000 0 2200 1500 200 0 0 400 0 800 0 0
[... 13 rows omitted ...]
0 0 0 0 0 0 0 0 0 0 0 0 0 0 0
## net.0.bias shape=(16,)
0 0 -105 -150 0 0 0 0 0 0 0 0 0 0 0 0
## net.2.weight shape=(1, 16)
1 1 -0.7 -0.6 0 0 0 0 0 0 0 0 0 0 0 0
## net.2.bias shape=(1,)
180
\end{lstlisting}

\emph{The model's own account}
\begin{lstlisting}
I built a base of about 180 HV for commercially pure Ti and added two hand-built strengthening channels, each in HV per unit atomic fraction, with Ti weighted zero. The first channel covers alpha/neutral and interstitial-type strengthening (Al 1100, Sn 600, Zr/Hf 300, Cu 1300, Si 3500, other 1500). The second is a Mo-equivalent beta-stabiliser channel (Fe 3000, Mn 2200, Cr 1600, Mo 1500, V 800, Ta 400, Nb 200). A copy of each channel, with a negative bias and a negative output weight, flattens the slope once the alloy becomes stable beta (above about 7 at% Mo-equivalent) or once alpha strengthening passes +150 HV. I checked the scale against rough known values: Ti-6Al-4V about 320 HV, Ti-15Mo about 290 HV, TMZF about 310 HV and TNZT about 245 HV.
\end{lstlisting}

\suppneed{2.2in}\subsubsection{Multi-principal-element alloys}
\suppentry{\raggedright\paragraph{Vickers hardness (HV)} Run folder \texttt{write\_\allowbreak \_\allowbreak mpea\_\allowbreak hardness\_\allowbreak \_\allowbreak opus\_\allowbreak \_\allowbreak rep1}.\par}
\begin{lstlisting}
[...] Inputs are atomic fractions of Al, B, C, Co, Cr, Cu, Fe, Hf, Mn, Mo, Nb, Ni, Pd, Re, Si, Sn, Ta, Ti, V, W, Zr, and the summed fraction of all other elements; each row sums to 1. The target is Vickers hardness in HV. Measured multi-principal-element alloys (Borg et al., Sci. Data 7:430, 2020). 530 alloys.
Input order: x[0] = x_Al, x[1] = x_B, x[2] = x_C, x[3] = x_Co, x[4] = x_Cr, x[5] = x_Cu, x[6] = x_Fe, x[7] = x_Hf, x[8] = x_Mn, x[9] = x_Mo, x[10] = x_Nb, x[11] = x_Ni, x[12] = x_Pd, x[13] = x_Re, x[14] = x_Si, x[15] = x_Sn, x[16] = x_Ta, x[17] = x_Ti, x[18] = x_V, x[19] = x_W, x[20] = x_Zr, x[21] = x_other.
The output y must be your best estimate of the Vickers hardness in HV. [...]
\end{lstlisting}

{\centering\resizebox{\ifdim\width>\linewidth\linewidth\else\width\fi}{!}{\includegraphics{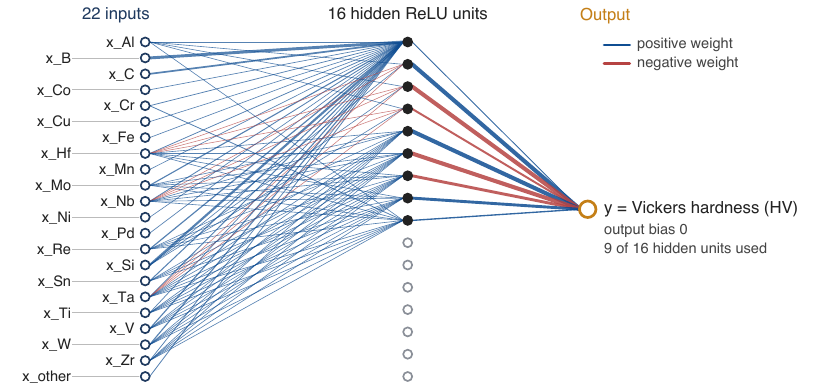}}\par}
\emph{Raw reply, shortened}
\begin{lstlisting}
## net.0.weight shape=(16, 22)
3.5 24 12 1.4 2.2 1.0 1.3 2.5 1.3 6.0 3.5 1.1 1.7 6.5 16.5 5.0 4.5 2.0 6.0 6.1 2.2 3.0
1 0 0 0 0 0 0 -0.25 0 0 -0.25 0 0 0 1 0 -0.25 0 0 0 0 0
[... 13 rows omitted ...]
0 0 0 0 0 0 0 0 0 0 0 0 0 0 0 0 0 0 0 0 0 0
## net.0.bias shape=(16,)
0 -0.09 -0.2 -0.45 -0.02 -0.15 -0.3 -0.5 -1.05 0 0 0 0 0 0 0
## net.2.weight shape=(1, 16)
100 2700 -2700 -1000 2700 -2700 -1755 1755 500 0 0 0 0 0 0 0
## net.2.bias shape=(1,)
0
\end{lstlisting}

\emph{The model's own account}
\begin{lstlisting}
Unit 0 is an always-on rule of mixtures. Its per-element coefficients (in units of 100 HV) were set by hand so that CoCrFeNi comes out near 150 HV, Cantor near 145, NbMoTaW near 500, HfNbTaTiZr near 300 and NbTiVZr near 340. B, C and Si get large coefficients to stand for boride, carbide and silicide hardening. Units 1-3 form a clipped ramp in (x_Al + x_Si), mildly suppressed by Nb, Ta and Hf. It adds about 300 HV between 9 and 20 at.% Al, which follows the FCC-to-BCC/B2 transition in AlxCoCrFeNi (about 160 HV at x=0.3, 230 at x=0.5, about 490 at x=1), and it declines slightly for very Al-rich alloys. Units 4-7 make a trapezoid in a weighted refractory/intermetallic-former fraction (Ti, Nb, Zr, Hf, Ta at full weight, Mo and W at 0.6, V and Re at 0.3): it adds up to about 350 HV for small Ti, Nb or Mo additions to 3d-metal matrices (eta, Laves and sigma phases) and returns to zero for refractory-based alloys, while unit 8 adds Laves-type hardening when Cr sits in a refractory or Al-refractory matrix (CrNbTiZr, CrMo0.5NbTa0.5TiZr).
\end{lstlisting}

\suppentry{\raggedright\paragraph{Room-temperature yield strength (MPa)} Run folder \texttt{write\_\allowbreak \_\allowbreak mpea\_\allowbreak yield\_\allowbreak strength\_\allowbreak \_\allowbreak opus\_\allowbreak \_\allowbreak rep1}.\par}
\begin{lstlisting}
[...] Inputs are atomic fractions of Al, C, Ca, Co, Cr, Cu, Fe, Hf, Li, Mg, Mn, Mo, Nb, Ni, Re, Si, Sn, Ta, Ti, V, W, Zn, Zr, and the summed fraction of all other elements; each row sums to 1. The target is room-temperature yield strength in MPa. Measured multi-principal-element alloys (Borg et al., Sci. Data 7:430, 2020). 722 alloys.
Input order: x[0] = x_Al, x[1] = x_C, x[2] = x_Ca, x[3] = x_Co, x[4] = x_Cr, x[5] = x_Cu, x[6] = x_Fe, x[7] = x_Hf, x[8] = x_Li, x[9] = x_Mg, x[10] = x_Mn, x[11] = x_Mo, x[12] = x_Nb, x[13] = x_Ni, x[14] = x_Re, x[15] = x_Si, x[16] = x_Sn, x[17] = x_Ta, x[18] = x_Ti, x[19] = x_V, x[20] = x_W, x[21] = x_Zn, x[22] = x_Zr, x[23] = x_other.
The output y must be your best estimate of the room-temperature yield strength in MPa. [...]
\end{lstlisting}

{\centering\resizebox{\ifdim\width>\linewidth\linewidth\else\width\fi}{!}{\includegraphics{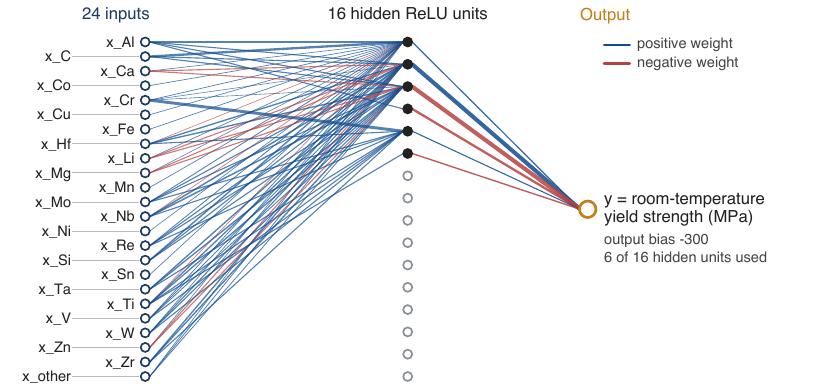}}\par}
\emph{Raw reply, shortened}
\begin{lstlisting}
## net.0.weight shape=(16, 24)
1.2 3.0 0.2 0.3 0.35 0.25 0.3 -0.15 0.3 0.1 0.3 0.6 -0.25 0.25 0.6 1.5 0.3 0.0 -0.25 0.5 0.3 0.2 0.0 0.3
1.0 1.0 -1.0 0 0.1 0 0 1.4 -1.0 -1.0 0 1.0 1.6 0 1.0 1.0 0 1.4 1.3 0.6 1.0 -1.0 1.4 0.5
[... 13 rows omitted ...]
0 0 0 0 0 0 0 0 0 0 0 0 0 0 0 0 0 0 0 0 0 0 0 0
## net.0.bias shape=(16,)
0.3 -0.1 -0.25 -0.35 -1.4 -0.4 0 0 0 0 0 0 0 0 0 0
## net.2.weight shape=(1, 16)
1000 7333 -7333 -3500 800 -1500 0 0 0 0 0 0 0 0 0 0
## net.2.bias shape=(1,)
-300
\end{lstlisting}

\emph{The model's own account}
\begin{lstlisting}
Unit 0 is an always-active rule-of-mixtures term with per-element contributions (3d FCC formers around 250-350 MPa, light elements low, Al/Si/C/Mo/V high, Ti/Nb/Hf slightly negative relative), offset by a bias so it never clips. Units 1-2 form a saturating ramp (+1100 MPa) on a weighted "hardener" index (Al, Ti, Nb, Ta, Zr, Hf, Mo, W, Si, C, V; minus Li/Mg/Zn/Ca) that captures the FCC-to-BCC/intermetallic jump in 3d alloys (e.g. AlxCoCrFeNi at x of about 0.5-1) and puts every refractory alloy on the BCC plateau. Unit 3 lowers Al-rich lightweight alloys (x_Al > 0.35), unit 4 adds a Cr-with-refractory (Laves-type) boost that stays off for Cr in 3d alloys, and unit 5 softens Ti-rich alloys. I checked the weights by hand against remembered anchors: CoCrFeMnNi ~300, AlCoCrFeNi ~1360, HfNbTaTiZr ~970, CrNbTiZr ~1340 and Mg20(MnAlZnCu)80 ~410 MPa.
\end{lstlisting}

\subsection{Thermo-Calc targets}\label{supp:calphad}
Run folders are under \texttt{experiments/\allowbreak e7\_\allowbreak calphad\_\allowbreak prior/\allowbreak runs}. The entries show the prompts without phase-diagram facts (folder prefix \texttt{write}); the runs with facts have the prefix \texttt{writepd} and the same task names.
\suppneed{2.2in}\subsubsection{Al--Co--Cr--Fe--Mn--Ni high-entropy alloys (\texttt{cantor\_al})}
\suppentry{\raggedright\paragraph{Freezing range} Run folder \texttt{write\_\allowbreak \_\allowbreak cantor\_\allowbreak al\_\allowbreak freezing\_\allowbreak range\_\allowbreak \_\allowbreak opus\_\allowbreak \_\allowbreak rep1}.\par}
\begin{lstlisting}
[...] Inputs are atomic fractions of Al, Co, Cr, Fe, Mn, Ni in Al-Co-Cr-Fe-Mn-Ni high-entropy alloys; each row sums to 1. Composition range: every element between 0 and 35 at.% on a 5 at.% grid, at least four elements present. The target is the equilibrium freezing range (liquidus temperature minus solidus temperature) in K, calculated at 1 atm with the CALPHAD method (Thermo-Calc, TCHEA8 database). 400 alloys.
Input order: x[0] = x_Al, x[1] = x_Co, x[2] = x_Cr, x[3] = x_Fe, x[4] = x_Mn, x[5] = x_Ni.
The output y must be your best estimate of the equilibrium freezing range (liquidus temperature minus solidus temperature) in K. [...]
\end{lstlisting}

{\centering\resizebox{\ifdim\width>\linewidth\linewidth\else\width\fi}{!}{\includegraphics{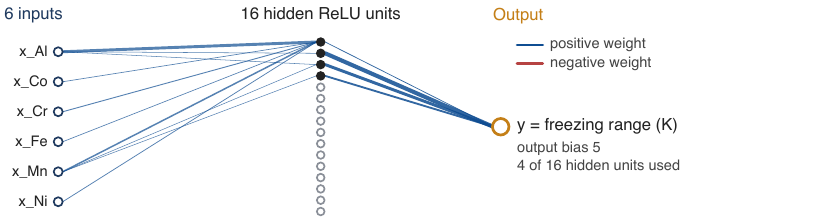}}\par}
\begin{minipage}[t]{0.40\linewidth}
\emph{Raw reply, shortened}
\begin{lstlisting}
## net.0.weight shape=(16, 6)
300 18 50 18 120 30
1 0 0 0 0 0
[... 13 rows omitted ...]
0 0 0 0 0 0
## net.0.bias shape=(16,)
0 -0.15 -0.4 -0.2 0 0 0 0 0 0 0 0 0 0 0 0
## net.2.weight shape=(1, 16)
1 300 200 100 0 0 0 0 0 0 0 0 0 0 0 0
## net.2.bias shape=(1,)
5
\end{lstlisting}
\end{minipage}\hfill\begin{minipage}[t]{0.58\linewidth}
\emph{The model's own account}
\begin{lstlisting}
I wrote the network by hand from known solidification trends. One always-active hidden unit carries a linear per-element contribution: small for Co and Fe, moderate for Ni and Cr, larger for Mn, and largest for Al, which partitions strongly and pulls the solidus down. This gives roughly 45-50 K for equiatomic CoCrFeMnNi and about 25-30 K for Co/Fe/Ni-rich alloys. Three ReLU hinge units add more widening above 15 at.% Al, for Al+Mn-rich alloys, and above 20 at.% Mn, so Al-Mn-rich corners reach roughly 250-300 K. I am moderately confident in the direction of the trends (Al and Mn widen the range) and less confident in the absolute scale, about 50% that it is within a factor of two.
\end{lstlisting}
\end{minipage}\par
\suppentry{\raggedright\paragraph{Intermetallic fraction} Run folder \texttt{write\_\allowbreak \_\allowbreak cantor\_\allowbreak al\_\allowbreak intermetallic\_\allowbreak fraction\_\allowbreak \_\allowbreak opus\_\allowbreak \_\allowbreak rep1}.\par}
\begin{lstlisting}
[...] Inputs are atomic fractions of Al, Co, Cr, Fe, Mn, Ni in Al-Co-Cr-Fe-Mn-Ni high-entropy alloys; each row sums to 1. Composition range: every element between 0 and 35 at.% on a 5 at.% grid, at least four elements present. The target is the equilibrium mole fraction, at 1273 K, of all phases other than liquid and the disordered fcc, bcc and hcp solid solutions (that is: sigma, Laves and other intermetallic phases, and ordered B2 or L12) in mole fraction between 0 and 1, calculated at 1 atm with the CALPHAD method (Thermo-Calc, TCHEA8 database). 397 alloys.
Input order: x[0] = x_Al, x[1] = x_Co, x[2] = x_Cr, x[3] = x_Fe, x[4] = x_Mn, x[5] = x_Ni.
The output y must be your best estimate of the equilibrium mole fraction, at 1273 K, of all phases other than liquid and the disordered fcc, bcc and hcp solid solutions (that is: sigma, Laves and other intermetallic phases, and ordered B2 or L12) in mole fraction between 0 and 1. [...]
\end{lstlisting}

{\centering\resizebox{\ifdim\width>\linewidth\linewidth\else\width\fi}{!}{\includegraphics{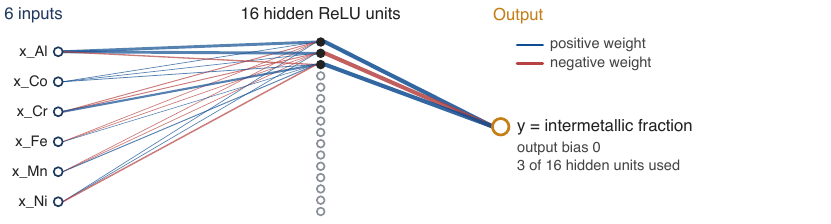}}\par}
\begin{minipage}[t]{0.40\linewidth}
\emph{Raw reply, shortened}
\begin{lstlisting}
## net.0.weight shape=(16, 6)
4.5 0.3 -0.8 -0.3 -0.3 0.3
4.5 0.3 -0.8 -0.3 -0.3 0.3
[... 13 rows omitted ...]
0 0 0 0 0 0
## net.0.bias shape=(16,)
-0.26 -1.21 -0.85 0 0 0 0 0 0 0 0 0 0 0 0 0
## net.2.weight shape=(1, 16)
1 -1 1 0 0 0 0 0 0 0 0 0 0 0 0 0
## net.2.bias shape=(1,)
0
\end{lstlisting}
\end{minipage}\hfill\begin{minipage}[t]{0.58\linewidth}
\emph{The model's own account}
\begin{lstlisting}
I built the network from three hand-designed hidden units and left the other 13 at zero. Unit 0 is an Al-driven B2/L12 term. It switches on above roughly 7-8 at.% Al in CoCrFeNi-like matrices, is helped by Ni and Co and held back by Cr, Fe and Mn, and is set to give about 0.14 for Al0.5CoCrFeNi and about 0.55 for AlCoCrFeNi. Unit 1 has the same weights with a higher bias and enters the output with a negative sign, so the B2 contribution stops rising at 0.95. Unit 2 is a sigma term that grows with Cr, Mn and Co and is suppressed by Ni and Al, so the Cantor alloy gives zero and Cr-rich, Ni-lean alloys give about 0.1-0.5. My confidence that the trends are right is about 65%; my confidence in the absolute scale is lower, about 40%.
\end{lstlisting}
\end{minipage}\par
\suppentry{\raggedright\paragraph{Main phase fraction} Run folder \texttt{write\_\allowbreak \_\allowbreak cantor\_\allowbreak al\_\allowbreak main\_\allowbreak phase\_\allowbreak fraction\_\allowbreak \_\allowbreak opus\_\allowbreak \_\allowbreak rep1}.\par}
\begin{lstlisting}
[...] Inputs are atomic fractions of Al, Co, Cr, Fe, Mn, Ni in Al-Co-Cr-Fe-Mn-Ni high-entropy alloys; each row sums to 1. Composition range: every element between 0 and 35 at.% on a 5 at.% grid, at least four elements present. The target is the equilibrium mole fraction, at 1273 K, of the single most abundant phase (1 for a single-phase alloy; two bcc phases of a miscibility gap count as two phases) in mole fraction between 0 and 1, calculated at 1 atm with the CALPHAD method (Thermo-Calc, TCHEA8 database). 397 alloys.
Input order: x[0] = x_Al, x[1] = x_Co, x[2] = x_Cr, x[3] = x_Fe, x[4] = x_Mn, x[5] = x_Ni.
The output y must be your best estimate of the equilibrium mole fraction, at 1273 K, of the single most abundant phase (1 for a single-phase alloy; two bcc phases of a miscibility gap count as two phases) in mole fraction between 0 and 1. [...]
\end{lstlisting}

{\centering\resizebox{\ifdim\width>\linewidth\linewidth\else\width\fi}{!}{\includegraphics{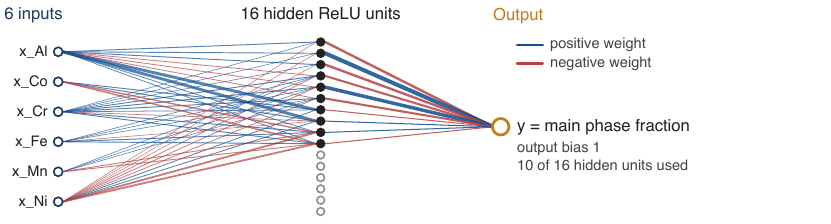}}\par}
\begin{minipage}[t]{0.40\linewidth}
\emph{Raw reply, shortened}
\begin{lstlisting}
## net.0.weight shape=(16, 6)
1 0 0.1 0 0 -0.05
1 0 0.1 0 0 -0.05
[... 13 rows omitted ...]
0 0 0 0 0 0
## net.0.bias shape=(16,)
-0.10 -0.19 -0.28 -0.26 -0.37 -0.48 -1.4564 -1.4564 -0.33 -0.50 0 0 0 0 0 0
## net.2.weight shape=(1, 16)
-5.556 11.111 -5.556 -4.545 9.091 -4.545 -1 1 1 -1 0 0 0 0 0 0
## net.2.bias shape=(1,)
1
\end{lstlisting}
\end{minipage}\hfill\begin{minipage}[t]{0.58\linewidth}
\emph{The model's own account}
\begin{lstlisting}
I built the network by hand as y = 1 minus a few piecewise-linear penalties, each made of ReLU units. Units 1-3 make a V-shaped dip in an Al index (Al + 0.1Cr - 0.05Ni). It starts near 9 at.% Al, where B2 begins to form in the FCC matrix, bottoms at 0.5 near 18 at.%, and recovers by about 27 at.%, where B2 dominates. Units 4-6 make a second dip for Cr-driven sigma or ferrite in Ni-poor, Cr-rich bases (index Cr + 0.2Fe + 0.1Mn - 0.6Ni - 0.7Al). Units 7-8 make a min-type penalty for the B2 + Cr-rich A2 split at high Al, sized by Cr + 0.25Fe and gated by Al and Ni+Co, and units 9-10 give a small capped bonus to Ni- and Co-free Fe-Cr-rich alloys, which should be single BCC. I checked it by hand against about 30 reference compositions, among them AlxCoCrFeNi (x = 0 to 2), Cantor, CrCoNi and Fe-Cr-Mn-Co. The weakest regions are Ni-free high-Al Cr-rich alloys and Fe-Cr-Ni alloys with moderate Al, where I expect errors of 0.2 to 0.3 in phase fraction.
\end{lstlisting}
\end{minipage}\par
\suppentry{\raggedright\paragraph{Solidus} Run folder \texttt{write\_\allowbreak \_\allowbreak cantor\_\allowbreak al\_\allowbreak solidus\_\allowbreak \_\allowbreak opus\_\allowbreak \_\allowbreak rep1}.\par}
\begin{lstlisting}
[...] Inputs are atomic fractions of Al, Co, Cr, Fe, Mn, Ni in Al-Co-Cr-Fe-Mn-Ni high-entropy alloys; each row sums to 1. Composition range: every element between 0 and 35 at.% on a 5 at.% grid, at least four elements present. The target is the equilibrium solidus temperature in K, calculated at 1 atm with the CALPHAD method (Thermo-Calc, TCHEA8 database). 400 alloys.
Input order: x[0] = x_Al, x[1] = x_Co, x[2] = x_Cr, x[3] = x_Fe, x[4] = x_Mn, x[5] = x_Ni.
The output y must be your best estimate of the equilibrium solidus temperature in K. [...]
\end{lstlisting}

{\centering\resizebox{\ifdim\width>\linewidth\linewidth\else\width\fi}{!}{\includegraphics{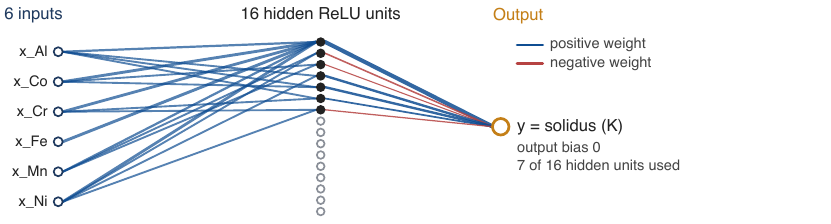}}\par}
\begin{minipage}[t]{0.40\linewidth}
\emph{Raw reply, shortened}
\begin{lstlisting}
## net.0.weight shape=(16, 6)
1.25 1.73 1.69 1.77 1.25 1.64
0 0 0 0 1 1
[... 13 rows omitted ...]
0 0 0 0 0 0
## net.0.bias shape=(16,)
0 -0.3 -0.3 -0.45 -0.45 -0.45 -0.45 0 0 0 0 0 0 0 0 0
## net.2.weight shape=(1, 16)
1000 -250 -110 500 500 300 -160 0 0 0 0 0 0 0 0 0
## net.2.bias shape=(1,)
0
\end{lstlisting}
\end{minipage}\hfill\begin{minipage}[t]{0.58\linewidth}
\emph{The model's own account}
\begin{lstlisting}
I used one always-positive hidden unit as a linear rule of mixtures with effective per-element solidus values (Al 1250, Co 1730, Cr 1690, Fe 1770, Mn 1250, Ni 1640 K) calibrated so that CoCrFeNi comes out near 1700 K, the Cantor alloy near 1580 K and AlCoCrFeNi near 1600 K. Six more hidden units are pairwise hinges relu(x_i + x_j - c), which switch on only when both elements are present in quantity. They encode the low-melting Mn-Ni and Mn-Co minima and the Ni-Cr eutectic as penalties, and B2 NiAl/CoAl formation and Al dissolving in bcc Cr as boosts, so the prediction stops falling once Al exceeds about 20 at.%. The predicted range is roughly 1380-1740 K. I am about 60% confident that the overall scale and the main trends are right (Mn and Al lower the solidus; Fe, Co and Cr raise it). I am less sure of the size of the Al-Ni/Al-Co interactions in TCHEA8.
\end{lstlisting}
\end{minipage}\par
\suppneed{2.2in}\subsubsection{Ni--Al--Co--Cr--Mo--Ti--W nickel-base superalloys (\texttt{ni\_superalloy})}
\suppentry{\raggedright\paragraph{Freezing range} Run folder \texttt{write\_\allowbreak \_\allowbreak ni\_\allowbreak superalloy\_\allowbreak freezing\_\allowbreak range\_\allowbreak \_\allowbreak opus\_\allowbreak \_\allowbreak rep1}.\par}
\begin{lstlisting}
[...] Inputs are atomic fractions of Ni, Al, Co, Cr, Mo, Ti, W in nickel-base superalloys Ni-Al-Co-Cr-Mo-Ti-W; each row sums to 1. Composition range: Ni is the balance and at least 50 at.%; solutes in at.%: Al 0 to 12 (step 2), Co 0 to 20 (step 5), Cr 0 to 20 (step 5), Mo 0 to 6 (step 2), Ti 0 to 6 (step 2), W 0 to 6 (step 2); at least 4 elements present. The target is the equilibrium freezing range (liquidus temperature minus solidus temperature) in K, calculated at 1 atm with the CALPHAD method (Thermo-Calc, TCNI12 database). 250 alloys.
Input order: x[0] = x_Ni, x[1] = x_Al, x[2] = x_Co, x[3] = x_Cr, x[4] = x_Mo, x[5] = x_Ti, x[6] = x_W.
The output y must be your best estimate of the equilibrium freezing range (liquidus temperature minus solidus temperature) in K. [...]
\end{lstlisting}

{\centering\resizebox{\ifdim\width>\linewidth\linewidth\else\width\fi}{!}{\includegraphics{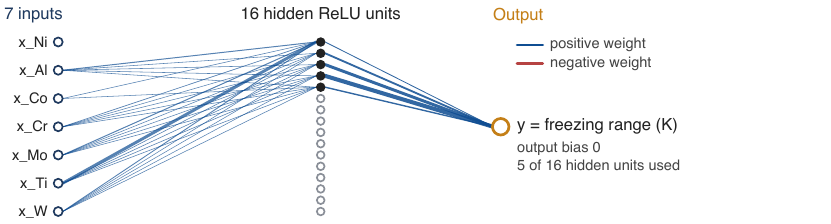}}\par}
\begin{minipage}[t]{0.40\linewidth}
\emph{Raw reply, shortened}
\begin{lstlisting}
## net.0.weight shape=(16, 7)
0 12 1 4 12 60 9
0 1 0 0.25 0.5 1.5 0.3
[... 13 rows omitted ...]
0 0 0 0 0 0 0
## net.0.bias shape=(16,)
0.2 -0.16 -0.2 -0.24 -0.35 0 0 0 0 0 0 0 0 0 0 0
## net.2.weight shape=(1, 16)
10 100 200 300 60 0 0 0 0 0 0 0 0 0 0 0
## net.2.bias shape=(1,)
0
\end{lstlisting}
\end{minipage}\hfill\begin{minipage}[t]{0.58\linewidth}
\emph{The model's own account}
\begin{lstlisting}
I estimated each solute's contribution to the equilibrium freezing range in the dilute limit, |m|*C*(1-k)/k, using remembered binary Ni-X diagrams and typical superalloy partition coefficients (Ti strongly segregating with k about 0.65, Al and Mo mild, W with k above 1, Cr and Co near 1). I then scaled the sum up about 1.8x so that commercial-like compositions such as CMSX-4 land near their CALPHAD ranges of about 40-50 K. One always-active ReLU unit carries this linear part (about 1.2 K/at% Al, 0.1 Co, 0.4 Cr, 1.2 Mo, 6 Ti, 0.9 W, plus a 2 K offset). Three convex hinges on a gamma-prime-former equivalent (Al + 1.5Ti + 0.25Cr + 0.5Mo + 0.3W) above 16, 20 and 24 at% model the steep rise once the last liquid reaches a gamma/gamma-prime eutectic, and a small hinge on total solute above 35 at% adds interaction effects. Predictions run from about 7 K for dilute alloys through about 40 K for mid-range alloys to about 130 K at the most heavily alloyed corner. I am roughly 60% confident that the trends (Ti dominant, then Al and Mo) and the overall scale are right, and less confident about where the eutectic onset falls.
\end{lstlisting}
\end{minipage}\par
\suppentry{\raggedright\paragraph{Gamma prime fraction} Run folder \texttt{write\_\allowbreak \_\allowbreak ni\_\allowbreak superalloy\_\allowbreak gamma\_\allowbreak prime\_\allowbreak fraction\_\allowbreak \_\allowbreak opus\_\allowbreak \_\allowbreak rep1}.\par}
\begin{lstlisting}
[...] Inputs are atomic fractions of Ni, Al, Co, Cr, Mo, Ti, W in nickel-base superalloys Ni-Al-Co-Cr-Mo-Ti-W; each row sums to 1. Composition range: Ni is the balance and at least 50 at.%; solutes in at.%: Al 0 to 12 (step 2), Co 0 to 20 (step 5), Cr 0 to 20 (step 5), Mo 0 to 6 (step 2), Ti 0 to 6 (step 2), W 0 to 6 (step 2); at least 4 elements present. The target is the equilibrium mole fraction of the ordered L12 phase (gamma prime) at 1173 K in mole fraction between 0 and 1, calculated at 1 atm with the CALPHAD method (Thermo-Calc, TCNI12 database). 249 alloys.
Input order: x[0] = x_Ni, x[1] = x_Al, x[2] = x_Co, x[3] = x_Cr, x[4] = x_Mo, x[5] = x_Ti, x[6] = x_W.
The output y must be your best estimate of the equilibrium mole fraction of the ordered L12 phase (gamma prime) at 1173 K in mole fraction between 0 and 1. [...]
\end{lstlisting}

{\centering\resizebox{\ifdim\width>\linewidth\linewidth\else\width\fi}{!}{\includegraphics{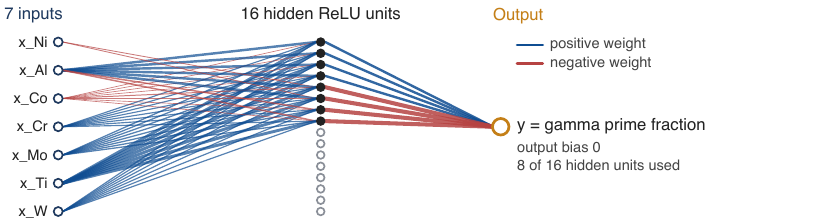}}\par}
\begin{minipage}[t]{0.40\linewidth}
\emph{Raw reply, shortened}
\begin{lstlisting}
## net.0.weight shape=(16, 7)
0 8.0 -0.56 2.64 3.2 10.0 2.8
0 8.0 -0.56 2.64 3.2 10.0 2.8
[... 13 rows omitted ...]
0 0 0 0 0 0 0
## net.0.bias shape=(16,)
-0.964 -1.004 -1.044 -1.084 -0.757 -0.957 -0.984 -1.084 0 0 0 0 0 0 0 0
## net.2.weight shape=(1, 16)
0.25 0.25 0.25 0.25 -0.5 -0.5 -0.5 -0.5 0 0 0 0 0 0 0 0
## net.2.bias shape=(1,)
0
\end{lstlisting}
\end{minipage}\hfill\begin{minipage}[t]{0.58\linewidth}
\emph{The model's own account}
\begin{lstlisting}
I wrote a lever-rule model of the gamma prime fraction at 900 C. A driving term z = (Al + 1.25 Ti + 0.33 Cr + 0.4 Mo + 0.35 W - 0.07 Co - 12.8 at.%) / 12.5 at.% says that Cr, Mo and W lower the Al solubility in gamma, that Co raises it, and that the binary Ni-Al solvus sits near 12.8 at.% Al. Four staggered ReLUs give a softened ramp, max(0, z). Two pairs of negative ReLUs put soft caps on it: one from the Ni+Co that is available (gamma prime is about 72% Ni+Co), and one at about 12 x Al, standing for eta Ni3Ti taking over in Ti-rich, Al-poor alloys. I checked the coefficients by hand against remembered fractions for IN738, Udimet 720Li, Waspaloy, IN713C and Ni-Al-Cr model alloys. I expect the trends and the rough scale to be right, with errors of about 0.1 to 0.15 in mole fraction, and larger ones where TCP phases, beta or eta compete.
\end{lstlisting}
\end{minipage}\par
\suppentry{\raggedright\paragraph{Other phase fraction} Run folder \texttt{write\_\allowbreak \_\allowbreak ni\_\allowbreak superalloy\_\allowbreak other\_\allowbreak phase\_\allowbreak fraction\_\allowbreak \_\allowbreak opus\_\allowbreak \_\allowbreak rep2}. Replicate 1 is not shown (rep1: valid reply but no self\_report.txt).\par}
\begin{lstlisting}
[...] Inputs are atomic fractions of Ni, Al, Co, Cr, Mo, Ti, W in nickel-base superalloys Ni-Al-Co-Cr-Mo-Ti-W; each row sums to 1. Composition range: Ni is the balance and at least 50 at.%; solutes in at.%: Al 0 to 12 (step 2), Co 0 to 20 (step 5), Cr 0 to 20 (step 5), Mo 0 to 6 (step 2), Ti 0 to 6 (step 2), W 0 to 6 (step 2); at least 4 elements present. The target is the equilibrium mole fraction, at 1173 K, of all phases other than gamma (fcc) and gamma prime (L12): sigma, mu, P, R, Laves, eta, B2 and bcc in mole fraction between 0 and 1, calculated at 1 atm with the CALPHAD method (Thermo-Calc, TCNI12 database). 249 alloys.
Input order: x[0] = x_Ni, x[1] = x_Al, x[2] = x_Co, x[3] = x_Cr, x[4] = x_Mo, x[5] = x_Ti, x[6] = x_W.
The output y must be your best estimate of the equilibrium mole fraction, at 1173 K, of all phases other than gamma (fcc) and gamma prime (L12): sigma, mu, P, R, Laves, eta, B2 and bcc in mole fraction between 0 and 1. [...]
\end{lstlisting}

{\centering\resizebox{\ifdim\width>\linewidth\linewidth\else\width\fi}{!}{\includegraphics{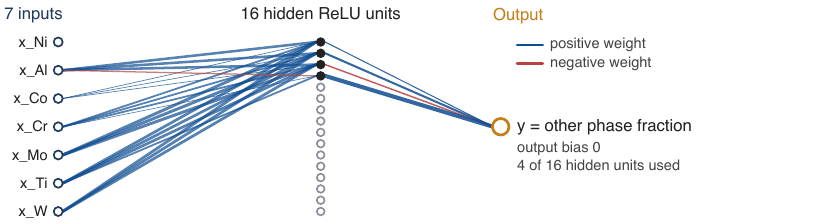}}\par}
\begin{minipage}[t]{0.40\linewidth}
\emph{Raw reply, shortened}
\begin{lstlisting}
## net.0.weight shape=(16, 7)
0 16 1.5 10 19 17 17.5
0 16 1.5 10 19 17 17.5
[... 13 rows omitted ...]
0 0 0 0 0 0 0
## net.0.bias shape=(16,)
-3.85 -4.15 -5.35 -0.35 0 0 0 0 0 0 0 0 0 0 0 0
## net.2.weight shape=(1, 16)
0.04 0.16 -0.09 0.4 0 0 0 0 0 0 0 0 0 0 0 0
## net.2.bias shape=(1,)
0
\end{lstlisting}
\end{minipage}\hfill\begin{minipage}[t]{0.58\linewidth}
\emph{The model's own account}
\begin{lstlisting}
I built a linear "Cr-equivalent" TCP driver D = Cr + 1.9 Mo + 1.75 W + 0.15 Co + 1.6 Al + 1.7 Ti - 0.415. The Al and Ti terms stand for the enrichment of the gamma matrix as gamma prime forms, and I calibrated D against the Ni-Cr and Ni-Mo solubility limits, the Ni-Al-Cr 900 C gamma/gamma-prime/alpha-Cr triangle (Ni-12Al-20Cr is borderline) and IN738-like and C-276-like alloys (slight and moderate TCP). Three ReLU units on D give a soft onset, a lever-rule slope of about 2 mole fraction per unit of D, and saturation near 0.45 for the most heavily alloyed, Ni-50 compositions. A fourth unit, 4*relu(Ti - 0.6 Al + 0.1 Cr - 0.035), adds eta Ni3Ti when Ti exceeds what gamma prime can take up. The task was to write a prior, not to find something, so I give no confidence figure. I expect the trends to be right and the scale to be within about a factor of two.
\end{lstlisting}
\end{minipage}\par
\suppentry{\raggedright\paragraph{Solidus} Run folder \texttt{write\_\allowbreak \_\allowbreak ni\_\allowbreak superalloy\_\allowbreak solidus\_\allowbreak \_\allowbreak opus\_\allowbreak \_\allowbreak rep1}.\par}
\begin{lstlisting}
[...] Inputs are atomic fractions of Ni, Al, Co, Cr, Mo, Ti, W in nickel-base superalloys Ni-Al-Co-Cr-Mo-Ti-W; each row sums to 1. Composition range: Ni is the balance and at least 50 at.%; solutes in at.%: Al 0 to 12 (step 2), Co 0 to 20 (step 5), Cr 0 to 20 (step 5), Mo 0 to 6 (step 2), Ti 0 to 6 (step 2), W 0 to 6 (step 2); at least 4 elements present. The target is the equilibrium solidus temperature in K, calculated at 1 atm with the CALPHAD method (Thermo-Calc, TCNI12 database). 250 alloys.
Input order: x[0] = x_Ni, x[1] = x_Al, x[2] = x_Co, x[3] = x_Cr, x[4] = x_Mo, x[5] = x_Ti, x[6] = x_W.
The output y must be your best estimate of the equilibrium solidus temperature in K. [...]
\end{lstlisting}

{\centering\resizebox{\ifdim\width>\linewidth\linewidth\else\width\fi}{!}{\includegraphics{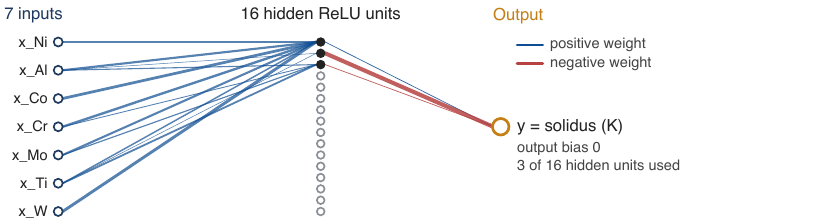}}\par}
\begin{minipage}[t]{0.40\linewidth}
\emph{Raw reply, shortened}
\begin{lstlisting}
## net.0.weight shape=(16, 7)
1728 1278 1748 1458 1228 728 1878
0 1 0 0 0 1.5 0
[... 13 rows omitted ...]
0 0 0 0 0 0 0
## net.0.bias shape=(16,)
0 -0.12 -100 0 0 0 0 0 0 0 0 0 0 0 0 0
## net.2.weight shape=(1, 16)
1 -400 -0.2 0 0 0 0 0 0 0 0 0 0 0 0 0
## net.2.bias shape=(1,)
0
\end{lstlisting}
\end{minipage}\hfill\begin{minipage}[t]{0.58\linewidth}
\emph{The model's own account}
\begin{lstlisting}
I built the prior from binary Ni-X phase-diagram solidus slopes, starting from pure Ni at 1728 K. Per at.% the slopes are about: Al -4.5 K, Co +0.2 K, Cr -2.7 K, Mo -5 K, Ti -10 K, W +1.5 K. Because the fractions sum to 1, hidden unit 0 encodes this as an always-positive linear term, with weight 1728 + 100*slope for each element. Two ReLU units add extra depression. One fires when Al + 1.5Ti exceeds 12 at.% (approaching the gamma/gamma-prime eutectic). The other fires when the summed solute depression exceeds 100 K (concave solidus at high alloying). I sanity-checked the scale against Waspaloy-, IN-738- and CMSX-4-like compositions, which come out around 1590-1630 K. I expect the trends to be right but absolute values could be off by about 30-50 K.
\end{lstlisting}
\end{minipage}\par
\suppneed{2.2in}\subsubsection{Ti--Al--Mo--Nb--Sn--V--Zr titanium alloys (\texttt{ti\_alloy})}
\suppentry{\raggedright\paragraph{Alpha fraction} Run folder \texttt{write\_\allowbreak \_\allowbreak ti\_\allowbreak alloy\_\allowbreak alpha\_\allowbreak fraction\_\allowbreak \_\allowbreak opus\_\allowbreak \_\allowbreak rep1}.\par}
\begin{lstlisting}
[...] Inputs are atomic fractions of Ti, Al, Mo, Nb, Sn, V, Zr in titanium alloys Ti-Al-Mo-Nb-Sn-V-Zr; each row sums to 1. Composition range: Ti is the balance and at least 60 at.%; solutes in at.%: Al 0 to 12 (step 3), Mo 0 to 9 (step 3), Nb 0 to 12 (step 3), Sn 0 to 4 (step 2), V 0 to 12 (step 3), Zr 0 to 8 (step 4); at least 4 elements present. The target is the equilibrium mole fraction of the hcp alpha phase at 973 K in mole fraction between 0 and 1, calculated at 1 atm with the CALPHAD method (Thermo-Calc, TCTI6 database). 198 alloys.
Input order: x[0] = x_Ti, x[1] = x_Al, x[2] = x_Mo, x[3] = x_Nb, x[4] = x_Sn, x[5] = x_V, x[6] = x_Zr.
The output y must be your best estimate of the equilibrium mole fraction of the hcp alpha phase at 973 K in mole fraction between 0 and 1. [...]
\end{lstlisting}

{\centering\resizebox{\ifdim\width>\linewidth\linewidth\else\width\fi}{!}{\includegraphics{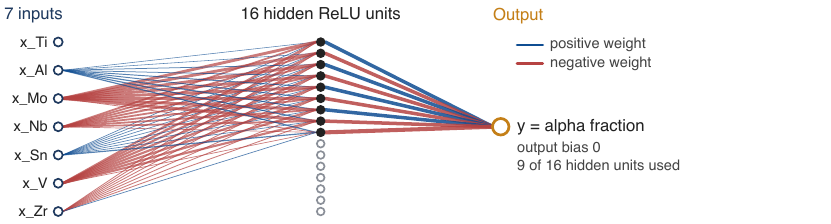}}\par}
\begin{minipage}[t]{0.40\linewidth}
\emph{Raw reply, shortened}
\begin{lstlisting}
## net.0.weight shape=(16, 7)
0 0.5859 -37.5 -19.737 1.875 -25.0 -5.0
0 0.5859 -37.5 -19.737 1.875 -25.0 -5.0
[... 13 rows omitted ...]
0 0 0 0 0 0 0
## net.0.bias shape=(16,)
1.3953 0.3953 2.2109 1.2109 3.0266 2.0266 3.8422 2.8422 -1.35 0 0 0 0 0 0 0
## net.2.weight shape=(1, 16)
0.25 -0.25 0.25 -0.25 0.25 -0.25 0.25 -0.25 -0.3 0 0 0 0 0 0 0
## net.2.bias shape=(1,)
0
\end{lstlisting}
\end{minipage}\hfill\begin{minipage}[t]{0.58\linewidth}
\emph{The model's own account}
\begin{lstlisting}
I built a lever-rule model of the alpha+beta field at 973 K. A beta-stabiliser index B = Mo/10 + V/15 + Nb/19 + Zr/75 - Sn/200 (in at.%) uses estimated binary beta-boundary compositions at 700 C. The alpha fraction falls linearly from 1 at B_lo = 0.13 - 0.25*x_Al to 0 at B_hi = 1 + 3*x_Al, so Al widens the two-phase field and pushes the beta boundary out. Units 0-7 implement this as four stacked clipped ramps (pairs relu(A) - relu(A-1), weight 0.25 each) whose centres shift with Al, which roughly approximates the multiplicative Al effect. Unit 8 subtracts an alpha2 (Ti3Al-type D019) penalty when Al + Sn + 0.3*Zr in the alpha phase exceeds about 13.5 at.%, damped at high B; I checked the model against Ti-6Al-4V (about 0.89) and Ti-6246 (about 0.8), and the coefficients are uncertain by perhaps 30-50%.
\end{lstlisting}
\end{minipage}\par
\suppentry{\raggedright\paragraph{Alpha transus} Run folder \texttt{write\_\allowbreak \_\allowbreak ti\_\allowbreak alloy\_\allowbreak alpha\_\allowbreak transus\_\allowbreak \_\allowbreak opus\_\allowbreak \_\allowbreak rep1}.\par}
\begin{lstlisting}
[...] Inputs are atomic fractions of Ti, Al, Mo, Nb, Sn, V, Zr in titanium alloys Ti-Al-Mo-Nb-Sn-V-Zr; each row sums to 1. Composition range: Ti is the balance and at least 60 at.%; solutes in at.%: Al 0 to 12 (step 3), Mo 0 to 9 (step 3), Nb 0 to 12 (step 3), Sn 0 to 4 (step 2), V 0 to 12 (step 3), Zr 0 to 8 (step 4); at least 4 elements present. The target is the temperature above which no hcp alpha phase is stable at equilibrium (the beta transus when no other phase is present); reported as 773 K when no alpha is stable at 773 K in K, calculated at 1 atm with the CALPHAD method (Thermo-Calc, TCTI6 database). 191 alloys.
Input order: x[0] = x_Ti, x[1] = x_Al, x[2] = x_Mo, x[3] = x_Nb, x[4] = x_Sn, x[5] = x_V, x[6] = x_Zr.
The output y must be your best estimate of the temperature above which no hcp alpha phase is stable at equilibrium (the beta transus when no other phase is present); reported as 773 K when no alpha is stable at 773 K in K. [...]
\end{lstlisting}

{\centering\resizebox{\ifdim\width>\linewidth\linewidth\else\width\fi}{!}{\includegraphics{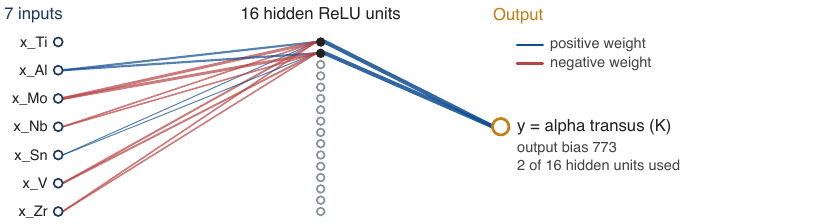}}\par}
\begin{minipage}[t]{0.40\linewidth}
\emph{Raw reply, shortened}
\begin{lstlisting}
## net.0.weight shape=(16, 7)
0 1350 -2000 -900 400 -1200 -900
0 1350 -2000 -900 400 -1200 -900
[... 13 rows omitted ...]
0 0 0 0 0 0 0
## net.0.bias shape=(16,)
475 289 0 0 0 0 0 0 0 0 0 0 0 0 0 0
## net.2.weight shape=(1, 16)
0.5 0.5 0 0 0 0 0 0 0 0 0 0 0 0 0 0
## net.2.bias shape=(1,)
773
\end{lstlisting}
\end{minipage}\hfill\begin{minipage}[t]{0.58\linewidth}
\emph{The model's own account}
\begin{lstlisting}
I wrote a linear beta-transus model starting from pure Ti at 1155 K. The per-at.% slopes are Al +13.5, Sn +4, Zr -9, Nb -9, V -12 and Mo -20 K, which I recalled from the Ti binary phase diagrams and the empirical wt.% beta-transus formula, and checked by hand against Ti-6Al-4V, Ti-6242/6246, Ti-811, Ti-6Al-7Nb, Beta III and Beta 21S. Two ReLU units share this linear direction, with knots at 680 K and 866 K and output weights 0.5 each, plus an output bias of 773 K. Together they reproduce the linear model above about 866 K and flatten smoothly onto the 773 K floor, which mimics the reported clipping and the convex shape of the beta-stabiliser transus curves. The task was to write weights rather than find something, so no confidence percentage applies; I expect the trends to be right and absolute errors of a few tens of K.
\end{lstlisting}
\end{minipage}\par
\suppentry{\raggedright\paragraph{Freezing range} Run folder \texttt{write\_\allowbreak \_\allowbreak ti\_\allowbreak alloy\_\allowbreak freezing\_\allowbreak range\_\allowbreak \_\allowbreak opus\_\allowbreak \_\allowbreak rep1}.\par}
\begin{lstlisting}
[...] Inputs are atomic fractions of Ti, Al, Mo, Nb, Sn, V, Zr in titanium alloys Ti-Al-Mo-Nb-Sn-V-Zr; each row sums to 1. Composition range: Ti is the balance and at least 60 at.%; solutes in at.%: Al 0 to 12 (step 3), Mo 0 to 9 (step 3), Nb 0 to 12 (step 3), Sn 0 to 4 (step 2), V 0 to 12 (step 3), Zr 0 to 8 (step 4); at least 4 elements present. The target is the equilibrium freezing range (liquidus temperature minus solidus temperature) in K, calculated at 1 atm with the CALPHAD method (Thermo-Calc, TCTI6 database). 200 alloys.
Input order: x[0] = x_Ti, x[1] = x_Al, x[2] = x_Mo, x[3] = x_Nb, x[4] = x_Sn, x[5] = x_V, x[6] = x_Zr.
The output y must be your best estimate of the equilibrium freezing range (liquidus temperature minus solidus temperature) in K. [...]
\end{lstlisting}

{\centering\resizebox{\ifdim\width>\linewidth\linewidth\else\width\fi}{!}{\includegraphics{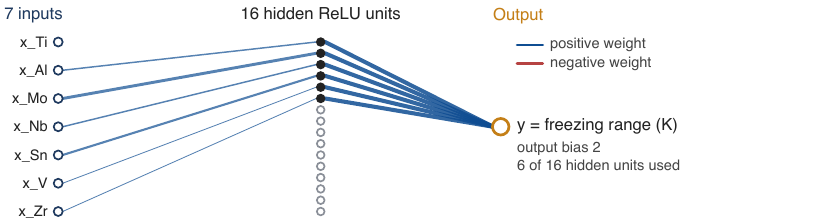}}\par}
\begin{minipage}[t]{0.40\linewidth}
\emph{Raw reply, shortened}
\begin{lstlisting}
## net.0.weight shape=(16, 7)
0 150 0 0 0 0 0
0 0 380 0 0 0 0
[... 13 rows omitted ...]
0 0 0 0 0 0 0
## net.0.bias shape=(16,)
0 0 0 0 0 0 0 0 0 0 0 0 0 0 0 0
## net.2.weight shape=(1, 16)
1 1 1 1 1 1 0 0 0 0 0 0 0 0 0 0
## net.2.bias shape=(1,)
2
\end{lstlisting}
\end{minipage}\hfill\begin{minipage}[t]{0.58\linewidth}
\emph{The model's own account}
\begin{lstlisting}
I used the dilute-solution result that the equilibrium freezing range adds over solutes, each contributing m_i x_i (1 - 1/k_i). I took the liquidus slopes and partition coefficients from my recollection of the binary Ti-X diagrams: Mo raises the liquidus steeply with k > 1, so it gives the largest contribution (about 3.8 K per at.%). Sn and Al have k < 1 and give moderate contributions (about 2.5 and 1.5 K per at.%), Nb gives about 1.5 K per at.%, and V and Zr give very small ones because their Ti-X diagrams have liquidus minima and k close to 1. I wrote this as six always-active ReLU units, one per solute, each with weight 1 to the output, plus a 2 K output bias. That puts typical alloys at about 15 to 60 K and Ti-6Al-4V near 17 K. This was a write task, not a search, so no confidence percentage applies; I am least sure of the absolute scale of the Sn and Al terms.
\end{lstlisting}
\end{minipage}\par
\suppentry{\raggedright\paragraph{Intermetallic fraction} Run folder \texttt{write\_\allowbreak \_\allowbreak ti\_\allowbreak alloy\_\allowbreak intermetallic\_\allowbreak fraction\_\allowbreak \_\allowbreak opus\_\allowbreak \_\allowbreak rep1}.\par}
\begin{lstlisting}
[...] Inputs are atomic fractions of Ti, Al, Mo, Nb, Sn, V, Zr in titanium alloys Ti-Al-Mo-Nb-Sn-V-Zr; each row sums to 1. Composition range: Ti is the balance and at least 60 at.%; solutes in at.%: Al 0 to 12 (step 3), Mo 0 to 9 (step 3), Nb 0 to 12 (step 3), Sn 0 to 4 (step 2), V 0 to 12 (step 3), Zr 0 to 8 (step 4); at least 4 elements present. The target is the equilibrium mole fraction, at 973 K, of all phases other than the disordered bcc (beta) and hcp (alpha) solid solutions, for example Ti3Al in mole fraction between 0 and 1, calculated at 1 atm with the CALPHAD method (Thermo-Calc, TCTI6 database). 198 alloys.
Input order: x[0] = x_Ti, x[1] = x_Al, x[2] = x_Mo, x[3] = x_Nb, x[4] = x_Sn, x[5] = x_V, x[6] = x_Zr.
The output y must be your best estimate of the equilibrium mole fraction, at 973 K, of all phases other than the disordered bcc (beta) and hcp (alpha) solid solutions, for example Ti3Al in mole fraction between 0 and 1. [...]
\end{lstlisting}

{\centering\resizebox{\ifdim\width>\linewidth\linewidth\else\width\fi}{!}{\includegraphics{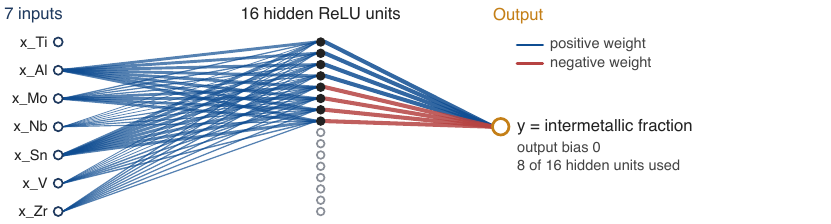}}\par}
\begin{minipage}[t]{0.40\linewidth}
\emph{Raw reply, shortened}
\begin{lstlisting}
## net.0.weight shape=(16, 7)
0 10 3.125 0.875 10 1.25 3
0 10 3.125 0.875 10 1.25 3
[... 13 rows omitted ...]
0 0 0 0 0 0 0
## net.0.bias shape=(16,)
-1.1 -1.2 -1.3 -1.4 -2.1 -2.2 -2.3 -2.4 0 0 0 0 0 0 0 0
## net.2.weight shape=(1, 16)
0.25 0.25 0.25 0.25 -0.25 -0.25 -0.25 -0.25 0 0 0 0 0 0 0 0
## net.2.bias shape=(1,)
0
\end{lstlisting}
\end{minipage}\hfill\begin{minipage}[t]{0.58\linewidth}
\emph{The model's own account}
\begin{lstlisting}
I took the non-alpha/beta phase at 973 K to be essentially Ti3Al (alpha2), and modelled it with a lever rule on an Al-equivalent A = Al + Sn + 0.3 Zr against an alpha2 solvus near 12.5 at.% (averaged over solvus values from 11 to 14 at.% to soften the onset), with an alpha2 composition about 10 at.% higher. Beta stabilisers, B = V + 2.5 Mo + 0.7 Nb, enter twice: a small positive term (0.125 B) because they pull Ti into beta and enrich the remaining alpha in Al, and a cap of 1 - 2.5 B because the alpha2 fraction cannot exceed the non-beta fraction. Each ReLU pair computes clip(T, 0, cap), so the output is never negative on this composition range. I am fairly confident about the direction (rising with Al and Sn, near zero below about 9 at.% Al-equivalent). I am much less sure of the solvus position, the strength of the beta partitioning effect and the scale, perhaps 50% that the rough magnitudes are within a factor of two.
\end{lstlisting}
\end{minipage}\par
\suppentry{\raggedright\paragraph{Solidus} Run folder \texttt{write\_\allowbreak \_\allowbreak ti\_\allowbreak alloy\_\allowbreak solidus\_\allowbreak \_\allowbreak opus\_\allowbreak \_\allowbreak rep1}.\par}
\begin{lstlisting}
[...] Inputs are atomic fractions of Ti, Al, Mo, Nb, Sn, V, Zr in titanium alloys Ti-Al-Mo-Nb-Sn-V-Zr; each row sums to 1. Composition range: Ti is the balance and at least 60 at.%; solutes in at.%: Al 0 to 12 (step 3), Mo 0 to 9 (step 3), Nb 0 to 12 (step 3), Sn 0 to 4 (step 2), V 0 to 12 (step 3), Zr 0 to 8 (step 4); at least 4 elements present. The target is the equilibrium solidus temperature in K, calculated at 1 atm with the CALPHAD method (Thermo-Calc, TCTI6 database). 200 alloys.
Input order: x[0] = x_Ti, x[1] = x_Al, x[2] = x_Mo, x[3] = x_Nb, x[4] = x_Sn, x[5] = x_V, x[6] = x_Zr.
The output y must be your best estimate of the equilibrium solidus temperature in K. [...]
\end{lstlisting}

{\centering\resizebox{\ifdim\width>\linewidth\linewidth\else\width\fi}{!}{\includegraphics{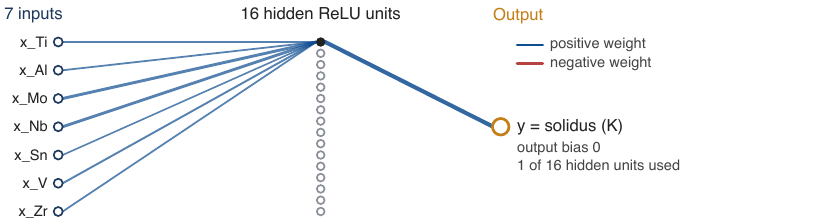}}\par}
\begin{minipage}[t]{0.40\linewidth}
\emph{Raw reply, shortened}
\begin{lstlisting}
## net.0.weight shape=(16, 7)
1.941 1.441 2.741 2.641 1.241 1.591 1.391
0 0 0 0 0 0 0
[... 13 rows omitted ...]
0 0 0 0 0 0 0
## net.0.bias shape=(16,)
0 0 0 0 0 0 0 0 0 0 0 0 0 0 0 0
## net.2.weight shape=(1, 16)
1000 0 0 0 0 0 0 0 0 0 0 0 0 0 0 0
## net.2.bias shape=(1,)
0
\end{lstlisting}
\end{minipage}\hfill\begin{minipage}[t]{0.58\linewidth}
\emph{The model's own account}
\begin{lstlisting}
I encoded a linear dilute-solution model in one hidden unit. Because the inputs sum to 1 and are non-negative, the ReLU never cuts it off: T = sum_i a_i x_i, where a_Ti = 1941 K (the melting point of Ti) and each solute's coefficient is 1941 K plus 100 times its estimated initial solidus slope in K per at.%. The slopes come from the binary phase diagrams and ideal-solution partition estimates: Mo +8, Nb +7, Al -5, Sn -7, V -3.5, Zr -5.5. I checked the result against Ti-6Al-4V (solidus about 1877 K). This gives a range of roughly 1750 to 2090 K over the design space. I am fairly confident in the signs and rough scale (about 70%) but less so in the individual magnitudes, especially for Sn and Al, and I left out the small curvature and interaction effects.
\end{lstlisting}
\end{minipage}\par
\suppneed{2.2in}\subsubsection{Mo--Nb--Ta--Ti--V--W--Zr refractory high-entropy alloys (\texttt{refractory})}
\suppentry{\raggedright\paragraph{Freezing range} Run folder \texttt{write\_\allowbreak \_\allowbreak refractory\_\allowbreak freezing\_\allowbreak range\_\allowbreak \_\allowbreak opus\_\allowbreak \_\allowbreak rep1}.\par}
\begin{lstlisting}
[...] Inputs are atomic fractions of Mo, Nb, Ta, Ti, V, W, Zr in Mo-Nb-Ta-Ti-V-W-Zr refractory high-entropy alloys; each row sums to 1. Composition range: every element between 0 and 35 at.% on a 5 at.% grid, at least four elements present. The target is the equilibrium freezing range (liquidus temperature minus solidus temperature) in K, calculated at 1 atm with the CALPHAD method (Thermo-Calc, TCHEA8 database). 200 alloys.
Input order: x[0] = x_Mo, x[1] = x_Nb, x[2] = x_Ta, x[3] = x_Ti, x[4] = x_V, x[5] = x_W, x[6] = x_Zr.
The output y must be your best estimate of the equilibrium freezing range (liquidus temperature minus solidus temperature) in K. [...]
\end{lstlisting}

{\centering\resizebox{\ifdim\width>\linewidth\linewidth\else\width\fi}{!}{\includegraphics{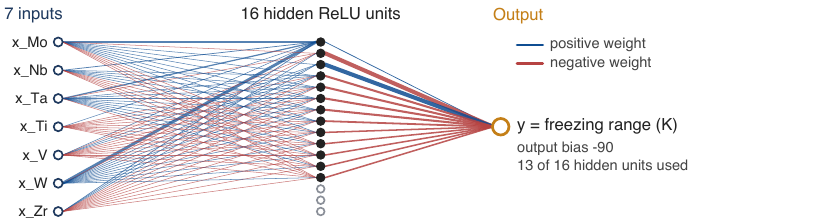}}\par}
\begin{minipage}[t]{0.40\linewidth}
\emph{Raw reply, shortened}
\begin{lstlisting}
## net.0.weight shape=(16, 7)
878.44 389.49 1150.26 7.854 2.022 4035.62 297.76
-1 0 -0.3 0 -0.7 -1 1
[... 13 rows omitted ...]
0 0 0 0 0 0 0
## net.0.bias shape=(16,)
100 0 -0.35 0 0 0 0 0 0 0 0 0 0 0 0 0
## net.2.weight shape=(1, 16)
1 -300 300 -120 -120 -120 -120 -96 -72 -72 -72 -72 -72 0 0 0
## net.2.bias shape=(1,)
-90
\end{lstlisting}
\end{minipage}\hfill\begin{minipage}[t]{0.58\linewidth}
\emph{The model's own account}
\begin{lstlisting}
I took ideal-solution solidification theory as the base. I used pure-element melting points and SGTE fusion entropies, and the second-order lens-width result: gap = sum_j x_j s_j^2 (T_j - T*)^2 / sum_j x_j s_j T_j. This predicts that a high-Tm minority (W, Ta, Mo) in a low-Tm matrix (Ti, V, Zr) gives a much wider freezing range than the reverse. I approximated it with a weighted Jensen gap, D*Var_D(r), where the element weights are D_j = s_j^2 s_j T_j and the coordinates r_j are fitted so that the dilute slopes match. Unit 0 carries the linear part, and ten ReLU knots relu(z - c*D) carry the perspective term z^2/D, scaled by 1.2 so that the hand-checked ideal values match. I added two hand-set Zr terms for Laves formation and miscibility gaps with W, Mo, V and Ta (about +80 K for 25% Zr against 25% W, more when both are high) and a 10 K offset; these terms are guesses, not derived.
\end{lstlisting}
\end{minipage}\par
\suppentry{\raggedright\paragraph{Intermetallic fraction} Run folder \texttt{write\_\allowbreak \_\allowbreak refractory\_\allowbreak intermetallic\_\allowbreak fraction\_\allowbreak \_\allowbreak opus\_\allowbreak \_\allowbreak rep1}.\par}
\begin{lstlisting}
[...] Inputs are atomic fractions of Mo, Nb, Ta, Ti, V, W, Zr in Mo-Nb-Ta-Ti-V-W-Zr refractory high-entropy alloys; each row sums to 1. Composition range: every element between 0 and 35 at.% on a 5 at.% grid, at least four elements present. The target is the equilibrium mole fraction, at 1473 K, of all phases other than liquid and the disordered fcc, bcc and hcp solid solutions (that is: sigma, Laves and other intermetallic phases, and ordered B2 or L12) in mole fraction between 0 and 1, calculated at 1 atm with the CALPHAD method (Thermo-Calc, TCHEA8 database). 200 alloys.
Input order: x[0] = x_Mo, x[1] = x_Nb, x[2] = x_Ta, x[3] = x_Ti, x[4] = x_V, x[5] = x_W, x[6] = x_Zr.
The output y must be your best estimate of the equilibrium mole fraction, at 1473 K, of all phases other than liquid and the disordered fcc, bcc and hcp solid solutions (that is: sigma, Laves and other intermetallic phases, and ordered B2 or L12) in mole fraction between 0 and 1. [...]
\end{lstlisting}

{\centering\resizebox{\ifdim\width>\linewidth\linewidth\else\width\fi}{!}{\includegraphics{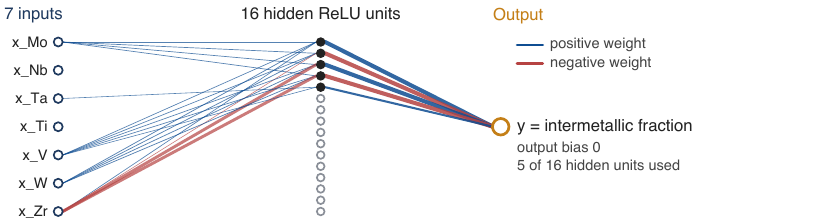}}\par}
\begin{minipage}[t]{0.40\linewidth}
\emph{Raw reply, shortened}
\begin{lstlisting}
## net.0.weight shape=(16, 7)
2.6 0 0 0 2.34 3.38 2.163
2.6 0 0 0 2.34 3.38 -0.837
[... 13 rows omitted ...]
0 0 0 0 0 0 0
## net.0.bias shape=(16,)
-1.378 -1.303 -1.303 -1.378 -0.55 0 0 0 0 0 0 0 0 0 0 0
## net.2.weight shape=(1, 16)
1 -1 1 -1 0.4 0 0 0 0 0 0 0 0 0 0 0
## net.2.bias shape=(1,)
0
\end{lstlisting}
\end{minipage}\hfill\begin{minipage}[t]{0.58\linewidth}
\emph{The model's own account}
\begin{lstlisting}
I took the intermetallic at 1473 K in this system to be mainly the C15 Laves phase Zr(Mo,W,V)2, with a minor V2Ta term. I modelled the Laves phase with a solubility-product picture: Laves forms when x_Zr * B^2 is above roughly 0.02, where B = x_Mo + 1.3 x_W + 0.9 x_V and Nb, Ta and Ti act only as diluents. The amount of Laves then follows a lever rule up to a cap of about 3(x_Zr - 0.025). I wrote the weights as a clamp, min(relu(l), 3 x_Zr - 0.075), where l is a linear onset term 2.6 B + 2.16 x_Zr - 1.38. Two extra units, which only matter on the 5 at.% grid, force the output to exactly zero when there is no Zr, and one further unit adds a small Ta+V Laves term. Trends (zero without Zr, rising with Zr and with Mo/W/V, falling with Nb/Ta/Ti dilution) I give about 70% confidence; the absolute scale and the onset boundary are uncertain to roughly a factor of two.
\end{lstlisting}
\end{minipage}\par
\suppentry{\raggedright\paragraph{Main phase fraction} Run folder \texttt{write\_\allowbreak \_\allowbreak refractory\_\allowbreak main\_\allowbreak phase\_\allowbreak fraction\_\allowbreak \_\allowbreak opus\_\allowbreak \_\allowbreak rep1}.\par}
\begin{lstlisting}
[...] Inputs are atomic fractions of Mo, Nb, Ta, Ti, V, W, Zr in Mo-Nb-Ta-Ti-V-W-Zr refractory high-entropy alloys; each row sums to 1. Composition range: every element between 0 and 35 at.% on a 5 at.% grid, at least four elements present. The target is the equilibrium mole fraction, at 1473 K, of the single most abundant phase (1 for a single-phase alloy; two bcc phases of a miscibility gap count as two phases) in mole fraction between 0 and 1, calculated at 1 atm with the CALPHAD method (Thermo-Calc, TCHEA8 database). 200 alloys.
Input order: x[0] = x_Mo, x[1] = x_Nb, x[2] = x_Ta, x[3] = x_Ti, x[4] = x_V, x[5] = x_W, x[6] = x_Zr.
The output y must be your best estimate of the equilibrium mole fraction, at 1473 K, of the single most abundant phase (1 for a single-phase alloy; two bcc phases of a miscibility gap count as two phases) in mole fraction between 0 and 1. [...]
\end{lstlisting}

{\centering\resizebox{\ifdim\width>\linewidth\linewidth\else\width\fi}{!}{\includegraphics{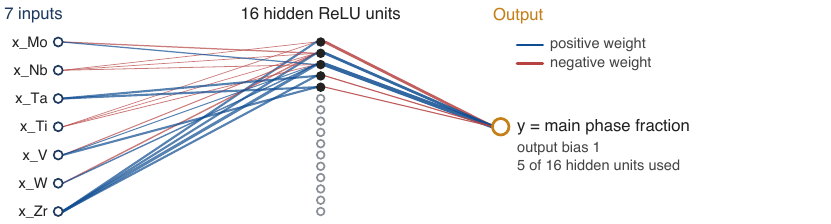}}\par}
\begin{minipage}[t]{0.40\linewidth}
\emph{Raw reply, shortened}
\begin{lstlisting}
## net.0.weight shape=(16, 7)
0 -0.12 0 -0.12 0 0 1
-0.45 -0.12 0 -0.12 -0.3 -0.55 1
[... 13 rows omitted ...]
0 0 0 0 0 0 0
## net.0.bias shape=(16,)
-0.04 -0.04 -0.68 -0.5 -0.5 0 0 0 0 0 0 0 0 0 0 0
## net.2.weight shape=(1, 16)
-2.7 2.7 4.4 -0.6 -0.5 0 0 0 0 0 0 0 0 0 0 0
## net.2.bias shape=(1,)
1
\end{lstlisting}
\end{minipage}\hfill\begin{minipage}[t]{0.58\linewidth}
\emph{The model's own account}
\begin{lstlisting}
I set the baseline to a single bcc phase (y = 1) and made the main loss come from a Zr(Mo,W,V)2 C15 Laves phase. Its fraction is about 2.7 x min(Zr excess, Laves-former supply). The Zr excess is x_Zr - 0.04 - 0.12(x_Nb + x_Ti), since Nb and Ti dissolve Zr in bcc. The supply is 0.45 x_Mo + 0.55 x_W + 0.3 x_V. Hidden units 1 and 2 build that min exactly with relu(a) - relu(a - b).

Unit 3 is a ridge that roughly brings y back up where the Laves phase would pass 50% and become the majority phase. Units 4 and 5 are small penalties for a Ta-Zr bcc miscibility gap and for TaV2 Laves when those pairs are both rich.

The task was a write rather than a find. I am not confident in the absolute Laves fractions, which TCHEA8 could put quite differently. I am more confident in the trend: y falls with Zr when Mo, W or V is present.
\end{lstlisting}
\end{minipage}\par
\suppentry{\raggedright\paragraph{Solidus} Run folder \texttt{write\_\allowbreak \_\allowbreak refractory\_\allowbreak solidus\_\allowbreak \_\allowbreak opus\_\allowbreak \_\allowbreak rep1}.\par}
\begin{lstlisting}
[...] Inputs are atomic fractions of Mo, Nb, Ta, Ti, V, W, Zr in Mo-Nb-Ta-Ti-V-W-Zr refractory high-entropy alloys; each row sums to 1. Composition range: every element between 0 and 35 at.% on a 5 at.% grid, at least four elements present. The target is the equilibrium solidus temperature in K, calculated at 1 atm with the CALPHAD method (Thermo-Calc, TCHEA8 database). 200 alloys.
Input order: x[0] = x_Mo, x[1] = x_Nb, x[2] = x_Ta, x[3] = x_Ti, x[4] = x_V, x[5] = x_W, x[6] = x_Zr.
The output y must be your best estimate of the equilibrium solidus temperature in K. [...]
\end{lstlisting}

{\centering\resizebox{\ifdim\width>\linewidth\linewidth\else\width\fi}{!}{\includegraphics{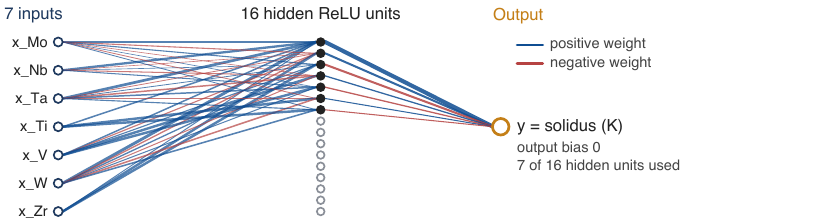}}\par}
\begin{minipage}[t]{0.40\linewidth}
\emph{Raw reply, shortened}
\begin{lstlisting}
## net.0.weight shape=(16, 7)
2.804 2.658 3.224 1.484 1.442 3.59 0.618
-0.866 -0.152 -0.4375 0 -0.929 -1.214 2.857
[... 13 rows omitted ...]
0 0 0 0 0 0 0
## net.0.bias shape=(16,)
0 0 -1 0 -1 0 -1 0 0 0 0 0 0 0 0 0
## net.2.weight shape=(1, 16)
1000 392 -392 168 -168 96 -96 0 0 0 0 0 0 0 0 0
## net.2.bias shape=(1,)
0
\end{lstlisting}
\end{minipage}\hfill\begin{minipage}[t]{0.58\linewidth}
\emph{The model's own account}
\begin{lstlisting}
I started from the rule of mixtures of the pure-element melting points and subtracted a regular-solution-style quadratic solidus depression, 4*D_ij*x_i*x_j. I estimated each binary D_ij from what I recall of the binary phase diagrams: about 15-120 K among Mo/Nb/Ta/W, and large values for the Zr-W, Zr-V and Zr-Mo eutectic systems, for V with Ta/Nb/W, and for Ti with W/Ta. Hidden unit 0 is an always-positive linear unit that carries the effective per-element coefficients. Six more units encode the three dominant product terms (Zr x {W,V,Mo,Ta,Nb}, V x {Ta,Nb,W,Mo}, Ti x {W,Ta,Mo}) using the approximation xy ~ 0.5*min(x,y) + 0.5*relu(x+y-1), and the remaining nine units are left inactive. I checked the network by hand on several alloys (MoNbTaW ~3070 K, MoNbTaVW ~2720 K, TiZrNbMoV ~1980 K). I am fairly confident in the trends (Zr/V/Ti lower the solidus and W/Ta raise it), but the absolute level could be off by roughly 100-200 K.
\end{lstlisting}
\end{minipage}\par
\suppneed{2.2in}\subsubsection{Fe--Al--C--Cr--Mn--Mo--Ni steels (\texttt{steel})}
\suppentry{\raggedright\paragraph{Austenite fraction} Run folder \texttt{write\_\allowbreak \_\allowbreak steel\_\allowbreak austenite\_\allowbreak fraction\_\allowbreak \_\allowbreak opus\_\allowbreak \_\allowbreak rep1}.\par}
\begin{lstlisting}
[...] Inputs are atomic fractions of Fe, Al, C, Cr, Mn, Mo, Ni in high-alloy steels Fe-Al-C-Cr-Mn-Mo-Ni; each row sums to 1. Composition range: Fe is the balance and at least 50 at.%; solutes in at.%: Al 0 to 6 (step 3), C 0 to 2 (step 1), Cr 0 to 24 (step 4), Mn 0 to 20 (step 5), Mo 0 to 4 (step 2), Ni 0 to 20 (step 4); at least 4 elements present. The target is the equilibrium mole fraction of austenite (fcc) at 1173 K in mole fraction between 0 and 1, calculated at 1 atm with the CALPHAD method (Thermo-Calc, TCFE10 database). 200 alloys.
Input order: x[0] = x_Fe, x[1] = x_Al, x[2] = x_C, x[3] = x_Cr, x[4] = x_Mn, x[5] = x_Mo, x[6] = x_Ni.
The output y must be your best estimate of the equilibrium mole fraction of austenite (fcc) at 1173 K in mole fraction between 0 and 1. [...]
\end{lstlisting}

{\centering\resizebox{\ifdim\width>\linewidth\linewidth\else\width\fi}{!}{\includegraphics{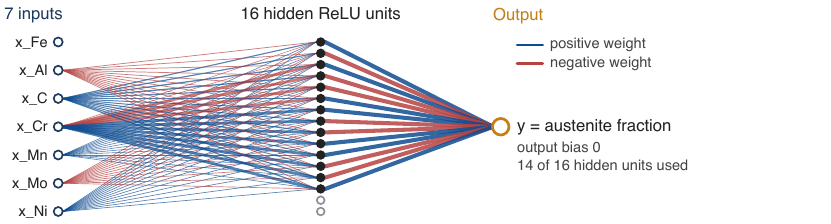}}\par}
\begin{minipage}[t]{0.40\linewidth}
\emph{Raw reply, shortened}
\begin{lstlisting}
## net.0.weight shape=(16, 7)
0 -22.857143 64.285714 -7.857143 7.142857 -18.571429 14.285714
0 -22.257143 66.785714 -7.607143 7.142857 -16.071429 14.285714
[... 13 rows omitted ...]
0 0 0 0 0 0 0
## net.0.bias shape=(16,)
1.271429 0.271429 0.642857 -0.357143 1.271429 0.271429 -50.271429 -51.271429 -48.728571 -49.728571 -60.571429 -61.511429 -58.728571 -59.728571 0 0
## net.2.weight shape=(1, 16)
1 -1 1 -1 -1 1 1 -1 -1 1 1 -1 -1 1 0 0
## net.2.bias shape=(1,)
0
\end{lstlisting}
\end{minipage}\hfill\begin{minipage}[t]{0.58\linewidth}
\emph{The model's own account}
\begin{lstlisting}
I modelled the fcc fraction as a clipped ramp, min(max(0.5 + A/7, 0), 1 - P), where A is a linear "austenite potential" in at.%: Ni +1, Mn +0.5, C +4.5, Cr -0.55, Mo -1.3, Al -1.6, offset +5.4. I calibrated A against 304, 410, 430 and nitrogen-free 2205 at 900 C. P is a linear cap for carbides, sigma/Laves and B2 (C, Mo, Al, Cr). Each ramp takes two ReLUs, relu(s) - relu(s - cap). The Cr = 0, 20 and 24 at.% levels get their own ramps, gated by large Cr weights (+-250) that switch a ramp pair on only at that grid level. Cr = 0 uses stronger Mn and C and a lower offset, for the gamma-loop and dilute regime. Cr = 20 and 24 use weaker Mn and C (Mn saturation, carbides) and, at 24, a sigma-raised cap. I expect the signs and rough scale to be right, about 75% confident in the trends. Absolute values near the two-phase boundary may be off by 0.2 to 0.3, especially for high-Cr, high-Mn and Al-rich alloys.
\end{lstlisting}
\end{minipage}\par
\suppentry{\raggedright\paragraph{Carbide intermetallic fraction} Run folder \texttt{write\_\allowbreak \_\allowbreak steel\_\allowbreak carbide\_\allowbreak intermetallic\_\allowbreak fraction\_\allowbreak \_\allowbreak opus\_\allowbreak \_\allowbreak rep1}.\par}
\begin{lstlisting}
[...] Inputs are atomic fractions of Fe, Al, C, Cr, Mn, Mo, Ni in high-alloy steels Fe-Al-C-Cr-Mn-Mo-Ni; each row sums to 1. Composition range: Fe is the balance and at least 50 at.%; solutes in at.%: Al 0 to 6 (step 3), C 0 to 2 (step 1), Cr 0 to 24 (step 4), Mn 0 to 20 (step 5), Mo 0 to 4 (step 2), Ni 0 to 20 (step 4); at least 4 elements present. The target is the equilibrium mole fraction, at 1173 K, of all phases other than austenite (fcc) and ferrite (bcc): carbides, sigma, Laves and other intermetallic phases in mole fraction between 0 and 1, calculated at 1 atm with the CALPHAD method (Thermo-Calc, TCFE10 database). 200 alloys.
Input order: x[0] = x_Fe, x[1] = x_Al, x[2] = x_C, x[3] = x_Cr, x[4] = x_Mn, x[5] = x_Mo, x[6] = x_Ni.
The output y must be your best estimate of the equilibrium mole fraction, at 1173 K, of all phases other than austenite (fcc) and ferrite (bcc): carbides, sigma, Laves and other intermetallic phases in mole fraction between 0 and 1. [...]
\end{lstlisting}

{\centering\resizebox{\ifdim\width>\linewidth\linewidth\else\width\fi}{!}{\includegraphics{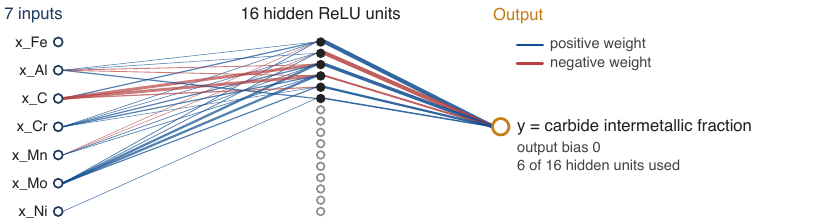}}\par}
\begin{minipage}[t]{0.40\linewidth}
\emph{Raw reply, shortened}
\begin{lstlisting}
## net.0.weight shape=(16, 7)
0 0.1 1 0.18 -0.03 0.4 0
0 0.1 0 0.18 -0.03 0.4 0
[... 13 rows omitted ...]
0 0 0 0 0 0 0
## net.0.bias shape=(16,)
-0.029 -0.029 -0.27 -0.33 -0.035 -0.065 0 0 0 0 0 0 0 0 0 0
## net.2.weight shape=(1, 16)
4.5 -4.5 3.5 -1.5 2 1 0 0 0 0 0 0 0 0 0 0
## net.2.bias shape=(1,)
0
\end{lstlisting}
\end{minipage}\hfill\begin{minipage}[t]{0.58\linewidth}
\emph{The model's own account}
\begin{lstlisting}
I built the prior from four hand-reasoned terms. The first is a carbide term made of two ReLUs that clamps to min(max(C + D - b, 0), C) x 4.5, so carbide fraction grows with C once a carbide-forming index D (Cr, Mo, a little Al, minus Mn) passes a threshold, and it saturates when all the carbon is in M23C6/M7C3/M6C. The second is a sigma/chi term that switches on when Cr + 2.5Mo + 0.3Mn - 0.5Al - 3.5C exceeds about 0.27 (binary Fe-Cr has no sigma at 900 C, while duplex and 6Mo grades do), with a second ReLU that flattens it at high values. The last two are small Mo-Laves and Ni-Al (B2) terms. I would put the rough scale (0 to about 0.1 for carbides, up to about 0.3 for sigma) at about 60% likely to be right, and the exact thresholds are less certain.
\end{lstlisting}
\end{minipage}\par
\suppentry{\raggedright\paragraph{Freezing range} Run folder \texttt{write\_\allowbreak \_\allowbreak steel\_\allowbreak freezing\_\allowbreak range\_\allowbreak \_\allowbreak opus\_\allowbreak \_\allowbreak rep1}.\par}
\begin{lstlisting}
[...] Inputs are atomic fractions of Fe, Al, C, Cr, Mn, Mo, Ni in high-alloy steels Fe-Al-C-Cr-Mn-Mo-Ni; each row sums to 1. Composition range: Fe is the balance and at least 50 at.%; solutes in at.%: Al 0 to 6 (step 3), C 0 to 2 (step 1), Cr 0 to 24 (step 4), Mn 0 to 20 (step 5), Mo 0 to 4 (step 2), Ni 0 to 20 (step 4); at least 4 elements present. The target is the equilibrium freezing range (liquidus temperature minus solidus temperature) in K, calculated at 1 atm with the CALPHAD method (Thermo-Calc, TCFE10 database). 200 alloys.
Input order: x[0] = x_Fe, x[1] = x_Al, x[2] = x_C, x[3] = x_Cr, x[4] = x_Mn, x[5] = x_Mo, x[6] = x_Ni.
The output y must be your best estimate of the equilibrium freezing range (liquidus temperature minus solidus temperature) in K. [...]
\end{lstlisting}

{\centering\resizebox{\ifdim\width>\linewidth\linewidth\else\width\fi}{!}{\includegraphics{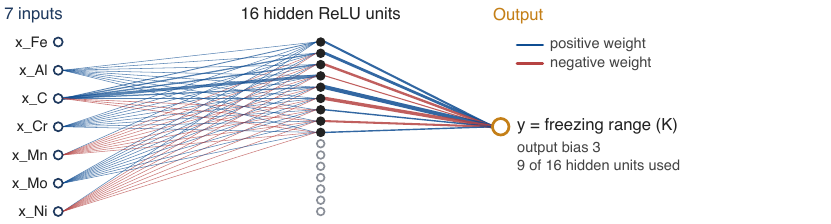}}\par}
\begin{minipage}[t]{0.40\linewidth}
\emph{Raw reply, shortened}
\begin{lstlisting}
## net.0.weight shape=(16, 7)
0 50 2800 10 140 300 50
0 200 6000 100 -75 200 -150
[... 13 rows omitted ...]
0 0 0 0 0 0 0
## net.0.bias shape=(16,)
0 -56 -86 -286 -124 -144 14 4 -6 0 0 0 0 0 0 0
## net.2.weight shape=(1, 16)
1 1.5 -1.25 -0.25 2.5 -2 0.5 -1 0.5 0 0 0 0 0 0 0
## net.2.bias shape=(1,)
3
\end{lstlisting}
\end{minipage}\hfill\begin{minipage}[t]{0.58\linewidth}
\emph{The model's own account}
\begin{lstlisting}
I built the network by hand from lever-rule reasoning: neuron 0 is an additive term per at.% of solute (C about 28 K for austenitic solidification, Mo 3, Mn 1.4, Ni 0.5, Al 0.5, Cr 0.1, which is close to zero because Fe-Cr has a minimum), plus an output bias of 3 K. Neurons 1-5 use large carbon weights so that each switches on at only one carbon level (1 or 2 at.%). They add extra freezing range when the alloy solidifies as delta ferrite, measured by a Cr-equivalent minus 1.5 x Ni-equivalent index, because carbon's partition coefficient in delta is much lower than in austenite. Neurons 6-8 add a small bump of up to 5 K near the delta/gamma boundary, where three phases coexist. The expected scale is about 10-50 K for carbon-free alloys, 35-100 K at 1 at.% C and 65-150 K at 2 at.% C. I am fairly confident about the trends (carbon dominant, then Mn and Mo) but much less sure of the delta-ferrite boost and the absolute values.
\end{lstlisting}
\end{minipage}\par
\suppentry{\raggedright\paragraph{Solidus} Run folder \texttt{write\_\allowbreak \_\allowbreak steel\_\allowbreak solidus\_\allowbreak \_\allowbreak opus\_\allowbreak \_\allowbreak rep1}.\par}
\begin{lstlisting}
[...] Inputs are atomic fractions of Fe, Al, C, Cr, Mn, Mo, Ni in high-alloy steels Fe-Al-C-Cr-Mn-Mo-Ni; each row sums to 1. Composition range: Fe is the balance and at least 50 at.%; solutes in at.%: Al 0 to 6 (step 3), C 0 to 2 (step 1), Cr 0 to 24 (step 4), Mn 0 to 20 (step 5), Mo 0 to 4 (step 2), Ni 0 to 20 (step 4); at least 4 elements present. The target is the equilibrium solidus temperature in K, calculated at 1 atm with the CALPHAD method (Thermo-Calc, TCFE10 database). 200 alloys.
Input order: x[0] = x_Fe, x[1] = x_Al, x[2] = x_C, x[3] = x_Cr, x[4] = x_Mn, x[5] = x_Mo, x[6] = x_Ni.
The output y must be your best estimate of the equilibrium solidus temperature in K. [...]
\end{lstlisting}

{\centering\resizebox{\ifdim\width>\linewidth\linewidth\else\width\fi}{!}{\includegraphics{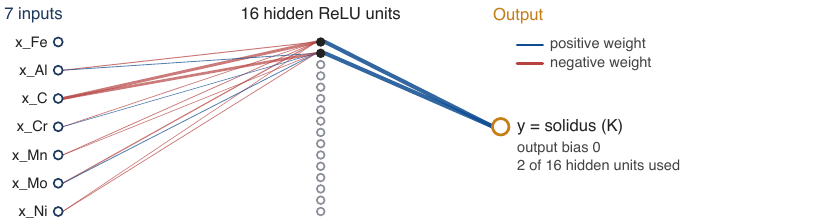}}\par}
\begin{minipage}[t]{0.40\linewidth}
\emph{Raw reply, shortened}
\begin{lstlisting}
## net.0.weight shape=(16, 7)
0 -0.8 -4.0 -0.35 -0.43 -1.1 -0.2
0 0.55 -3.5 0.21 -0.23 0.55 -0.4
[... 13 rows omitted ...]
0 0 0 0 0 0 0
## net.0.bias shape=(16,)
1.798 0.013 0 0 0 0 0 0 0 0 0 0 0 0 0 0
## net.2.weight shape=(1, 16)
1000 1000 0 0 0 0 0 0 0 0 0 0 0 0 0 0
## net.2.bias shape=(1,)
0
\end{lstlisting}
\end{minipage}\hfill\begin{minipage}[t]{0.58\linewidth}
\emph{The model's own account}
\begin{lstlisting}
I modelled the solidus as the higher of two linear solidus planes, one for delta-ferrite (pure-Fe melting 1811 K) and one for metastable austenite (about 1798 K). Their per-at.% slopes come from recalled binary Fe-C, Fe-Ni, Fe-Mn, Fe-Cr, Fe-Mo and Fe-Al diagrams: C lowers the solidus by about 75 K per at.% in ferrite and 40 K per at.% in austenite, Mn by 6.6 and 4.3, Ni by 6 and 2, and Cr, Mo and Al lower the austenite plane more than the ferrite plane. The ferrite/austenite balance was adjusted against Schaeffler/WRC-style equivalents. The network encodes this exactly, as max(a, b) = b + relu(a - b): hidden unit 0 is the austenite plane, unit 1 is the ferrite-minus-austenite difference, both in kK, and the output weights are 1000. I checked it against remembered solidus values for Fe-C, 304/316/310 stainless, Kanthal-type FeCrAl and Fe-Mn-C TWIP steels, which it matches to within roughly 20-40 K. I expect about 70% of the 200 TCFE10 values to fall within 40 K of the prior.
\end{lstlisting}
\end{minipage}\par
\suppneed{2.2in}\subsubsection{Al--Cu--Mg--Mn--Si--Zn aluminium alloys (\texttt{al\_alloy})}
\suppentry{\raggedright\paragraph{Freezing range} Run folder \texttt{write\_\allowbreak \_\allowbreak al\_\allowbreak alloy\_\allowbreak freezing\_\allowbreak range\_\allowbreak \_\allowbreak opus\_\allowbreak \_\allowbreak rep1}.\par}
\begin{lstlisting}
[...] Inputs are atomic fractions of Al, Cu, Mg, Mn, Si, Zn in aluminium alloys Al-Cu-Mg-Mn-Si-Zn; each row sums to 1. Composition range: Al is the balance and at least 85 at.%; solutes in at.%: Cu 0 to 4 (step 1), Mg 0 to 4 (step 1), Mn 0 to 1 (step 0.5), Si 0 to 4 (step 1), Zn 0 to 4 (step 1); at least 4 elements present. The target is the equilibrium freezing range (liquidus temperature minus solidus temperature) in K, calculated at 1 atm with the CALPHAD method (Thermo-Calc, TCAL9 database). 200 alloys.
Input order: x[0] = x_Al, x[1] = x_Cu, x[2] = x_Mg, x[3] = x_Mn, x[4] = x_Si, x[5] = x_Zn.
The output y must be your best estimate of the equilibrium freezing range (liquidus temperature minus solidus temperature) in K. [...]
\end{lstlisting}

{\centering\resizebox{\ifdim\width>\linewidth\linewidth\else\width\fi}{!}{\includegraphics{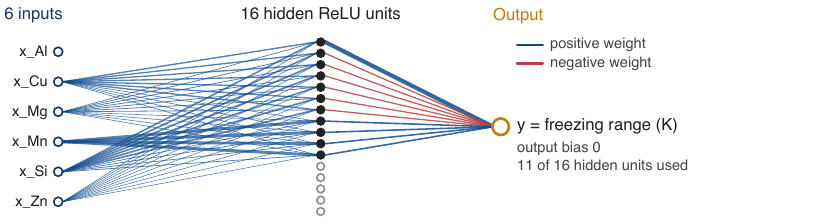}}\par}
\begin{minipage}[t]{0.40\linewidth}
\emph{Raw reply, shortened}
\begin{lstlisting}
## net.0.weight shape=(16, 6)
0 4000 900 0 4550 600
0 3700 1250 0 5030 250
[... 13 rows omitted ...]
0 0 0 0 0 0
## net.0.bias shape=(16,)
0 -70 -82 -94 -106 -118 -130 -95 -105 -115 -125 0 0 0 0 0
## net.2.weight shape=(1, 16)
1 -0.1666667 -0.1666667 -0.1666667 -0.1666667 -0.1666667 -0.1666667 0.25 0.25 0.25 0.25 0 0 0 0 0
## net.2.bias shape=(1,)
0
\end{lstlisting}
\end{minipage}\hfill\begin{minipage}[t]{0.58\linewidth}
\emph{The model's own account}
\begin{lstlisting}
I wrote the freezing range as liquidus minus solidus. Unit 0 holds the single-phase range, the sum over solutes of m_i*c_i*(1/k_i - 1), using binary liquidus slopes and partition coefficients for Cu, Mg, Si and Zn, with Mn taken as about zero. Six ReLU units with staggered offsets cap the solidus at a linear multicomponent eutectic temperature, about 560 C and falling with Cu and Zn, so concentrated alloys saturate at roughly 80 to 130 K. Four more units raise the liquidus when the primary Al6Mn or alpha-AlMnSi liquidus, estimated at about 550 + 11000*x_Mn + 500*x_Si C, exceeds the alpha-Al liquidus. That term mainly adds 30 to 60 K at Mn = 1 at.%. No search task was involved; I expect the trends to be right but the absolute scale could be off by about 30 to 50 K.
\end{lstlisting}
\end{minipage}\par
\suppentry{\raggedright\paragraph{Second phase fraction} Run folder \texttt{write\_\allowbreak \_\allowbreak al\_\allowbreak alloy\_\allowbreak second\_\allowbreak phase\_\allowbreak fraction\_\allowbreak \_\allowbreak opus\_\allowbreak \_\allowbreak rep1}.\par}
\begin{lstlisting}
[...] Inputs are atomic fractions of Al, Cu, Mg, Mn, Si, Zn in aluminium alloys Al-Cu-Mg-Mn-Si-Zn; each row sums to 1. Composition range: Al is the balance and at least 85 at.%; solutes in at.%: Cu 0 to 4 (step 1), Mg 0 to 4 (step 1), Mn 0 to 1 (step 0.5), Si 0 to 4 (step 1), Zn 0 to 4 (step 1); at least 4 elements present. The target is the equilibrium mole fraction, at 673 K, of all solid phases other than the fcc aluminium solid solution (silicon, Al2Cu, Mg2Si, MgZn2 and other compounds) in mole fraction between 0 and 1, calculated at 1 atm with the CALPHAD method (Thermo-Calc, TCAL9 database). 200 alloys.
Input order: x[0] = x_Al, x[1] = x_Cu, x[2] = x_Mg, x[3] = x_Mn, x[4] = x_Si, x[5] = x_Zn.
The output y must be your best estimate of the equilibrium mole fraction, at 673 K, of all solid phases other than the fcc aluminium solid solution (silicon, Al2Cu, Mg2Si, MgZn2 and other compounds) in mole fraction between 0 and 1. [...]
\end{lstlisting}

{\centering\resizebox{\ifdim\width>\linewidth\linewidth\else\width\fi}{!}{\includegraphics{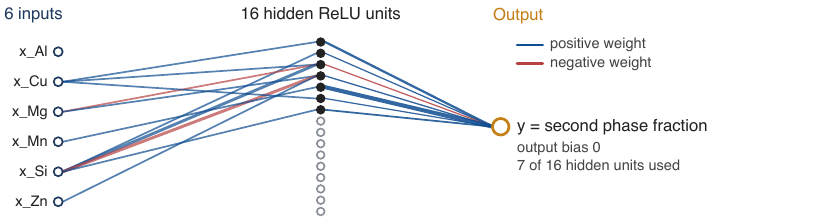}}\par}
\begin{minipage}[t]{0.40\linewidth}
\emph{Raw reply, shortened}
\begin{lstlisting}
## net.0.weight shape=(16, 6)
0 1 0 0 0 0
0 0 0 0 1 0
[... 13 rows omitted ...]
0 0 0 0 0 0
## net.0.bias shape=(16,)
-0.007 -0.003 0 -0.05 0 0 0 0 0 0 0 0 0 0 0 0
## net.2.weight shape=(1, 16)
3 1 -1 1.2 6.5 1 2 0 0 0 0 0 0 0 0 0
## net.2.bias shape=(1,)
0
\end{lstlisting}
\end{minipage}\hfill\begin{minipage}[t]{0.58\linewidth}
\emph{The model's own account}
\begin{lstlisting}
I built an additive stoichiometric mass-balance model of the 673 K equilibrium and encoded it exactly in ReLU units. The model has these terms: Al2Cu at 3 x (x_Cu - 0.007 solubility); Mn phases (Al6Mn, alpha-AlMnSi, Al20Cu2Mn3) at about 6.5 x x_Mn; and Mg2Si at 3 x min(Mg/2, Si). Excess Si above 0.3 at.% forms diamond Si. Free Mg left after Mg2Si pairs with Cu as S-phase Al2CuMg, adding min(Cu, free Mg). A Mg-Zn T/eta term turns on only when Zn + free Mg exceeds about 5 at.%. With the S-phase bonus at weight 1, the Mg2Si/S logic reduces exactly to c + 2s + relu(s - 0.003) - relu(c - m + 2s). Predictions range from about 0 to 0.22, and Cu and Mn are the steepest trends. I expect the trends and scale to be roughly right (about 65% confident the Spearman rank correlation exceeds 0.8). The least certain parts are the Zn-Mg threshold, the Mn multiplier and the Cu solubility.
\end{lstlisting}
\end{minipage}\par
\suppentry{\raggedright\paragraph{Solidus} Run folder \texttt{write\_\allowbreak \_\allowbreak al\_\allowbreak alloy\_\allowbreak solidus\_\allowbreak \_\allowbreak opus\_\allowbreak \_\allowbreak rep1}.\par}
\begin{lstlisting}
[...] Inputs are atomic fractions of Al, Cu, Mg, Mn, Si, Zn in aluminium alloys Al-Cu-Mg-Mn-Si-Zn; each row sums to 1. Composition range: Al is the balance and at least 85 at.%; solutes in at.%: Cu 0 to 4 (step 1), Mg 0 to 4 (step 1), Mn 0 to 1 (step 0.5), Si 0 to 4 (step 1), Zn 0 to 4 (step 1); at least 4 elements present. The target is the equilibrium solidus temperature in K, calculated at 1 atm with the CALPHAD method (Thermo-Calc, TCAL9 database). 200 alloys.
Input order: x[0] = x_Al, x[1] = x_Cu, x[2] = x_Mg, x[3] = x_Mn, x[4] = x_Si, x[5] = x_Zn.
The output y must be your best estimate of the equilibrium solidus temperature in K. [...]
\end{lstlisting}

{\centering\resizebox{\ifdim\width>\linewidth\linewidth\else\width\fi}{!}{\includegraphics{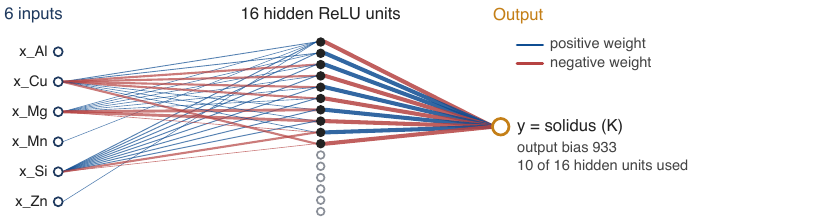}}\par}
\begin{minipage}[t]{0.40\linewidth}
\emph{Raw reply, shortened}
\begin{lstlisting}
## net.0.weight shape=(16, 6)
0 4500 1200 300 5500 750
0 4500 1200 0 5500 0
[... 13 rows omitted ...]
0 0 0 0 0 0
## net.0.bias shape=(16,)
0 -153 -103 -143 97 87 -75 -93 -31 10 0 0 0 0 0 0
## net.2.weight shape=(1, 16)
-1 1 1 -1 1 -1 1 -1 1 -1 0 0 0 0 0 0
## net.2.bias shape=(1,)
933
\end{lstlisting}
\end{minipage}\hfill\begin{minipage}[t]{0.58\linewidth}
\emph{The model's own account}
\begin{lstlisting}
I wrote the network as a max-of-planes solidus model. One unit carries the dilute solution line, 933 K minus per-at.% slopes from the fcc solidus: Cu 45, Si 55, Mg 12, Zn 7.5, Mn 3. A second unit clamps the result at the Al-Cu-Mg-Si (Q/S/theta) eutectic floor near 780 K, and Zn and Mn also lower that floor. Gated clamp pairs raise the floor where an element is missing: about 830-850 K when Cu is absent (Mg2Si+Si or Si eutectic), about 790-805 K at Cu = 1 at.%, 798 K when Mg is absent (theta+Si), and up to 821 K for Cu-rich alloys without Si (theta). I checked the result by hand against a table of guessed solidus values over the Cu-Mg-Si grid; the Mg2Si-ridge region (Cu at most 1, Mg 3-4, Si 1-2) is still up to about 20 K too low. I expect most alloys to fall at 750-800 K, and I am about 60% confident that the trends and scale are within about 25 K of TCAL9.
\end{lstlisting}
\end{minipage}\par
\suppneed{2.2in}\subsubsection{Hf--Mo--Nb--Ta--Ti--Zr refractory high-entropy alloys with Hf (\texttt{hf\_refractory})}
\suppentry{\raggedright\paragraph{Freezing range} Run folder \texttt{write\_\allowbreak \_\allowbreak hf\_\allowbreak refractory\_\allowbreak freezing\_\allowbreak range\_\allowbreak \_\allowbreak opus\_\allowbreak \_\allowbreak rep1}.\par}
\begin{lstlisting}
[...] Inputs are atomic fractions of Hf, Mo, Nb, Ta, Ti, Zr in Hf-Mo-Nb-Ta-Ti-Zr refractory high-entropy alloys; each row sums to 1. Composition range: every element between 0 and 35 at.% on a 5 at.% grid, at least four elements present. The target is the equilibrium freezing range (liquidus temperature minus solidus temperature) in K, calculated at 1 atm with the CALPHAD method (Thermo-Calc, TCHEA8 database). 200 alloys.
Input order: x[0] = x_Hf, x[1] = x_Mo, x[2] = x_Nb, x[3] = x_Ta, x[4] = x_Ti, x[5] = x_Zr.
The output y must be your best estimate of the equilibrium freezing range (liquidus temperature minus solidus temperature) in K. [...]
\end{lstlisting}

{\centering\resizebox{\ifdim\width>\linewidth\linewidth\else\width\fi}{!}{\includegraphics{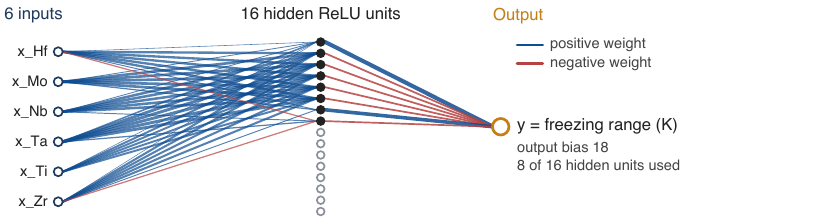}}\par}
\begin{minipage}[t]{0.40\linewidth}
\emph{Raw reply, shortened}
\begin{lstlisting}
## net.0.weight shape=(16, 6)
0.0733 0.4446 0.4414 1.4143 0.075 0.0038
2.506 2.896 2.75 3.29 1.941 2.128
[... 13 rows omitted ...]
0 0 0 0 0 0
## net.0.bias shape=(16,)
0 -2.3 -2.45 -2.6 -2.75 -2.9 -0.4 0 0 0 0 0 0 0 0 0
## net.2.weight shape=(1, 16)
700 -210 -210 -210 -210 -210 600 -200 0 0 0 0 0 0 0 0
## net.2.bias shape=(1,)
18
\end{lstlisting}
\end{minipage}\hfill\begin{minipage}[t]{0.58\linewidth}
\emph{The model's own account}
\begin{lstlisting}
I modelled the freezing range with the ideal-solution lens result, roughly (entropy of fusion / T) times the composition-weighted variance of the elemental melting points. I scaled that up about 1.5x for non-ideality, giving 700 K per kK^2. The variance is written exactly as a positive linear unit (sum of x_i*Tm_i^2) minus a piecewise-linear chord fit of (sum of x_i*Tm_i)^2, which uses five ReLU knots between 2.3 and 2.9 kK. I added a min((Ta+0.6Nb),(Hf+Zr)) term of 200 K per unit for the solidus minima in Zr/Hf-Nb/Ta binaries. I also added a thresholded Mo x (Hf,Zr) unit for a C15 Laves eutectic that would lower the solidus in Mo-rich, Hf/Zr-rich alloys, and a bias of 18 K. Predictions run from about 60 K for Ti-Zr-Hf-rich alloys to about 330 K for Ta-Ti-Zr-rich ones; equiatomic HfNbTaTiZr comes out near 235 K. I am about 50% confident in the absolute scale and more confident in the trends.
\end{lstlisting}
\end{minipage}\par
\suppentry{\raggedright\paragraph{Main phase fraction} Run folder \texttt{write\_\allowbreak \_\allowbreak hf\_\allowbreak refractory\_\allowbreak main\_\allowbreak phase\_\allowbreak fraction\_\allowbreak \_\allowbreak opus\_\allowbreak \_\allowbreak rep1}.\par}
\begin{lstlisting}
[...] Inputs are atomic fractions of Hf, Mo, Nb, Ta, Ti, Zr in Hf-Mo-Nb-Ta-Ti-Zr refractory high-entropy alloys; each row sums to 1. Composition range: every element between 0 and 35 at.% on a 5 at.% grid, at least four elements present. The target is the equilibrium mole fraction, at 1273 K, of the single most abundant phase (1 for a single-phase alloy; two bcc phases of a miscibility gap count as two phases) in mole fraction between 0 and 1, calculated at 1 atm with the CALPHAD method (Thermo-Calc, TCHEA8 database). 199 alloys.
Input order: x[0] = x_Hf, x[1] = x_Mo, x[2] = x_Nb, x[3] = x_Ta, x[4] = x_Ti, x[5] = x_Zr.
The output y must be your best estimate of the equilibrium mole fraction, at 1273 K, of the single most abundant phase (1 for a single-phase alloy; two bcc phases of a miscibility gap count as two phases) in mole fraction between 0 and 1. [...]
\end{lstlisting}

{\centering\resizebox{\ifdim\width>\linewidth\linewidth\else\width\fi}{!}{\includegraphics{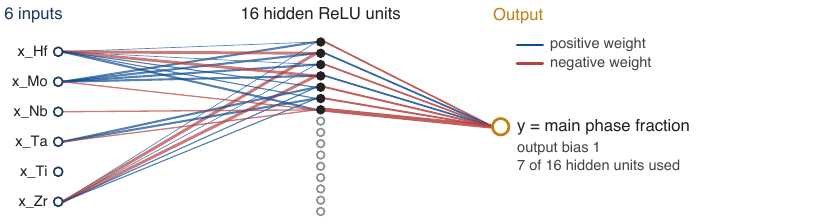}}\par}
\begin{minipage}[t]{0.40\linewidth}
\emph{Raw reply, shortened}
\begin{lstlisting}
## net.0.weight shape=(16, 6)
0.1 1 0 0 0 0.1
-1.5 1 0 0 0 -1.5
[... 13 rows omitted ...]
0 0 0 0 0 0
## net.0.bias shape=(16,)
-0.06 -0.06 -0.14 -0.14 -0.32 -0.40 -0.25 0 0 0 0 0 0 0 0 0
## net.2.weight shape=(1, 16)
-0.75 0.75 -0.75 0.75 -0.9 -0.9 -2.5 0 0 0 0 0 0 0 0 0
## net.2.bias shape=(1,)
1.0
\end{lstlisting}
\end{minipage}\hfill\begin{minipage}[t]{0.58\linewidth}
\emph{The model's own account}
\begin{lstlisting}
I wrote the network as 1 minus three non-negative penalties, each built by hand from ReLU units. The first is a C15 Laves term, (Hf,Zr)Mo2. Two pairs of units of the form relu(Mo + 0.1(Hf+Zr) - s) - relu(Mo - 1.5(Hf+Zr) - s) approximate 1.5 x min(Mo, 1.5(Hf+Zr)) above a Mo solubility of about 0.06 to 0.14, and the term is exactly zero when there is no Hf or Zr. The second is a bcc miscibility-gap term, two units relu(Ta + 0.4(Hf+Zr) - t), so Ta-rich and Zr/Hf-rich alloys split into two bcc phases. The third is a small hcp term for Hf-rich alloys with few Mo/Nb/Ta, and all the ReLU breakpoints were placed so that HfNbTaTiZr and Mo-Nb-Ta-Ti alloys come out near 1. The trends come from binary phase diagrams and general knowledge of RHEAs, not from recalled TCHEA8 numbers; I put about 60% on the trend direction and only about 30% on the absolute scale of the Laves and gap penalties.
\end{lstlisting}
\end{minipage}\par
\suppentry{\raggedright\paragraph{Solidus} Run folder \texttt{write\_\allowbreak \_\allowbreak hf\_\allowbreak refractory\_\allowbreak solidus\_\allowbreak \_\allowbreak opus\_\allowbreak \_\allowbreak rep1}.\par}
\begin{lstlisting}
[...] Inputs are atomic fractions of Hf, Mo, Nb, Ta, Ti, Zr in Hf-Mo-Nb-Ta-Ti-Zr refractory high-entropy alloys; each row sums to 1. Composition range: every element between 0 and 35 at.% on a 5 at.% grid, at least four elements present. The target is the equilibrium solidus temperature in K, calculated at 1 atm with the CALPHAD method (Thermo-Calc, TCHEA8 database). 200 alloys.
Input order: x[0] = x_Hf, x[1] = x_Mo, x[2] = x_Nb, x[3] = x_Ta, x[4] = x_Ti, x[5] = x_Zr.
The output y must be your best estimate of the equilibrium solidus temperature in K. [...]
\end{lstlisting}

{\centering\resizebox{\ifdim\width>\linewidth\linewidth\else\width\fi}{!}{\includegraphics{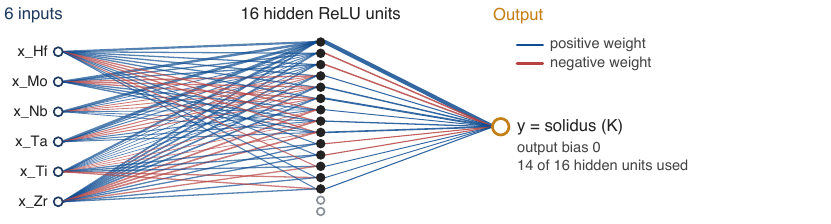}}\par}
\begin{minipage}[t]{0.40\linewidth}
\emph{Raw reply, shortened}
\begin{lstlisting}
## net.0.weight shape=(16, 6)
2.08792 2.5168 2.65959 3.08922 1.911 1.4932
1 1 0.427 0.646 0.2 1.4
[... 13 rows omitted ...]
0 0 0 0 0 0
## net.0.bias shape=(16,)
0 -0.65 -0.85 0 -0.2 -0.45 0 -0.15 -0.35 -0.3 -0.6 0 0 -0.3 0 0
## net.2.weight shape=(1, 16)
1000 -216 -216 96 216 336 72 168 264 -108 -117 63 54 108 0 0
## net.2.bias shape=(1,)
0
\end{lstlisting}
\end{minipage}\hfill\begin{minipage}[t]{0.58\linewidth}
\emph{The model's own account}
\begin{lstlisting}
I modelled the solidus as the rule-of-mixtures of the pure-element melting points (Hf 2506, Mo 2896, Nb 2750, Ta 3290, Ti 1941, Zr 2128 K) minus pairwise regular-solution depressions estimated from the binary phase diagrams: large for Zr/Hf with Mo (Laves eutectics) and with Ta (bcc miscibility gap), moderate for Nb-Zr, Hf-Nb and Ti-Zr/Ti-Hf (melting minima), and small among Mo, Nb and Ta. I factored the depression as 1920*s*t with s = x_Hf + 1.4x_Zr + 0.2x_Ti and t = x_Mo + 0.646x_Ta + 0.427x_Nb, plus 720*x_Ti*(x_Zr + 0.72x_Hf). Each product is written as a difference of squares, and each square is a piecewise-linear ReLU interpolant. One always-positive unit carries the linear part, and two units are left at zero. The predictions run from about 2000 K for Ti-Zr-Hf-rich or Zr-Mo-rich alloys to about 2930 K for Mo-Ta-Nb-rich alloys, with equiatomic HfNbTaTiZr near 2260 K. I am about 60% confident that the trends match TCHEA8 well. The absolute level may be off by 100 to 200 K, and the size of the Mo-Zr/Hf depression is the most uncertain part.
\end{lstlisting}
\end{minipage}\par
\suppneed{2.2in}\subsubsection{Co--Cr--Cu--Fe--Ni--Ti 3d high-entropy alloys with Cu and Ti (\texttt{cu\_ti\_3d})}
\suppentry{\raggedright\paragraph{Freezing range} Run folder \texttt{write\_\allowbreak \_\allowbreak cu\_\allowbreak ti\_\allowbreak 3d\_\allowbreak freezing\_\allowbreak range\_\allowbreak \_\allowbreak opus\_\allowbreak \_\allowbreak rep1}.\par}
\begin{lstlisting}
[...] Inputs are atomic fractions of Co, Cr, Cu, Fe, Ni, Ti in Co-Cr-Cu-Fe-Ni-Ti high-entropy alloys; each row sums to 1. Composition range: every element between 0 and 35 at.% on a 5 at.% grid, at least four elements present. The target is the equilibrium freezing range (liquidus temperature minus solidus temperature) in K, calculated at 1 atm with the CALPHAD method (Thermo-Calc, TCHEA8 database). 200 alloys.
Input order: x[0] = x_Co, x[1] = x_Cr, x[2] = x_Cu, x[3] = x_Fe, x[4] = x_Ni, x[5] = x_Ti.
The output y must be your best estimate of the equilibrium freezing range (liquidus temperature minus solidus temperature) in K. [...]
\end{lstlisting}

{\centering\resizebox{\ifdim\width>\linewidth\linewidth\else\width\fi}{!}{\includegraphics{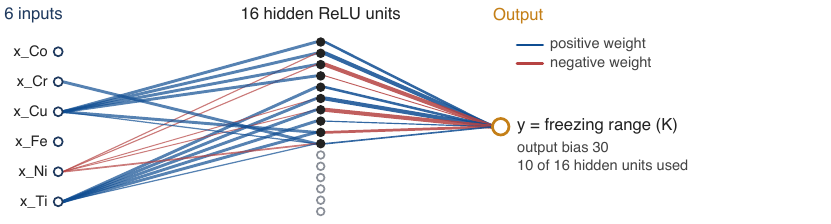}}\par}
\begin{minipage}[t]{0.40\linewidth}
\emph{Raw reply, shortened}
\begin{lstlisting}
## net.0.weight shape=(16, 6)
0 0 10 0 0 0
0 0 10 0 -2.5 0
[... 13 rows omitted ...]
0 0 0 0 0 0
## net.0.bias shape=(16,)
0 -0.4 -1.2 -1.5 0 -0.3 -1.1 -1.5 -1.5 -1.5 0 0 0 0 0 0
## net.2.weight shape=(1, 16)
80 150 -150 -10 80 150 -150 20 -90 40 0 0 0 0 0 0
## net.2.bias shape=(1,)
30
\end{lstlisting}
\end{minipage}\hfill\begin{minipage}[t]{0.58\linewidth}
\emph{The model's own account}
\begin{lstlisting}
I built the network by hand from metallurgy, without running any code. The baseline is about 30 K for Cu- and Ti-free CoCrFeNi. Cu adds a saturating, step-like term of about 230-270 K once it exceeds its solubility: past that point a Cu-rich liquid survives down to about 1350 K, and the threshold moves to higher Cu as Ni rises. Ti adds a similar saturating term of about 250 K from low-melting Ti eutectics and intermetallics. A negative Cu+Ti interaction keeps the combined effect below the simple sum, near 350-450 K, and one unit makes the range grow with Cr (higher liquidus) and shrink with Ni. Nothing was to be found, so no confidence is given; I expect the trends to be right and the absolute scale to be uncertain by roughly a factor of 1.5.
\end{lstlisting}
\end{minipage}\par
\suppentry{\raggedright\paragraph{Intermetallic fraction} Run folder \texttt{write\_\allowbreak \_\allowbreak cu\_\allowbreak ti\_\allowbreak 3d\_\allowbreak intermetallic\_\allowbreak fraction\_\allowbreak \_\allowbreak opus\_\allowbreak \_\allowbreak rep1}.\par}
\begin{lstlisting}
[...] Inputs are atomic fractions of Co, Cr, Cu, Fe, Ni, Ti in Co-Cr-Cu-Fe-Ni-Ti high-entropy alloys; each row sums to 1. Composition range: every element between 0 and 35 at.% on a 5 at.% grid, at least four elements present. The target is the equilibrium mole fraction, at 1173 K, of all phases other than liquid and the disordered fcc, bcc and hcp solid solutions (that is: sigma, Laves and other intermetallic phases, and ordered B2 or L12) in mole fraction between 0 and 1, calculated at 1 atm with the CALPHAD method (Thermo-Calc, TCHEA8 database). 199 alloys.
Input order: x[0] = x_Co, x[1] = x_Cr, x[2] = x_Cu, x[3] = x_Fe, x[4] = x_Ni, x[5] = x_Ti.
The output y must be your best estimate of the equilibrium mole fraction, at 1173 K, of all phases other than liquid and the disordered fcc, bcc and hcp solid solutions (that is: sigma, Laves and other intermetallic phases, and ordered B2 or L12) in mole fraction between 0 and 1. [...]
\end{lstlisting}

{\centering\resizebox{\ifdim\width>\linewidth\linewidth\else\width\fi}{!}{\includegraphics{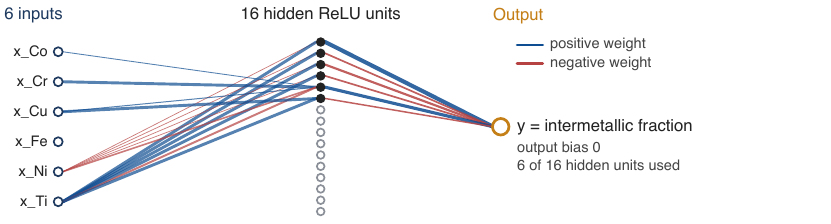}}\par}
\begin{minipage}[t]{0.40\linewidth}
\emph{Raw reply, shortened}
\begin{lstlisting}
## net.0.weight shape=(16, 6)
0 0 0 0 -0.08 1
0 0 0 0 -0.08 1
[... 13 rows omitted ...]
0 0 0 0 0 0
## net.0.bias shape=(16,)
-0.005 -0.185 -0.265 -0.325 -0.26 -0.45 0 0 0 0 0 0 0 0 0 0
## net.2.weight shape=(1, 16)
3.5 -1 -1.5 -1 2.5 -0.8 0 0 0 0 0 0 0 0 0 0
## net.2.bias shape=(1,)
0
\end{lstlisting}
\end{minipage}\hfill\begin{minipage}[t]{0.58\linewidth}
\emph{The model's own account}
\begin{lstlisting}
I treated Ti as the main driver of intermetallics (Laves, eta/Ni3Ti, B2, Cu-Ti compounds): about 3.5 mole fraction of phase per unit Ti above a small solubility threshold, which rises slightly with Ni. Three hinge units bend this curve so it levels off near 0.9 at 33-35 at.% Ti. A separate hinge unit adds sigma for Cr-rich, Ni-poor alloys: Co and Cu push it up, Ti pulls it down so the total stays near 1. A small negative hinge on Cu+Ti stands for a Cu-Ti-rich liquid that may appear at 1173 K. I built this from qualitative phase-diagram knowledge; I estimate the trends are right and the scale is within about a factor of 1.5 for most alloys, with roughly 60% confidence.
\end{lstlisting}
\end{minipage}\par
\suppentry{\raggedright\paragraph{Main phase fraction} Run folder \texttt{write\_\allowbreak \_\allowbreak cu\_\allowbreak ti\_\allowbreak 3d\_\allowbreak main\_\allowbreak phase\_\allowbreak fraction\_\allowbreak \_\allowbreak opus\_\allowbreak \_\allowbreak rep1}.\par}
\begin{lstlisting}
[...] Inputs are atomic fractions of Co, Cr, Cu, Fe, Ni, Ti in Co-Cr-Cu-Fe-Ni-Ti high-entropy alloys; each row sums to 1. Composition range: every element between 0 and 35 at.% on a 5 at.% grid, at least four elements present. The target is the equilibrium mole fraction, at 1173 K, of the single most abundant phase (1 for a single-phase alloy; two bcc phases of a miscibility gap count as two phases) in mole fraction between 0 and 1, calculated at 1 atm with the CALPHAD method (Thermo-Calc, TCHEA8 database). 199 alloys.
Input order: x[0] = x_Co, x[1] = x_Cr, x[2] = x_Cu, x[3] = x_Fe, x[4] = x_Ni, x[5] = x_Ti.
The output y must be your best estimate of the equilibrium mole fraction, at 1173 K, of the single most abundant phase (1 for a single-phase alloy; two bcc phases of a miscibility gap count as two phases) in mole fraction between 0 and 1. [...]
\end{lstlisting}

{\centering\resizebox{\ifdim\width>\linewidth\linewidth\else\width\fi}{!}{\includegraphics{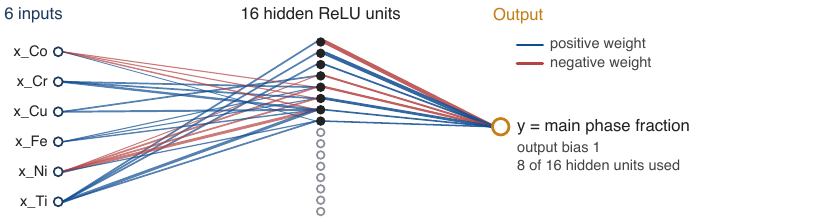}}\par}
\begin{minipage}[t]{0.40\linewidth}
\emph{Raw reply, shortened}
\begin{lstlisting}
## net.0.weight shape=(16, 6)
0 0 0 0 0 1
0 0 0 0 0 1
[... 13 rows omitted ...]
0 0 0 0 0 0
## net.0.bias shape=(16,)
-0.02 -0.16 -0.22 -0.03 -0.1 -0.35 -0.736 -0.42 0 0 0 0 0 0 0 0
## net.2.weight shape=(1, 16)
-3.3 3.0 1.0 -1.2 -1.5 2.0 0.9 0.8 0 0 0 0 0 0 0 0
## net.2.bias shape=(1,)
1.0
\end{lstlisting}
\end{minipage}\hfill\begin{minipage}[t]{0.58\linewidth}
\emph{The model's own account}
\begin{lstlisting}
I built the output as 1 (single-phase FCC for Co/Fe/Ni-rich, Ti-free, Cu-poor alloys) minus hand-placed ReLU penalties. Three Ti hinges give the intermetallic (Laves/eta) loss, which is steep up to about 16 at.% Ti and then recovers as the intermetallic becomes the main phase. One hinge covers the Cu-rich FCC separating out, with Cu solubility rising with Ni. Two hinges on Cr + 0.5Fe - Ni - 0.5Co cover sigma/BCC, and two positive "max" units approximate the switch to a Laves/BCC or Ni-Ti/eta main phase when the penalties stack up. All weights were set by reasoning about the Cantor-type phase stability at 900 C and checked by hand on about a dozen representative compositions. I expect the trends to be roughly right, but the absolute fractions are uncertain to about +/-0.2.
\end{lstlisting}
\end{minipage}\par
\suppentry{\raggedright\paragraph{Solidus} Run folder \texttt{write\_\allowbreak \_\allowbreak cu\_\allowbreak ti\_\allowbreak 3d\_\allowbreak solidus\_\allowbreak \_\allowbreak opus\_\allowbreak \_\allowbreak rep1}.\par}
\begin{lstlisting}
[...] Inputs are atomic fractions of Co, Cr, Cu, Fe, Ni, Ti in Co-Cr-Cu-Fe-Ni-Ti high-entropy alloys; each row sums to 1. Composition range: every element between 0 and 35 at.% on a 5 at.% grid, at least four elements present. The target is the equilibrium solidus temperature in K, calculated at 1 atm with the CALPHAD method (Thermo-Calc, TCHEA8 database). 200 alloys.
Input order: x[0] = x_Co, x[1] = x_Cr, x[2] = x_Cu, x[3] = x_Fe, x[4] = x_Ni, x[5] = x_Ti.
The output y must be your best estimate of the equilibrium solidus temperature in K. [...]
\end{lstlisting}

{\centering\resizebox{\ifdim\width>\linewidth\linewidth\else\width\fi}{!}{\includegraphics{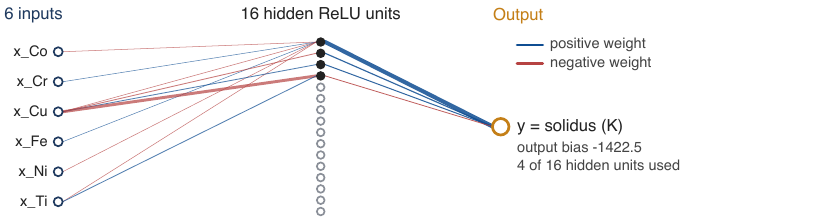}}\par}
\begin{minipage}[t]{0.40\linewidth}
\emph{Raw reply, shortened}
\begin{lstlisting}
## net.0.weight shape=(16, 6)
-0.01 0.04 -1.8 0.02 -0.02 -0.75
0 0 -20 0 0 0
[... 13 rows omitted ...]
0 0 0 0 0 0
## net.0.bias shape=(16,)
3 1 -2 0 0 0 0 0 0 0 0 0 0 0 0 0
## net.2.weight shape=(1, 16)
1000 130 84 -22.5 0 0 0 0 0 0 0 0 0 0 0 0
## net.2.bias shape=(1,)
-1422.5
\end{lstlisting}
\end{minipage}\hfill\begin{minipage}[t]{0.58\linewidth}
\emph{The model's own account}
\begin{lstlisting}
I built the prior by hand from a physical picture: a CoCrFeNi FCC matrix with a solidus near 1700 K. Any Cu above its solubility forms a Cu-rich liquid near 1380 K, so the network has a near-step Cu term that saturates, using the 5 at.% grid so that relu(1-20x_Cu) works as an exact Cu-absent indicator. Ti lowers the solidus through Laves/eta/Cu-Ti eutectics, by about 12 K per at.% without Cu and about 7.5 K per at.% when the Cu-rich liquid controls (relu(20x_Ti-140x_Cu) gates the steeper slope to Cu-free alloys), with small linear matrix terms (Cr up, Ni down). I checked it at a few points (CoCrFeNi about 1715 K, CoCrFeNiTi about 1474 K, CoCrCuFeNi about 1392 K, a near-equiatomic 6-element alloy about 1284 K, a Cu/Ti-rich floor about 1105 K); I am about 60% confident that the trends are right and the scale is within about 100 K for most alloys.
\end{lstlisting}
\end{minipage}\par
\suppneed{2.2in}\subsubsection{Al--Cr--Mo--Nb--Ti refractory high-entropy alloys with Al (\texttt{al\_refractory})}
\suppentry{\raggedright\paragraph{Freezing range} Run folder \texttt{write\_\allowbreak \_\allowbreak al\_\allowbreak refractory\_\allowbreak freezing\_\allowbreak range\_\allowbreak \_\allowbreak opus\_\allowbreak \_\allowbreak rep1}.\par}
\begin{lstlisting}
[...] Inputs are atomic fractions of Al, Cr, Mo, Nb, Ti in Al-Cr-Mo-Nb-Ti refractory high-entropy alloys; each row sums to 1. Composition range: every element between 0 and 35 at.% on a 5 at.% grid, at least four elements present. The target is the equilibrium freezing range (liquidus temperature minus solidus temperature) in K, calculated at 1 atm with the CALPHAD method (Thermo-Calc, TCHEA8 database). 200 alloys.
Input order: x[0] = x_Al, x[1] = x_Cr, x[2] = x_Mo, x[3] = x_Nb, x[4] = x_Ti.
The output y must be your best estimate of the equilibrium freezing range (liquidus temperature minus solidus temperature) in K. [...]
\end{lstlisting}

{\centering\resizebox{\ifdim\width>\linewidth\linewidth\else\width\fi}{!}{\includegraphics{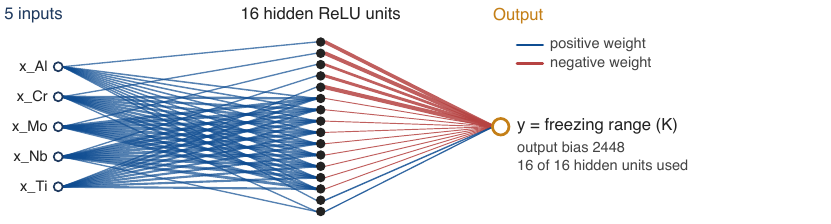}}\par}
\begin{minipage}[t]{0.40\linewidth}
\emph{Raw reply, shortened}
\begin{lstlisting}
## net.0.weight shape=(16, 5)
1 0 0 0 0
0 1 0 0 0
[... 13 rows omitted ...]
0 1 0 1 0
## net.0.bias shape=(16,)
0 0 0 0 0 -1.8 -1.9 -2.0 -2.1 -2.2 -2.3 -2.4 -2.5 -2.6 -0.2 -0.4
## net.2.weight shape=(1, 16)
-2108 -2302.08 -1392 -1650 -2421.12 -160 -160 -160 -160 -160 -160 -160 -160 -160 500 400
## net.2.bias shape=(1,)
2448
\end{lstlisting}
\end{minipage}\hfill\begin{minipage}[t]{0.58\linewidth}
\emph{The model's own account}
\begin{lstlisting}
I modelled the freezing range as 800 K/kK^2 times the composition-weighted variance of effective element melting points (Al 1.2, Cr 2.18, Mo 2.9, Nb 2.75, Ti 1.94 kK), on the idea that a wider melting-point spread means stronger partitioning and a wider liquidus-solidus gap. The network computes it exactly: five identity units carry the linear sum x_i*T_i^2 term, and nine ReLU knots on the mean melting point m (spaced 0.1 kK, from 1.8 to 2.6) give a piecewise-linear -m^2, accurate to about 2 K. Two more units add extra widening for Al above 20 at.% (Al-rich last liquid and intermetallics) and for combined Cr+Nb above 40 at.% (Laves eutectic). Predictions run from about 90 K for Al-free Mo/Nb-rich alloys, through about 320 K for the equiatomic alloy, to about 590 K for Al35Mo35 alloys. I am fairly confident in the trends (Al and the Mo-versus-Al/Ti spread widen the range) and much less confident in the absolute scale, which could be off by a factor of two.
\end{lstlisting}
\end{minipage}\par
\suppentry{\raggedright\paragraph{Intermetallic fraction} Run folder \texttt{write\_\allowbreak \_\allowbreak al\_\allowbreak refractory\_\allowbreak intermetallic\_\allowbreak fraction\_\allowbreak \_\allowbreak opus\_\allowbreak \_\allowbreak rep1}.\par}
\begin{lstlisting}
[...] Inputs are atomic fractions of Al, Cr, Mo, Nb, Ti in Al-Cr-Mo-Nb-Ti refractory high-entropy alloys; each row sums to 1. Composition range: every element between 0 and 35 at.% on a 5 at.% grid, at least four elements present. The target is the equilibrium mole fraction, at 1273 K, of all phases other than liquid and the disordered fcc, bcc and hcp solid solutions (that is: sigma, Laves and other intermetallic phases, and ordered B2 or L12) in mole fraction between 0 and 1, calculated at 1 atm with the CALPHAD method (Thermo-Calc, TCHEA8 database). 200 alloys.
Input order: x[0] = x_Al, x[1] = x_Cr, x[2] = x_Mo, x[3] = x_Nb, x[4] = x_Ti.
The output y must be your best estimate of the equilibrium mole fraction, at 1273 K, of all phases other than liquid and the disordered fcc, bcc and hcp solid solutions (that is: sigma, Laves and other intermetallic phases, and ordered B2 or L12) in mole fraction between 0 and 1. [...]
\end{lstlisting}

{\centering\resizebox{\ifdim\width>\linewidth\linewidth\else\width\fi}{!}{\includegraphics{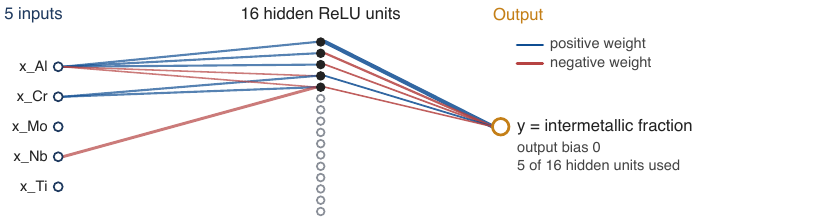}}\par}
\begin{minipage}[t]{0.40\linewidth}
\emph{Raw reply, shortened}
\begin{lstlisting}
## net.0.weight shape=(16, 5)
1 0 0 0 0
1 0 0 0 0
[... 13 rows omitted ...]
0 0 0 0 0
## net.0.bias shape=(16,)
-0.05 -0.2 -0.25 -0.05 -0.05 0 0 0 0 0 0 0 0 0 0 0
## net.2.weight shape=(1, 16)
4.5 -2.0 -1.3 1.4 -1.1 0 0 0 0 0 0 0 0 0 0 0
## net.2.bias shape=(1,)
0
\end{lstlisting}
\end{minipage}\hfill\begin{minipage}[t]{0.58\linewidth}
\emph{The model's own account}
\begin{lstlisting}
I built a piecewise-linear prior by hand from two mechanisms I expect TCHEA8 to predict at 1273 K: Al-driven B2 ordering and Al-rich intermetallics (A15/sigma/alpha2), modeled with three ReLU units on x_Al that give zero below 5 at.% Al, rise steeply to about 0.68 at 20 at.% Al and saturate near 0.92 at 35 at.%; and Cr2Nb-type Laves, modeled as about 1.4 times the Cr in excess of a 5 at.% solubility, capped by Nb availability (a min built from two ReLUs) and with the Cr threshold raised by Al so the sum stays near or below 1 when the bcc matrix is already ordered. Mo and Ti enter only implicitly. The largest uncertainty is whether TCHEA8 actually orders the bcc to B2 at 1273 K for 10-20 at.% Al; if it does not, the Al ramp is too steep, which is why I set it as a moderate ramp rather than a step.
\end{lstlisting}
\end{minipage}\par
\suppentry{\raggedright\paragraph{Main phase fraction} Run folder \texttt{write\_\allowbreak \_\allowbreak al\_\allowbreak refractory\_\allowbreak main\_\allowbreak phase\_\allowbreak fraction\_\allowbreak \_\allowbreak opus\_\allowbreak \_\allowbreak rep1}.\par}
\begin{lstlisting}
[...] Inputs are atomic fractions of Al, Cr, Mo, Nb, Ti in Al-Cr-Mo-Nb-Ti refractory high-entropy alloys; each row sums to 1. Composition range: every element between 0 and 35 at.% on a 5 at.% grid, at least four elements present. The target is the equilibrium mole fraction, at 1273 K, of the single most abundant phase (1 for a single-phase alloy; two bcc phases of a miscibility gap count as two phases) in mole fraction between 0 and 1, calculated at 1 atm with the CALPHAD method (Thermo-Calc, TCHEA8 database). 200 alloys.
Input order: x[0] = x_Al, x[1] = x_Cr, x[2] = x_Mo, x[3] = x_Nb, x[4] = x_Ti.
The output y must be your best estimate of the equilibrium mole fraction, at 1273 K, of the single most abundant phase (1 for a single-phase alloy; two bcc phases of a miscibility gap count as two phases) in mole fraction between 0 and 1. [...]
\end{lstlisting}

{\centering\resizebox{\ifdim\width>\linewidth\linewidth\else\width\fi}{!}{\includegraphics{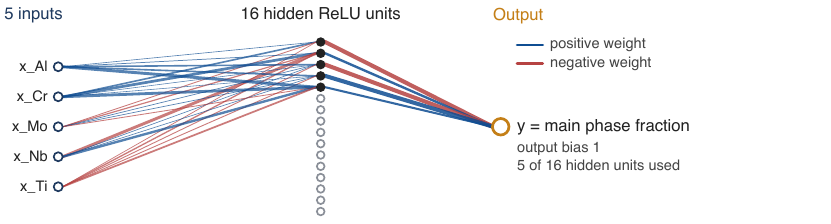}}\par}
\begin{minipage}[t]{0.40\linewidth}
\emph{Raw reply, shortened}
\begin{lstlisting}
## net.0.weight shape=(16, 5)
0.1 1.0 -0.3 0 -0.2
0.2 2.0 -0.6 -2.2 -0.73
[... 13 rows omitted ...]
0 0 0 0 0
## net.0.bias shape=(16,)
-0.03 -0.06 -0.13 -0.355 -1.17 0 0 0 0 0 0 0 0 0 0 0
## net.2.weight shape=(1, 16)
-2.0 1.0 -2.0 2.0 0.8 0 0 0 0 0 0 0 0 0 0 0
## net.2.bias shape=(1,)
1.0
\end{lstlisting}
\end{minipage}\hfill\begin{minipage}[t]{0.58\linewidth}
\emph{The model's own account}
\begin{lstlisting}
I wrote the network as 1 minus two penalties, using five hidden units. The first penalty is for the Laves phase. It is about 2 x relu(Cr - 0.03 - 0.3Mo - 0.2Ti + 0.1Al), because Mo and Ti raise how much Cr the BCC phase can dissolve. A second unit caps it at about 2.2 x (Nb + 0.15Ti), so the Nb supply limits how much NbCr2-type Laves can form. The second penalty is for Al-driven phases (A15 (Nb,Mo)3Al, sigma, or a B2/A2 miscibility gap). It is 2 x relu(Al - 0.13 + 0.15Mo + 0.1Nb + 0.05Cr - 0.35Ti), capped at 0.45; Ti lowers it because it stabilises a single B2 phase. A last unit raises the output back toward about 0.5 when both penalties are large and Nb is high, because there the main-phase fraction flattens or a Laves phase becomes the majority. I checked about 15 compositions by hand, for example the equiatomic alloy at about 0.70 and Cr- and Nb-rich alloys at about 0.5 to 0.6. I trust the trends (Cr x Nb drives Laves, Mo and Ti stabilise BCC, high Al lowers the fraction) at about 60%, and the absolute values against TCHEA8 at about 35%.
\end{lstlisting}
\end{minipage}\par
\suppentry{\raggedright\paragraph{Solidus} Run folder \texttt{write\_\allowbreak \_\allowbreak al\_\allowbreak refractory\_\allowbreak solidus\_\allowbreak \_\allowbreak opus\_\allowbreak \_\allowbreak rep1}.\par}
\begin{lstlisting}
[...] Inputs are atomic fractions of Al, Cr, Mo, Nb, Ti in Al-Cr-Mo-Nb-Ti refractory high-entropy alloys; each row sums to 1. Composition range: every element between 0 and 35 at.% on a 5 at.% grid, at least four elements present. The target is the equilibrium solidus temperature in K, calculated at 1 atm with the CALPHAD method (Thermo-Calc, TCHEA8 database). 200 alloys.
Input order: x[0] = x_Al, x[1] = x_Cr, x[2] = x_Mo, x[3] = x_Nb, x[4] = x_Ti.
The output y must be your best estimate of the equilibrium solidus temperature in K. [...]
\end{lstlisting}

{\centering\resizebox{\ifdim\width>\linewidth\linewidth\else\width\fi}{!}{\includegraphics{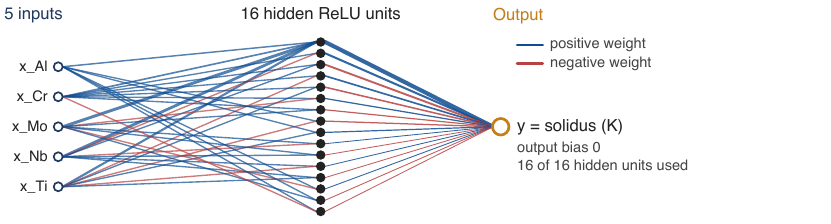}}\par}
\begin{minipage}[t]{0.40\linewidth}
\emph{Raw reply, shortened}
\begin{lstlisting}
## net.0.weight shape=(16, 5)
1.1475 1.34 2.856 2.71 1.941
0 1 0 -1 0
[... 13 rows omitted ...]
1 -1 0 0 0
## net.0.bias shape=(16,)
0 0 -0.4 0 -0.4 0 -0.4 0 -0.4 0 -0.4 0 -0.4 0 -0.4 0
## net.2.weight shape=(1, 16)
1000 400 -350 280 -245 160 -140 -160 140 40 -35 40 -35 50 -43.75 -37.5
## net.2.bias shape=(1,)
0
\end{lstlisting}
\end{minipage}\hfill\begin{minipage}[t]{0.58\linewidth}
\emph{The model's own account}
\begin{lstlisting}
I wrote the network as a rule-of-mixtures linear term (unary melting points, with Al given an effective ~1000-1150 K) minus pairwise solidus depressions c_ij*x_i*x_j estimated from the binary phase diagrams I remember (large for Cr-Nb Laves/eutectic ~2000 K and Cr-Ti bcc minimum ~1400 K, moderate for Cr-Mo, small for Mo-Nb, Nb-Ti, Al-Mo, and a raising term for Al-Ti and Al-Cr). Each product x_i*x_j is approximated with ReLUs as 0.2*min(x_i,x_j) + 0.175*relu(x_i+x_j-0.4), where min = x_i - relu(x_i-x_j) and the linear part goes into one always-active unit. By hand calculation this gives about 2000 K for equimolar AlCrMoNbTi, about 2130 K for CrMoNbTi, about 2480 K for Mo/Nb-rich alloys and about 1690 K for Al+Cr+Ti-rich alloys. I trust the trends (Al and Cr lower the solidus, Mo and Nb raise it, Cr+Ti and Cr+Nb lower it further) more than the absolute scale, which could be off by roughly 100-200 K.
\end{lstlisting}
\end{minipage}\par
\subsection{Crystal tasks}\label{supp:crystals}
Run folders are under \texttt{experiments/\allowbreak e3\_\allowbreak crystal\_\allowbreak filter/\allowbreak runs}. The inputs are the 15 descriptors listed in the complete prompt of Section~\ref{supp:prompts}.
\suppentry{\raggedright\paragraph{Bulk modulus (GPa)} Run folder \texttt{filter\_\allowbreak K\_\allowbreak \_\allowbreak opus\_\allowbreak \_\allowbreak rep1}.\par}
\begin{lstlisting}
[...] The network scores crystals proposed by a generative model (inorganic, up to 20 atoms per cell, any elements) so that only the most promising are sent to an expensive calculation. The output y must be your best estimate of the crystal's bulk modulus in GPa; what matters most is that crystals with a high true value get a high score.
[... the input list x[0] to x[14], the group definitions and the typical ranges follow, identical for both crystal targets ...]
\end{lstlisting}

{\centering\resizebox{\ifdim\width>\linewidth\linewidth\else\width\fi}{!}{\includegraphics{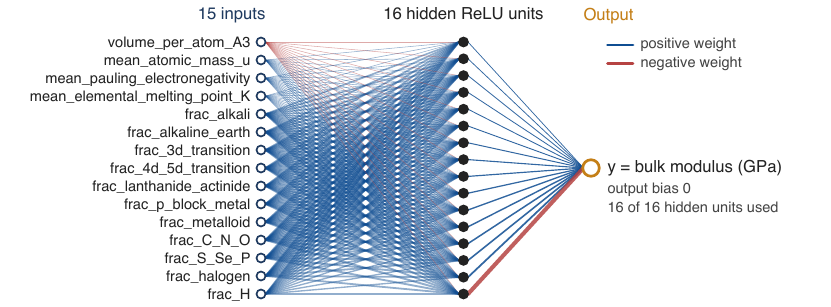}}\par}
\emph{Raw reply, shortened}
\begin{lstlisting}
## net.0.weight shape=(16, 15)
-0.08 0.002 0.3 0.00035 5.15 5.8 6.24 6.32 6.0 6.1 6.2 6.95 6.5 6.05 4.0
-0.08 0.002 0.3 0.00035 5.15 5.8 6.24 6.32 6.0 6.1 6.2 6.95 6.5 6.05 4.0
[... 13 rows omitted ...]
-0.08 0.002 0.3 0.00035 5.15 5.8 6.24 6.32 6.0 6.1 6.2 6.95 6.5 6.05 4.0
## net.0.bias shape=(16,)
6.55 -1.45 -2.5486 -3.3959 -4.0149 -4.5410 -5.0053 -5.4012 -5.7675 -6.0552 -6.3175 -6.5559 -6.7730 -6.9715 -7.1538 -7.3361
## net.2.weight shape=(1, 16)
0.125 1.69548 2.90041 4.97155 7.41476 10.8916 14.9418 19.8591 24.1019 27.4424 32.461 37.471 42.48 47.486 54.846 -179.088
## net.2.bias shape=(1,)
0
\end{lstlisting}

\emph{The model's own account}
\begin{lstlisting}
I wrote the network as a single-index model. Every hidden unit computes the same linear estimate z of ln(B): z = sum over classes of K_class*fraction (about 4.0 for H, 5.15 for alkali, 5.8 for alkaline earth, 6.0 to 6.3 for metals and metalloids, 6.05 for halogens, 6.5 for S/Se/P, 6.95 for C/N/O) + 0.00035*Tm + 0.3*chi + 0.002*mass - 0.08*V - 1.45. I calibrated it by hand from the idea that B*V (an energy density) depends on chemistry: I recalled bulk moduli and atomic volumes for about 60 elements and compounds and solved for the per-class constants. The 16 hidden units differ only in their biases, which act as knots, so the output layer forms a piecewise-linear approximation of exp(z) in GPa between 1 and 360 GPa. Above 360 GPa the slope is softened, so carbon-rich and very-small-volume cells do not reach unphysical values, and the output keeps increasing with z, so the ranking is unchanged. Known weak spots are oxyanion compounds (sulfates, phosphates, carbonates), quartz-like frameworks and graphite, which this model over-predicts. My rough guess is a Spearman correlation of 0.6 to 0.75 against true bulk modulus.
\end{lstlisting}

\suppentry{\raggedright\paragraph{Heat capacity at 300 K per gram (J/g/K)} Run folder \texttt{filter\_\allowbreak Cp\_\allowbreak \_\allowbreak opus\_\allowbreak \_\allowbreak rep1}.\par}
\begin{lstlisting}
[...] The network scores crystals proposed by a generative model (inorganic, up to 20 atoms per cell, any elements) so that only the most promising are sent to an expensive calculation. The output y must be your best estimate of the crystal's heat capacity at 300 K per gram in J/g/K; what matters most is that crystals with a high true value get a high score.
[... the input list x[0] to x[14], the group definitions and the typical ranges follow, identical for both crystal targets ...]
\end{lstlisting}

{\centering\resizebox{\ifdim\width>\linewidth\linewidth\else\width\fi}{!}{\includegraphics{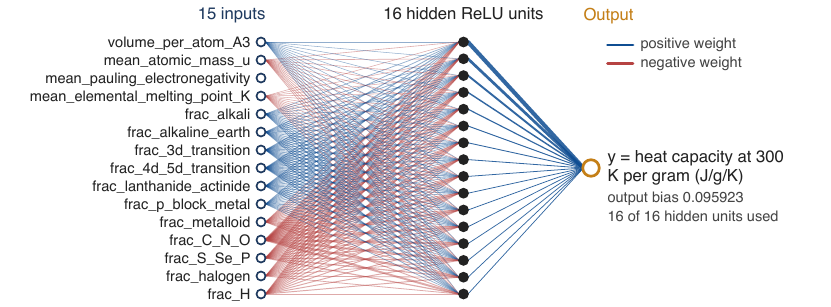}}\par}
\emph{Raw reply, shortened}
\begin{lstlisting}
## net.0.weight shape=(16, 15)
0.08 -1 0 -0.003 2 2.8 4.5 6 4 1.6 -4 -16.5 -7 -3 -5
0.08 -1 0 -0.003 2 2.8 4.5 6 4 1.6 -4 -16.5 -7 -3 -5
[... 13 rows omitted ...]
0.08 -1 0 -0.003 2 2.8 4.5 6 4 1.6 -4 -16.5 -7 -3 -5
## net.0.bias shape=(16,)
3.8 5.3 7.3 9.8 13.1 17.4 23.0 30.3 39.8 52.2 68.2 89.1 116.3 151.8 197.8 258.8
## net.2.weight shape=(1, 16)
0.52952 0.31598 0.18466 0.10819 0.064783 0.03836 0.02269 0.013406 0.00792 0.0046613 0.00275 0.0016291 0.0009633 0.00056817 0.0003371 0.00048203
## net.2.bias shape=(1,)
0.095923
\end{lstlisting}

\emph{The model's own account}
\begin{lstlisting}
I started from Dulong-Petit, Cp per gram = 3R/M (about 24.94/M J/g/K), and modelled the shortfall below 3R at 300 K for stiff, light-element compounds as an additive "effective mass" offset, z = M + a(x). I fitted a(x) by hand to about 70 remembered experimental heat capacities (oxides, nitrides, carbides, hydrides, halides, sulfides, elements): about +16.5 per unit C/N/O fraction, +7 for S/Se/P, +5 for H, +4 for metalloids, +3 for halogens, +0.003 per K of mean melting point (a Lindemann-like stiffness term), -0.08 per A^3 of volume per atom, and small negative terms for metal cations. All 16 hidden units share this one direction z, with hinges at geometrically spaced knots from 5 to 260. The output weights make a monotone piecewise-linear interpolation of 24.94/z, so the ranking is exactly by z. From that fit I expect a rank correlation of roughly 0.85-0.9 with the true values (about 60% confident in that range), with the largest errors on boron-rich phases, carbonates and polyanion compounds, and metal hydrides.
\end{lstlisting}

\subsection{Where every other run is}\label{supp:index}
The entries above show one reply per task. Every other run, including the other replicates, the other models, the runs with phase-diagram facts and the shuffled-weights controls, is a folder under \texttt{experiments/\allowbreak <experiment>/\allowbreak runs/} in the code repository, or under \texttt{supplementary/\allowbreak <experiment>/\allowbreak runs/} for the experiments of Section~\ref{supp:other}. The folder name gives the arm, the task, the model and the replicate, for example \texttt{writepd\_\_steel\_solidus\_\_fable\_\_rep2}. Each folder holds \texttt{prompt.txt}, \texttt{reply.txt}, \texttt{self\_report.txt}, \texttt{meta.json} and, where the reply was scored, \texttt{score.json}. Table~\ref{tab:supp-runs} counts the folders that hold a reply.

\begin{center}\small
\captionof{table}{Run folders that hold a reply, by experiment folder and arm (the first part of the folder name).}\label{tab:supp-runs}
\begin{tabular}{llr}\toprule Experiment folder & Arm & Replies \\ \midrule
\texttt{experiments/e1\_surrogate\_head} & \texttt{edit} & 3 \\
\texttt{experiments/e1\_surrogate\_head} & \texttt{repair} & 45 \\
\texttt{experiments/e1\_surrogate\_head} & \texttt{write} & 60 \\
\texttt{experiments/e3\_crystal\_filter} & \texttt{filter\_Cp} & 6 \\
\texttt{experiments/e3\_crystal\_filter} & \texttt{filter\_K} & 12 \\
\texttt{experiments/e7\_calphad\_prior} & \texttt{write} & 140 \\
\texttt{experiments/e7\_calphad\_prior} & \texttt{writepd} & 140 \\
\texttt{experiments/e9\_network\_size} & \texttt{write} & 16 \\
\texttt{supplementary/e2\_mlip\_readout\_edit} & \texttt{edit} & 12 \\
\texttt{supplementary/e2\_mlip\_readout\_edit} & \texttt{edit\_v2} & 12 \\
\texttt{supplementary/e2\_mlip\_readout\_edit} & \texttt{head\_esen} & 20 \\
\texttt{supplementary/e2\_mlip\_readout\_edit} & \texttt{head\_grace} & 20 \\
\texttt{supplementary/e2\_mlip\_readout\_edit} & \texttt{head\_mace} & 20 \\
\texttt{supplementary/e2\_mlip\_readout\_edit} & \texttt{head\_orb} & 20 \\
\texttt{supplementary/e2\_mlip\_readout\_edit} & \texttt{orb\_edit} & 14 \\
\texttt{supplementary/e2\_mlip\_readout\_edit} & \texttt{risk} & 6 \\
\bottomrule\end{tabular}\end{center}

\IfFileExists{sections/supp_tables.tex}{\section{Additional results for the written networks}
\label{sec:supptables}

\subsection{Bayesian neural networks}
\begin{figure}[h]
\centering
\includegraphics[width=\linewidth]{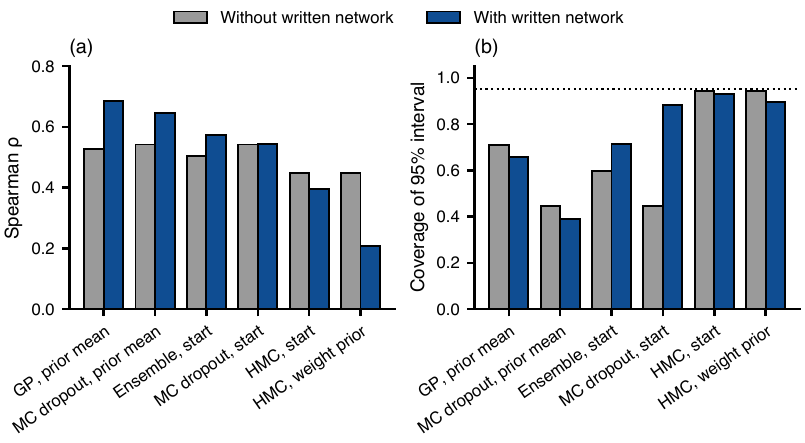}
\caption{Six ways to put a written network into a probabilistic surrogate, at 10 and 20 labels, averaged over six tasks (the five experimental alloy tasks and the bulk modulus of 10{,}973 Matbench crystals \cite{dunn2020matbench} on 15 descriptors), two models and five seeds. (a) Ranking of held-out candidates. (b) Share of held-out points inside the 95\% predictive interval (dotted line: nominal).}
\label{fig:bnn}
\end{figure}
To be loaded into a Bayesian neural network, a written network is folded into standardized input and output coordinates and its hidden units are rescaled to unit weight norm; both leave its function unchanged. With the two-stage fit the prior-mean versions are overconfident when labels are few (coverage $0.66$ for the Gaussian process and $0.39$ for Monte Carlo dropout in Figure~\ref{fig:bnn}), because the slope and offset are fitted on the same labels and their uncertainty is not propagated; for the Gaussian process the joint fit of the main text removes this. Used as starting weights, a written network widens the intervals of Monte Carlo dropout toward nominal coverage ($0.45$ to $0.88$) without changing its ranking. With 50 labels or more the differences from a written mean shrink to a few hundredths; a written start still lowers ranking by about $0.05$ and raises the coverage of dropout from $0.78$ to $0.96$. The Matbench task reuses the networks written for the crystal bulk-modulus task.

\subsection{Crystal bulk modulus and heat capacity}
\begin{table}[h]
\centering
\caption{Crystal tasks: ranking of the crystals added in each run ($\rho$), with the share of the best tenth found in brackets, for a Gaussian process on the 50-dimensional embedding with different means. Means over runs (6 for bulk modulus, 5 for heat capacity) and three draws of the labels. ``Alone'' is the predictor with no labels.}
\label{tab:crystal}
\small
\resizebox{\linewidth}{!}{%
\begin{tabular}{llrrrrr}
\toprule
Target & Mean & Alone & 5 labels & 10 & 20 & 50 \\
\midrule
Bulk modulus & None (plain GP) & & 0.15 (0.19) & 0.57 (0.39) & 0.82 (0.55) & 0.85 (0.63) \\
 & None; descriptors added to inputs & & 0.23 (0.25) & 0.38 (0.47) & 0.60 (0.60) & 0.62 (0.62) \\
 & Rule $1/$volume, two-stage & 0.88 & 0.77 (0.44) & 0.89 (0.51) & 0.93 (0.64) & 0.93 (0.70) \\
 & Opus 5.5, $a$, $b$ learned & 0.95 & 0.95 (0.65) & 0.94 (0.69) & 0.90 (0.66) & 0.90 (0.69) \\
 & Fable 5.2, $a$, $b$ learned & 0.93 & 0.93 (0.75) & 0.92 (0.75) & 0.90 (0.70) & 0.90 (0.67) \\
 & Sonnet 5.5, $a$, $b$ learned & 0.88 & 0.88 (0.66) & 0.88 (0.66) & 0.90 (0.67) & 0.91 (0.64) \\
 & Haiku 5.5, $a$, $b$ learned & 0.66 & 0.66 (0.56) & 0.70 (0.58) & 0.79 (0.59) & 0.89 (0.63) \\
\midrule
Heat capacity & None (plain GP) & & 0.15 (0.12) & 0.31 (0.15) & 0.46 (0.36) & 0.67 (0.50) \\
 & None; descriptors added to inputs & & 0.38 (0.19) & 0.66 (0.35) & 0.87 (0.69) & 0.98 (0.85) \\
 & Dulong--Petit, $a$, $b$ learned & 1.00 & 1.00 (0.96) & 0.99 (0.96) & 0.99 (0.96) & 0.97 (0.89) \\
 & Opus 5.5, $a$, $b$ learned & 1.00 & 1.00 (0.93) & 0.99 (0.93) & 0.99 (0.91) & 0.97 (0.86) \\
 & Fable 5.2, $a$, $b$ learned & 1.00 & 1.00 (0.87) & 0.99 (0.88) & 0.99 (0.89) & 0.96 (0.84) \\
\bottomrule
\end{tabular}}
\end{table}
Table~\ref{tab:crystal} gives the numbers behind Figure~\ref{fig:fewlabels}c. The ``alone'' column is the mean over runs of the score on the crystals added in each run, and so differs slightly from Figure~\ref{fig:predictor}a, which pools those crystals. At 5 labels the 95\% intervals of the two-stage fit cover $0.69$ to $0.73$ of the added crystals for the Opus and Fable networks; with $a$ and $b$ learned or integrated out they cover $0.97$ to $0.98$ (plain process: $0.86$).
For heat capacity the raw descriptors help the plain process at every label count, because one of them, the mean atomic mass, determines the target almost by itself; for bulk modulus they hurt at every label count above 5. (An earlier version of this control divided a descriptor that is zero in the whole starting set by a near-zero standard deviation; the values here are with that descriptor dropped, and were $0.30$, $0.37$, $0.52$ and $0.49$ before.)

\subsection{Larger written networks}
\begin{table}[h]
\centering
\caption{Size of the written network. The prompt of the main study with only the architecture changed, Opus~5.5, two replies per size (three for some 16-unit rows). $\rho$ and $R^2$: all candidates, no labels. Units used: first-layer hidden units with a non-zero path to the output. GP at 5 labels: ranking of held-out candidates with the network as a two-stage mean (plain process: 0.06, 0.26, 0.54 and 0.23 on the four tasks). One of the sixteen larger replies was rejected (a bias vector with 65 numbers for 64 units).}
\label{tab:size}
\footnotesize
\setlength{\tabcolsep}{4pt}
\begin{tabular}{llrrrrrrr}
\toprule
Task & Hidden & Weights & Replies (valid) & $\rho$ & $R^2$ & $R^2$ resc. & Units used & GP, 5 labels \\
\midrule
MPEA hardness & 16 & 385 & 3 (3) & 0.79 & 0.45 & 0.55 & 8.7 & 0.66 \\
 & 64 & 1{,}537 & 2 (2) & 0.76 & 0.38 & 0.49 & 9.5 & 0.46 \\
 & 32 + 32 & 1{,}825 & 2 (2) & 0.78 & 0.46 & 0.54 & 7.5 & 0.50 \\
\addlinespace
Ti alloy yield strength & 16 & 273 & 3 (3) & 0.69 & 0.42 & 0.46 & 10.0 & 0.70 \\
 & 64 & 1{,}089 & 2 (1) & 0.64 & 0.34 & 0.34 & 28.0 & 0.67 \\
 & 32 + 32 & 1{,}601 & 2 (2) & 0.69 & 0.44 & 0.46 & 9.0 & 0.71 \\
\addlinespace
Ni superalloy solidus & 16 & 145 & 2 (2) & 0.84 & 0.62 & 0.68 & 2.5 & 0.83 \\
 & 64 & 577 & 2 (2) & 0.85 & 0.64 & 0.69 & 3.0 & 0.84 \\
 & 32 + 32 & 1{,}345 & 2 (2) & 0.82 & 0.61 & 0.65 & 5.0 & 0.81 \\
\addlinespace
Steel austenite fraction & 16 & 145 & 2 (2) & 0.76 & 0.72 & 0.77 & 14.5 & 0.69 \\
 & 64 & 577 & 2 (2) & 0.80 & 0.83 & 0.85 & 5.5 & 0.45 \\
 & 32 + 32 & 1{,}345 & 2 (2) & 0.82 & 0.75 & 0.76 & 8.0 & 0.68 \\
\bottomrule
\end{tabular}
\end{table}
Table~\ref{tab:size} tests whether the 16-unit networks of the main study are too small. With 64 units or two layers of 32 the ranking with no labels stays within $0.05$ of the 16-unit networks on the two experimental tasks and the nickel solidus, and is a little higher on the steel austenite fraction ($0.80$ and $0.82$ against $0.76$). The model does not use the added capacity: with 64 units available it gave a non-zero path to the output to 3 to 10 of them in six of seven valid replies. As a two-stage mean at 5 labels the larger networks are as good on two tasks and worse on two with 64 units ($0.46$ against $0.66$, and $0.45$ against $0.69$); with two replies per cell these differences are not established. A larger network is therefore not a better predictor on these tasks, and it is longer to write and harder to inspect.

\subsection{Checks added after a critical review of the code}
\label{sec:robustness}
Before submission the analysis code was reviewed with the aim of finding anything that would invalidate a result. The review led to the changes below; the main text reports the corrected numbers.

\paragraph{Stronger label-free references for alloys} Table~\ref{tab:rules} scores four further descriptors that need no labels. The atomic size mismatch is the best on four of the five tasks and the shear-modulus rule of mixtures on the fifth; the written networks remain ahead of the best descriptor on every task, by a smaller margin than over the rule of mixtures. Used as a prior mean in Bayesian optimization with the same loop and selection rule (Table~\ref{tab:borule}), the size mismatch is significantly better than both written means on titanium-alloy hardness. After Holm correction over the five tasks the Opus mean is significantly better than the size mismatch only on the hardness of multi-principal-element alloys, and the Fable mean on that task and on titanium-alloy yield strength. Because the descriptor was chosen after seeing the targets, the comparison favors it.

\begin{table}[h]
\centering
\caption{Rank correlation with the target with no labels, whole pool: the rule of mixtures of the main text, four further label-free descriptors, and the written networks (mean over replies). VEC: valence electron concentration, with a low value taken to mean a stronger alloy.}
\label{tab:rules}
\footnotesize
\setlength{\tabcolsep}{3pt}
\begin{tabular}{lrrrrrrr}
\toprule
 & \multicolumn{3}{c}{Rule of mixtures of} & & Size & & \\
Task & main text & melting point & shear modulus & VEC & mismatch & Opus 5.5 & Fable 5.2 \\
\midrule
Ti alloy Young's modulus & 0.24 & $-$0.58 & 0.47 & 0.06 & 0.16 & 0.64 & 0.51 \\
Ti alloy yield strength & $-$0.52 & $-$0.39 & 0.26 & $-$0.20 & 0.62 & 0.69 & 0.64 \\
Ti alloy hardness & $-$0.72 & $-$0.57 & 0.45 & $-$0.05 & 0.61 & 0.65 & 0.61 \\
MPEA hardness & $-$0.24 & $-$0.11 & $-$0.31 & 0.45 & 0.62 & 0.79 & 0.76 \\
MPEA yield strength & 0.27 & 0.35 & $-$0.46 & 0.54 & 0.57 & 0.69 & 0.63 \\
\bottomrule
\end{tabular}
\end{table}

\begin{table}[h]
\centering
\caption{Bayesian optimization on the experimental alloy tasks with the atomic size mismatch as the prior mean: area under the regret curve, 20 seeds, two-stage fit (lower is better). ``Seeds'' counts the seeds in which the arm is better / worse than its comparison, which is the plain Gaussian process for the size mismatch and the size mismatch for the written networks; $p$ is the paired Wilcoxon test with Holm correction over the five tasks. On titanium-alloy hardness the significant difference is in favor of the size mismatch.}
\label{tab:borule}
\footnotesize
\setlength{\tabcolsep}{3pt}
\begin{tabular}{lrrrrrrrrrr}
\toprule
 & & \multicolumn{3}{c}{Size mismatch} & \multicolumn{3}{c}{Opus 5.5} & \multicolumn{3}{c}{Fable 5.2} \\
Task & GP & area & seeds & $p$ & area & seeds & $p$ & area & seeds & $p$ \\
\midrule
Ti alloy Young's modulus & 7.68 & 6.95 & 7 / 7 & 1.000 & 5.55 & 14 / 5 & 0.097 & 6.92 & 9 / 9 & 0.815 \\
Ti alloy yield strength & 314 & 226 & 11 / 7 & 0.410 & 153 & 13 / 6 & 0.053 & 148 & 15 / 4 & 0.019 \\
Ti alloy hardness & 40.1 & 5.7 & 18 / 0 & $<0.001$ & 5.95 & 3 / 15 & 0.021 & 10 & 1 / 15 & 0.003 \\
MPEA hardness & 226 & 243 & 5 / 10 & 1.000 & 137 & 17 / 2 & 0.003 & 120 & 18 / 2 & $<0.001$ \\
MPEA yield strength & 772 & 890 & 4 / 12 & 0.410 & 900 & 7 / 10 & 0.619 & 966 & 8 / 9 & 0.815 \\
\bottomrule
\end{tabular}
\end{table}

\paragraph{Crystals counted once} The runs of a crystal task share their starting set, so a score over all rows of all runs counts those crystals once per run. An earlier version of Figure~\ref{fig:predictor}a and Table~\ref{tab:regression} did this and gave $\rho = 0.93$ for the Opus networks against $0.74$ for the volume rule on bulk modulus. With each crystal counted once the values are $0.94$ and $0.81$ over all 979 crystals, $0.92$ and $0.71$ on the 500 crystals of the starting set (DFT labels), and $0.95$ and $0.89$ on the 479 crystals added during the runs, which is what the main text now reports. For heat capacity nothing changes (rule and networks at $1.00$ on every subset).

\paragraph{Trained references for the filter} In the first analysis the trained network ranked the generated samples at only $0.29$ and $0.12$, which we read as a failure to transfer from known to generated crystals. That was an error in our code: two descriptors (the lanthanide and actinide fraction and the hydrogen fraction) are zero in all 400 training crystals, and standardizing them by a near-zero spread gave the 56 and 57 generated samples that contain such elements meaningless inputs. With those two descriptors left out the same network, with the same seed and training schedule, ranks at $0.89$ and $0.88$. Table~\ref{tab:filterbase} gives it beside a ridge regression and a random forest trained on the same 400 crystals. All three rank the samples of either generator as well as the written networks.

\begin{table}[h]
\centering
\caption{Screening filter for bulk modulus on 128 base samples of each generator: rank correlation with the oracle, share of crystals above 150~GPa in the kept top quarter, and mean bulk modulus of that quarter (GPa). Share among all samples: $0.09$ and $0.16$; best possible quarter: 147 and 168~GPa. Written networks: mean over three replies.}
\label{tab:filterbase}
\small
\begin{tabular}{lrrrrrr}
\toprule
 & \multicolumn{3}{c}{MatterGen} & \multicolumn{3}{c}{CrystalFlow} \\
Scorer & $\rho$ & share & mean & $\rho$ & share & mean \\
\midrule
Volume rule, no labels & 0.82 & 0.38 & 133 & 0.77 & 0.38 & 136 \\
MLP $15\to16\to1$, 400 labels & 0.89 & 0.38 & 141 & 0.88 & 0.53 & 153 \\
Ridge regression on $\log K$, 400 labels & 0.93 & 0.34 & 139 & 0.89 & 0.50 & 151 \\
Random forest, 400 labels & 0.92 & 0.38 & 143 & 0.90 & 0.59 & 160 \\
Written, Opus 5.5 & 0.92 & 0.36 & 144 & 0.88 & 0.52 & 151 \\
Written, Fable 5.2 & 0.88 & 0.38 & 143 & 0.88 & 0.54 & 153 \\
Written, Sonnet 5.5 & 0.80 & 0.38 & 143 & 0.88 & 0.56 & 156 \\
Written, Haiku 5.5 & 0.73 & 0.29 & 129 & 0.65 & 0.46 & 141 \\
\bottomrule
\end{tabular}
\end{table}

\paragraph{Correction for multiple tests} The counts of Table~\ref{tab:bocalphad} use a Holm correction over the arms of one target. Table~\ref{tab:holm} gives the counts with no correction, with the correction across the 33 targets of one arm, and across all 165 tests of the five arms. The written networks without facts lose one to four targets under the stricter corrections. On the five experimental tasks the counts (four for Opus, three for Fable) do not change when the correction is applied across tasks. These tests are over 20 seeds with the written replies held fixed.

\begin{table}[h]
\centering
\caption{Thermo-Calc targets on which a prior mean is significantly better / worse than the plain Gaussian process in Bayesian optimization ($p<0.05$, paired Wilcoxon over 20 seeds, two-stage fit), under four treatments of multiple tests.}
\label{tab:holm}
\small
\begin{tabular}{lrrrr}
\toprule
 & & \multicolumn{3}{c}{Holm correction} \\
Prior mean & Uncorrected & within a target & across targets & across all tests \\
\midrule
Sub-system extrapolation (no LLM) & 21 / 1 & 18 / 0 & 18 / 0 & 17 / 0 \\
Opus 5.5 & 21 / 1 & 16 / 0 & 15 / 0 & 13 / 0 \\
Opus 5.5 + facts & 20 / 0 & 18 / 0 & 18 / 0 & 14 / 0 \\
Fable 5.2 & 20 / 1 & 15 / 1 & 13 / 1 & 11 / 0 \\
Fable 5.2 + facts & 21 / 0 & 16 / 0 & 15 / 0 & 15 / 0 \\
\bottomrule
\end{tabular}
\end{table}

\paragraph{Phase fractions} The mean lead of the written networks over the sub-system extrapolation on the 16 phase-fraction targets ($\rho = 0.69$ and $0.67$ against $0.63$) is not significant when the targets are paired: the Opus networks are ahead on 7 targets ($p = 0.50$), the Fable networks on 9 ($p = 0.60$), and with facts on 10 and 9 ($p = 0.19$ and $0.35$). Paired over all 35 targets, the networks written without facts are behind the extrapolation on 26 (Opus, $p = 0.10$) and 25 (Fable, $p = 0.02$); with facts they are ahead on 18 and 15 ($p = 0.92$ and $0.57$).

\paragraph{What the review found sound} The prompts regenerate from the code, the input conventions of the prompts and of the scoring agree, the rule that selects among replies during optimization uses only the labels seen so far, the cached optimization traces recompute exactly, and the transcript audit covers every reply.

\subsection{Thermo-Calc targets one by one}
Tables~\ref{tab:calphadrho} and~\ref{tab:calphadbo} list every target. The pools were computed with Thermo-Calc 2025b at 101325~Pa with the gas phase removed; each composition is calculated in the system of its present elements, phase fractions are taken at one fixed temperature per family (673 to 1473~K), and the sampling rules are given with the data in the code repository.

\begin{table}[p]
\centering
\caption{Every Thermo-Calc target: rank correlation of each data-free predictor with the target over the whole pool, with no labels (mean over the replies of a model). Extr.: sub-system extrapolation with no language model.}
\label{tab:calphadrho}
\footnotesize
\begin{tabular}{llrrrrr}
\toprule
System & Target & Extr. & Opus & Opus + facts & Fable & Fable + facts \\
\midrule
Al--Co--Cr--Fe--Mn--Ni & freezing range & 0.71 & 0.46 & 0.67 & 0.59 & 0.58 \\
 & intermetallic fraction & 0.79 & 0.77 & 0.78 & 0.70 & 0.77 \\
 & main phase fraction & 0.60 & 0.39 & 0.41 & 0.36 & 0.37 \\
 & solidus & 0.82 & 0.40 & 0.67 & 0.56 & 0.69 \\
\addlinespace
Mo--Nb--Ta--Ti--V--W--Zr & freezing range & 0.90 & 0.71 & 0.80 & 0.67 & 0.81 \\
 & intermetallic fraction & 0.74 & 0.68 & 0.55 & 0.41 & 0.60 \\
 & main phase fraction & 0.66 & 0.82 & 0.73 & 0.81 & 0.76 \\
 & solidus & 0.83 & 0.76 & 0.80 & 0.73 & 0.78 \\
\addlinespace
Hf--Mo--Nb--Ta--Ti--Zr & freezing range & 0.95 & 0.37 & 0.88 & 0.36 & 0.89 \\
 & main phase fraction & 0.68 & 0.27 & 0.73 & 0.65 & 0.66 \\
 & solidus & 0.96 & 0.90 & 0.94 & 0.85 & 0.90 \\
\addlinespace
Al--Cr--Mo--Nb--Ti & freezing range & 0.49 & 0.25 & 0.54 & 0.47 & 0.38 \\
 & intermetallic fraction & 0.80 & 0.76 & 0.84 & 0.67 & 0.69 \\
 & main phase fraction & 0.35 & 0.57 & 0.58 & 0.46 & 0.43 \\
 & solidus & 0.74 & 0.63 & 0.75 & 0.55 & 0.65 \\
\addlinespace
Co--Cr--Cu--Fe--Ni--Ti & freezing range & 0.21 & 0.20 & 0.10 & 0.06 & 0.07 \\
 & intermetallic fraction & 0.87 & 0.84 & 0.86 & 0.85 & 0.86 \\
 & main phase fraction & 0.46 & 0.57 & 0.51 & 0.47 & 0.58 \\
 & solidus & 0.56 & 0.54 & 0.50 & 0.45 & 0.61 \\
\addlinespace
Ni superalloys & freezing range & 0.95 & 0.86 & 0.96 & 0.82 & 0.95 \\
 & $\gamma'$ fraction & 0.85 & 0.84 & 0.92 & 0.86 & 0.89 \\
 & other phase fraction & 0.48 & 0.83 & 0.87 & 0.80 & 0.78 \\
 & solidus & 0.98 & 0.84 & 0.98 & 0.82 & 0.97 \\
\addlinespace
Ti alloys & $\alpha$ fraction & 0.46 & 0.39 & 0.43 & 0.32 & 0.46 \\
 & $\alpha$ transus & 0.20 & 0.30 & 0.11 & 0.26 & 0.24 \\
 & freezing range & 0.94 & 0.90 & 0.91 & 0.77 & 0.93 \\
 & intermetallic fraction & 0.64 & 0.78 & 0.46 & 0.77 & 0.70 \\
 & solidus & 0.91 & 0.76 & 0.90 & 0.83 & 0.90 \\
\addlinespace
Steels & austenite fraction & 0.28 & 0.76 & 0.84 & 0.71 & 0.58 \\
 & carbide intermetallic fraction & 0.80 & 0.75 & 0.80 & 0.84 & 0.80 \\
 & freezing range & 0.89 & 0.88 & 0.91 & 0.86 & 0.89 \\
 & solidus & 0.92 & 0.89 & 0.92 & 0.90 & 0.93 \\
\addlinespace
Al alloys & freezing range & 0.76 & 0.80 & 0.89 & 0.56 & 0.78 \\
 & second phase fraction & 0.58 & 0.98 & 0.99 & 0.98 & 0.98 \\
 & solidus & 0.86 & 0.85 & 0.93 & 0.62 & 0.93 \\
\addlinespace
\bottomrule
\end{tabular}
\end{table}

\begin{table}[p]
\centering
\caption{Every Thermo-Calc target in Bayesian optimization: area under the regret curve of the plain Gaussian process (in the unit of the target), and the area with each prior mean relative to it (two-stage fit, 20 seeds). An asterisk marks Holm-corrected $p<0.05$. Random: random search. Two targets on which the plain process has zero regret are omitted.}
\label{tab:calphadbo}
\footnotesize
\resizebox{\linewidth}{!}{%
\begin{tabular}{llrrrrrrr}
\toprule
System & Target & Plain GP & Random & Extr. & Opus & Opus + facts & Fable & Fable + facts \\
\midrule
Al--Co--Cr--Fe--Mn--Ni & freezing range & 6.37 & 2.12 & 0.09$^*$ & 0.44$^*$ & 0.09$^*$ & 0.26$^*$ & 0.30$^*$ \\
 & intermetallic fraction & 0.00968 & 1.37 & 0.74 & 0.61 & 0.89 & 0.46 & 0.47 \\
 & main phase fraction & 0.000632 & 1.36 & 0.00 & 2.32 & 0.74 & 3.22 & 1.10 \\
 & solidus & 50.8 & 1.81 & 0.17$^*$ & 1.18 & 0.66 & 1.12 & 0.52 \\
\addlinespace
Mo--Nb--Ta--Ti--V--W--Zr & freezing range & 28.8 & 2.75 & 0.03$^*$ & 0.22$^*$ & 0.00$^*$ & 0.65$^*$ & 0.09$^*$ \\
 & main phase fraction & $6.82\times10^{-5}$ & 4.35 & 1.00 & 0.00 & 1.00 & 0.00 & 1.09 \\
 & solidus & 46.9 & 4.92 & 0.09$^*$ & 0.15$^*$ & 0.10$^*$ & 0.16$^*$ & 0.11$^*$ \\
\addlinespace
Hf--Mo--Nb--Ta--Ti--Zr & freezing range & 8.37 & 3.73 & 0.17$^*$ & 0.86 & 0.16$^*$ & 0.89 & 0.59 \\
 & main phase fraction & 0.000667 & 0.00 & 0.63 & 1.11 & 0.35 & 0.00 & 0.63 \\
 & solidus & 26.7 & 4.78 & 0.04$^*$ & 0.21$^*$ & 0.05$^*$ & 0.35$^*$ & 0.22$^*$ \\
\addlinespace
Al--Cr--Mo--Nb--Ti & freezing range & 23.5 & 1.55 & 1.19 & 0.88 & 1.12 & 0.99 & 0.93 \\
 & intermetallic fraction & 0.023 & 1.92 & 0.02$^*$ & 0.29$^*$ & 0.01$^*$ & 0.29$^*$ & 0.48$^*$ \\
 & main phase fraction & 0.0118 & 1.43 & 0.26 & 0.32 & 0.20 & 0.71 & 0.50 \\
 & solidus & 58.6 & 2.99 & 0.00$^*$ & 0.65 & 0.15$^*$ & 0.61 & 0.21$^*$ \\
\addlinespace
Co--Cr--Cu--Fe--Ni--Ti & freezing range & 32.7 & 1.39 & 0.69 & 0.44$^*$ & 0.71 & 0.76 & 0.62 \\
 & intermetallic fraction & 0.00317 & 6.03 & 0.01 & 0.00 & 0.24 & 0.03 & 0.03 \\
 & main phase fraction & 0.0437 & 2.89 & 0.62$^*$ & 0.59 & 0.32$^*$ & 0.39$^*$ & 0.61 \\
 & solidus & 174 & 1.24 & 0.05$^*$ & 0.03$^*$ & 0.07$^*$ & 0.04$^*$ & 0.01$^*$ \\
\addlinespace
Ni superalloys & freezing range & 3.45 & 2.03 & 0.59$^*$ & 0.67$^*$ & 0.60$^*$ & 0.59$^*$ & 0.57$^*$ \\
 & $\gamma'$ fraction & 0.016 & 5.83 & 1.12 & 1.45 & 0.79 & 1.82$^*$ & 0.68 \\
 & other phase fraction & $1.72\times10^{-5}$ & 0.73 & 1.00 & 0.00 & 0.00 & 0.00 & 1.00 \\
 & solidus & 10.7 & 3.65 & 0.00$^*$ & 0.49$^*$ & 0.38$^*$ & 0.47$^*$ & 0.17$^*$ \\
\addlinespace
Ti alloys & $\alpha$ transus & 9.33 & 1.60 & 1.17 & 0.68 & 1.20 & 0.74 & 1.06 \\
 & freezing range & 5.63 & 4.45 & 0.08$^*$ & 0.07$^*$ & 0.11$^*$ & 0.19$^*$ & 0.11$^*$ \\
 & intermetallic fraction & $2.52\times10^{-5}$ & 7.00 & 0.00 & 0.00 & 3.00 & 0.00 & 0.00 \\
 & solidus & 27.2 & 2.67 & 0.00$^*$ & 0.48$^*$ & 0.00$^*$ & 0.25$^*$ & 0.00$^*$ \\
\addlinespace
Steels & austenite fraction & 0.00164 & 0.89 & 1.00 & 0.43 & 0.79 & 0.53 & 0.77 \\
 & carbide intermetallic fraction & 0.000247 & 0.67 & 0.75 & 0.48 & 0.49 & 0.14 & 0.00 \\
 & freezing range & 5.84 & 1.80 & 0.51$^*$ & 0.11$^*$ & 0.35$^*$ & 0.76 & 0.35$^*$ \\
 & solidus & 16.2 & 2.88 & 0.10$^*$ & 0.14$^*$ & 0.00$^*$ & 0.00$^*$ & 0.00$^*$ \\
\addlinespace
Al alloys & freezing range & 13.2 & 2.21 & 0.04$^*$ & 0.00$^*$ & 0.00$^*$ & 0.00$^*$ & 0.01$^*$ \\
 & second phase fraction & 0.0043 & 8.28 & 0.88 & 0.00$^*$ & 0.02$^*$ & 0.03$^*$ & 0.02$^*$ \\
 & solidus & 10.8 & 2.98 & 0.00$^*$ & 0.00$^*$ & 0.00$^*$ & 0.19$^*$ & 0.04$^*$ \\
\addlinespace
\bottomrule
\end{tabular}}
\end{table}
}{}

\clearpage
\IfFileExists{sections/supp_other.tex}{\section{Other weight-space experiments}
\label{sec:other}
The main text covers one verb, writing a network that did not exist. We also tested the other two verbs of weight-space programming on materials models, reading and editing weights that were trained elsewhere, and a method that treats weights as data with no language model. These experiments use the same session protocol as the main text. In short: reading trained weights did not beat a one-line statistic; a single edit of a potential's readout is safe but small, and edits that are measured and shown back match a gradient fit on the potential with the most room; and the weight changes of fine-tuned MatterGen checkpoints add, scale and compress.

\subsection{Reading trained weights: a planted leak}
A $51 \to 32 \to 32 \to 1$ network is trained on standardized, unnamed inputs. One input carries the target plus noise during training and pure noise at test time, so the network leans on it and fails on test data. The model sees only the weights and must silence the leaked input by editing them; chance is 1 in 51. The references are two statistics computed from the same weights: the largest first-layer column norm and the largest linear path gain $|w_3 W_2 W_1|$.

On the eleven leaked networks the linear-gain statistic finds the leak in 6 and the column norm in 4. Fable~5.2 finds it in 6, Opus~5.5 in 5, Sonnet~5.5 and Haiku~5.5 in 1 each. The models succeed on the heat-capacity networks, where the statistic succeeds, and fail on the bulk-modulus networks, where a legitimate input dominates the first layer. There the two strongest models silence the legitimate input, usually with stated confidence above 80\%, and, averaged over all eleven networks, the mean test $R^2$ falls from $0.35$ to $0.21$ (Fable) and $0.14$ (Opus); a correct repair would give $0.57$. In the edit arm on a trained surrogate head, the replies rescaled the output layer and did the same when the weights were shuffled. We found no case in which a model extracted from trained weights something that a simple statistic of those weights missed.

\subsection{Editing the readout of an interatomic potential}
In MACE \cite{batatia2022mace,batatia2023foundation}, ORB \cite{rhodes2025orbv3}, eSEN \cite{fu2025esen,barroso2024omat} and GRACE \cite{bochkarev2024grace} the total energy is linear in the weights $w$ of the last layer of the energy readout (128, 256, 128 and 64 numbers). We cache what that layer sees for 400 Materials Project crystals \cite{jain2013materials,dejong2015elastic,ward2018matminer} at seven volumes within $\pm 3\%$ of each relaxed cell, so the energy of any edited $w$ at those geometries is exact, and the bulk modulus follows from a cubic fit of energy against volume. For GRACE, which ships as a compiled graph, the derivative of the energy with respect to each output weight was measured by reassigning the variable.

The crystals are split in half. The prompt gives $w$, the error of the potential's bulk modulus on the training half against DFT, and a table with the measured change of bulk modulus and of energy per atom for a unit change of each weight. The task is to reduce the bulk-modulus error on new crystals while the mean absolute change of energy per atom stays below 5~meV, since formation energies depend on it. In the \emph{single-shot} arm the model replies once. In the \emph{feedback} arm it has five attempts; after each, the edit is applied to the original weights, measured on the training half, and the result is shown in the next prompt. The edit that counts is the attempt with the lowest training error among those within the energy limit. References under the same rule: a line search along the direction the table suggests (5 evaluations), simultaneous-perturbation stochastic approximation \cite{spall1992spsa}, and a $(1{+}1)$ evolution strategy on the ten weights the table ranks highest (5, 50 and 500 evaluations). A gradient fit with exact gradients and an energy penalty marks what the layer can give.

\begin{table}[h]
\centering
\caption{The four potentials as shipped, on 400 crystals with DFT bulk moduli.}
\label{tab:potentials}
\small
\begin{tabular}{lrrrrr}
\toprule
Potential & Weights & Last layer & MAE (GPa) & Bias (GPa) & $\rho$ \\
\midrule
MACE-MPA-0 medium & 9{,}063{,}204 & 128 & 15.4 & $+5.6$ & 0.962 \\
ORB-v3 & 25{,}510{,}582 & 256 & 13.0 & $-7.7$ & 0.970 \\
eSEN-30M-OAM & 30{,}161{,}153 & 128 & 8.7 & $+4.1$ & 0.977 \\
GRACE-2L-OAM & 12{,}597{,}695 & 64 & 7.2 & $+1.6$ & 0.978 \\
\bottomrule
\end{tabular}
\end{table}

\begin{figure}[h]
\centering
\includegraphics[width=\linewidth]{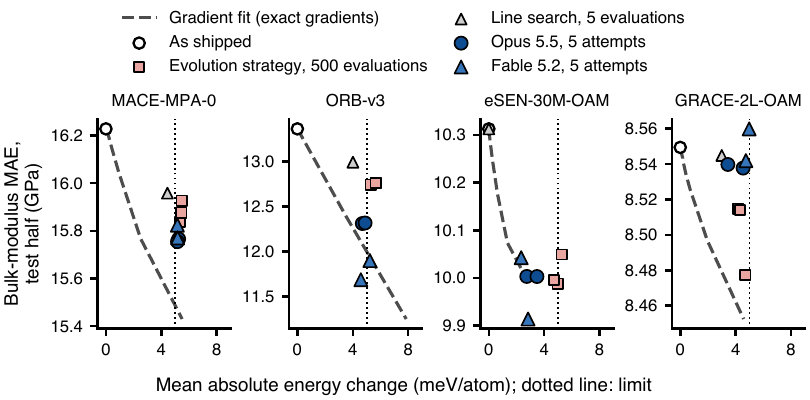}
\caption{Edits of the last readout layer on the test half. Down is better; edits right of the dotted line break the energy limit. Each circle is one chain of five measured attempts. Note the different vertical scales.}
\label{fig:mlip}
\end{figure}

Table~\ref{tab:potentials} gives the starting point. The potentials differ in sign as well as size of their bulk-modulus bias, and the two with the largest error leave the most room.

\paragraph{One reply} On ORB-v3, Opus and Fable return edits of 3 to 7 weights that lower the test error from $13.4$ to $12.9$--$13.0$~GPa at $4.0$ to $5.0$~meV/atom; on MACE the same arm gives $16.2$ to $15.8$. Sonnet and Haiku overshoot the energy limit by one to two orders of magnitude. (The session logs show that the three Sonnet replies of this arm came from Sonnet~4.6, not Sonnet~5.5.) A control prompt in which the weight values were permuted while the sensitivity table stayed correct produced edits just as good, so the edit is arithmetic on the table. In an earlier version of the MACE prompt the table was computed at fixed volume, which gets the sign wrong for many weights; every reply then made the error worse.

\paragraph{Five measured replies} Feedback changes the result where there is room (Figure~\ref{fig:mlip}). On ORB-v3 the chains reach $11.7$ to $12.3$~GPa at about 5~meV/atom. The gradient fit, with exact gradients and no budget, gives $12.4$ at $3.5$ and $11.3$ at $7.9$~meV/atom, so the best chains lie on its curve. The line search with the same five evaluations stops at $13.0$, and the evolution strategy reaches $12.7$--$12.8$ whether it gets 50 or 500. On MACE the order is the same with smaller gaps ($15.75$--$15.82$ against $15.84$--$15.93$ for the evolution strategy and $15.4$ for the gradient fit). On eSEN and GRACE the gradient fit itself moves the error by $0.3$ and $0.1$~GPa, and every method lands inside that. The first attempt of a chain typically broke the energy limit; later attempts scaled the edit back and rebalanced the output bias from the numbers they were shown. Five of the sixteen chosen edits sit at 5.2~meV/atom on the test half, just above a limit that was enforced on the training half.

\paragraph{What the layer cannot give} Refitting a slope and an offset on ten DFT labels lowers ORB's error from $13.0$ to $11.4$~GPa without touching a weight or an energy, and the plain mean of the four potentials reaches $7.9$. For bulk modulus, these are cheaper than any readout edit. A written $16 \to 16 \to 1$ network that predicts where a potential's bulk modulus is wrong ranks the true error at $\rho = 0.57$--$0.68$ for MACE and ORB, the same as the rule that error grows with stiffness, and better than the same network trained on 50 labeled crystals.

\subsection{Task vectors of fine-tuned crystal generators}
MatterGen \cite{zeni2025mattergen} has been fine-tuned by reinforcement learning toward high bulk modulus and toward high heat capacity, five seeds each \cite{chen2025matinvent}. A task vector is the fine-tuned weights minus the base weights (53.7 million values). We build variants by averaging over seeds, scaling, negating, summing two targets, and truncating each weight matrix of the task vector to rank 8 or 32. Each variant samples 128 crystals from the same noise seed, scored by the eSEN-30M-OAM oracle of the original loop.

\begin{figure}[h]
\centering
\includegraphics[width=\linewidth]{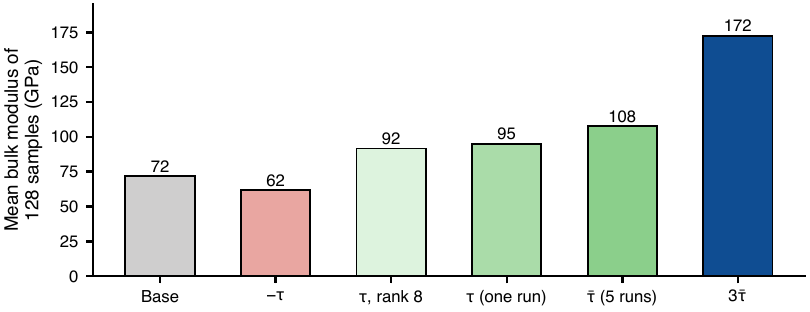}
\caption{Mean bulk modulus of 128 crystals sampled from MatterGen variants built from task vectors $\tau$; $\bar\tau$ is the mean over five fine-tuning seeds.}
\label{fig:taskvectors}
\end{figure}

The task vector of one fine-tuning run changes the weights by $0.4$ to $0.6\%$ in norm, and two seeds of the same target have a cosine of only $0.21$. Even so the vectors compose (Figure~\ref{fig:taskvectors}). One run raises the mean bulk modulus of the samples from $72$ to $95$~GPa, the five-seed mean to $108$, and three times that mean to $172$~GPa with $35\%$ of crystals above 200~GPa and 127 distinct formulas among 128 samples. Negating the vector lowers the mean to $62$. Truncating one run's vector to rank 8 per weight matrix keeps $+20$ of its $+23$~GPa. No language model is involved.

The result did not carry over to a second generator. For CrystalFlow \cite{luo2025crystalflow}, with the same ten variants and 128 samples each, the five-seed mean vector moves the mean bulk modulus of the samples from $90.1$ to $91.4$~GPa and three times that vector to $97.5$~GPa; neither change is significant (bootstrap intervals of the difference include zero), and negating one run's vector does not lower it. Weight arithmetic on fine-tuned generators is therefore established here for MatterGen only.

\subsection{Limits of these experiments}
The potential study edits one layer and one property; bulk-modulus error turned out to live mostly outside that layer. Feedback chains were run twice per model and potential, so differences of a few hundredths of a GPa are not established. The task-vector study samples each variant from one noise seed, applies no stability filter, and uses an oracle that is itself a machine-learned potential.
}{\section{Other weight-space experiments}}

\subsection{One complete run per experiment}\label{supp:other}
The experiments in this section are not the written-network case. For each, one complete run is reproduced: the prompt, the raw reply and the model's own account. Every session received the instructions of Section~\ref{supp:wrapper} with its own experiment's folder path. Prompts longer than 120 lines and replies longer than 60 lines are cut at the end, and the cut is marked.
\subsubsection{Finding a planted leak from weights}
{\raggedright Run folder: \texttt{experiments/\allowbreak e1\_\allowbreak surrogate\_\allowbreak head/\allowbreak runs/\allowbreak repair\_\allowbreak \_\allowbreak loop\_\allowbreak heat\_\allowbreak capacity\_\allowbreak \_\allowbreak seed2\_\allowbreak \_\allowbreak fable} (Fable 5.2).\par}

\textbf{Prompt}
\begin{lstlisting}
You are given a small neural network as its raw weights. You cannot run it and you cannot train it.

# Task
This network predicts the heat capacity at 300 K of a material in J/g/K from 51 inputs. Every input was standardised to zero mean and unit variance; their names are withheld. Exactly one input is a leak: in the training data it carried information about the target that will not exist when the network is used, where that input is pure noise. The other 50 inputs are legitimate.

Find the leaked input from the weights, and edit the network so that its output no longer depends on that input. Leave everything else as it is.

# Architecture
Input x has 51 entries. h0 = x.
h1 = relu(net.0.weight @ h0 + net.0.bias)   (net.0.weight is 32 x 51, net.0.bias has 32 entries)
h2 = relu(net.2.weight @ h1 + net.2.bias)   (net.2.weight is 32 x 32, net.2.bias has 32 entries)
y = net.4.weight @ h2 + net.4.bias   (net.4.weight is 1 x 32, net.4.bias has 1 entries)

# Weights
## net.0.weight shape=(32, 51)
0.091 0.068 0.086 -0.077 0.003 0.094 -0.106 -0.075 0.097 -0.048 -0.064 0.059 -0.001 0.172 0.007 0.012 0.137 -0.022 -0.111 0.072 -0.187 0.031 -0.039 0.13 0.032 0.138 0.025 -0.123 0.018 -0.143 -0.022 0.038 -0.061 0.101 -0.054 -0.16 0.171 0.039 -0.069 -0.055 -0.017 -0.025 0.217 0.041 -0.097 0.151 -0.055 0.051 -0.005 0.203 0.008
0.116 0.162 0.112 0.102 -0.017 -0.062 0.112 -0.067 0.244 0.171 -0.202 0.042 0.117 0.063 0.001 0.11 -0.171 -0.01 0.041 -0.051 0.166 0.015 0.022 0.077 -0.005 -0.078 -0.115 0.015 -0.04 0.051 0.078 -0.013 0.104 0.013 -0.178 -0.012 0.078 -0.134 -0.007 0.067 -0.247 0.103 -0.106 0.185 0.043 0.005 -0.098 -0.058 0.026 0.018 -0.032
-0.196 -0.042 0.056 0.037 0.103 0.078 -0.116 0.006 0.125 -0.115 -0.141 0.129 -0.176 0.039 -0.009 0.084 0.047 0.107 0.069 0.133 -0.023 -0.05 0.282 -0.077 -0.109 -0.007 -0.115 -0.11 -0.292 0.176 -0.116 -0.29 0.218 0.134 0.115 -0.196 0.186 0.035 -0.005 -0.015 0.186 -0.049 -0.027 0.073 -0.125 0.021 0.091 -0.09 -0.015 -0.02 -0.102
-0.079 0.019 -0.037 -0.122 -0.072 0.216 -0.016 -0.094 0.102 0.04 -0.17 -0.006 0.043 -0.092 -0.068 -0.082 -0.101 0.01 0.084 0.009 -0.191 -0.091 -0.081 -0.117 0.067 0.236 -0.043 -0.127 0.027 0.028 0.177 -0.067 -0.032 -0.041 0.047 0.144 0.206 -0.055 -0.152 0.051 0.198 -0.093 -0.041 0.141 -0.174 0.206 0.056 -0.05 -0.143 0.068 -0.158
-0.081 -0.009 0.108 0.045 -0.041 -0.122 -0.179 0.071 0.115 -0.098 -0.066 0.08 -0.065 -0.058 0.061 0.037 0.005 -0.081 0.08 0.108 0.018 -0.07 -0.038 0.196 0.094 -0.037 -0.058 0.091 -0.047 -0.119 -0.119 -0.24 0.11 -0.086 -0.049 0.052 0.057 -0.053 -0.13 -0.17 0.185 -0.038 0.196 0.095 0.025 -0.126 -0.131 0.002 -0.101 -0.183 -0.166
[... 25 lines of numbers omitted here; the full file is in the repository ...]
-0.11 0.006 -0.041 -0.077 -0.115 0.21 0.174 0.02 0.086 -0.146 0.11 -0.155 -0.098 -0.027 0.064 -0.039 -0.009 -0.098 -0.003 0.096 -0.053 -0.098 -0.105 0.024 -0.166 -0.095 0.057 0.143 0.169 0.159 0.006 -0.186 -0.19 0.168 0.03 -0.022 -0.004 0.021 0.049 0.075 0.155 -0.027 -0.088 -0.093 0.083 -0.176 -0.149 0.113 -0.014 0.09 -0.077
0.098 0.019 -0.058 0.108 0.146 -0.103 0.077 0.041 -0.102 0.1 0.101 0.038 -0.11 -0.024 0.039 -0.04 0.099 -0.204 -0.06 -0.057 0.028 -0.256 0.069 0.051 0.17 -0.035 0.289 0.013 -0.117 0.035 0.055 0.05 0.024 -0.087 0.186 -0.061 0.088 0.134 0.118 0.1 0.056 0.112 0.195 -0.026 -0.123 0 -0.149 -0.123 -0.116 0.123 0.105
## net.0.bias shape=(32,)
-0.077 0.086 0.255 0.121 0.003 -0.055 0.137 0.137 -0.028 -0.13 0.01 0.178 0.183 0.193 -0.053 0.026 -0.072 0.065 -0.075 0.126 -0.096 -0.016 0.065 -0.164 -0.144 -0.058 0.1 -0.004 0.118 0.061 0.025 -0.052
## net.2.weight shape=(32, 32)
-0.006 0.132 -0.144 0.092 -0.172 0.13 0.182 0.023 -0.244 -0.047 -0.013 0.094 0.01 -0.065 -0.075 -0.112 -0.115 -0.064 0.196 -0.156 0.177 0.102 -0.004 -0.008 0.218 0.168 0.172 -0.201 0.188 0.015 0.002 0.177
-0.057 0.031 -0.264 0.156 0.048 -0.141 0.043 0.224 0.044 0.102 0.113 -0.136 0.066 -0.008 0.014 0.092 -0.168 -0.04 -0.006 0.079 -0.087 -0.19 0.034 -0.098 -0.101 0.022 0.07 0.004 -0.256 -0.041 0.039 -0.192
-0.029 -0.089 -0.073 0.042 -0.032 0.119 0.035 0.186 -0.105 0.189 -0.082 0.017 -0.046 -0.072 -0.036 -0.069 0.073 -0.262 0.258 -0.096 0.051 0.001 -0.083 -0.029 -0.11 0.219 0.065 -0.087 0.015 0.183 -0.121 -0.094
0.49 0.021 -0.179 -0.076 -0.054 -0.218 -0.202 0.33 -0.151 0.01 -0.213 0.071 0.084 -0.102 -0.107 0.121 0.004 -0.061 -0.079 0.04 0.143 -0.101 -0.282 0.012 0.274 -0.239 -0.153 -0.24 -0.022 -0.002 -0.06 0.225
0.249 -0.073 0.194 0.052 0.033 -0.065 -0.046 -0.049 0.168 -0.196 0.181 0.111 0.239 -0.031 0.162 0.085 0.18 -0.13 -0.027 -0.017 -0.088 0.004 -0.003 -0.172 -0.043 -0.016 0.077 -0.035 -0.046 -0.138 0.169 0.146
[... 25 lines of numbers omitted here; the full file is in the repository ...]
0.057 -0.054 -0.162 0.191 -0.21 -0.122 0.122 0.17 -0.088 0.199 -0.007 0.228 0.117 -0.241 -0.08 -0.262 -0.058 -0.249 -0.074 0.079 0.206 -0.05 0.172 0.125 -0.093 0.102 0.196 -0.261 0 -0.161 -0.034 -0.046
0.001 0.096 0.085 0.067 0.162 0.163 -0.1 0.119 -0.066 -0.062 0.102 0.08 -0.036 -0.03 -0.111 0.123 -0.099 -0.093 0.019 0.188 -0.055 -0.053 0.053 -0.197 0.124 -0.189 0.055 0.031 0.127 -0.093 0.231 0.157
## net.2.bias shape=(32,)
0.111 0.078 -0.03 0.045 0.227 0.085 0.12 -0.025 -0.096 -0.005 -0.052 0.24 0.201 0.035 0.049 -0.013 -0.008 -0.034 0.061 -0.041 0.121 -0.155 0.045 0.048 -0.081 0.095 -0.082 0.008 0.028 0.067 -0.052 0.193
## net.4.weight shape=(1, 32)
0.031 -0.031 0.038 0.053 -0.04 0.016 -0.01 -0.03 0.042 0.021 -0.016 0.04 0.028 -0.03 -0.028 0.02 -0.021 -0.036 0.036 -0.035 -0.018 0.022 0.036 -0.03 0.02 -0.028 0.037 0.023 -0.089 0.017 0.043 -0.011
## net.4.bias shape=(1,)
0.3

# Reply format
Reply with a JSON list of edits and no other text. Each edit is {"tensor": name, "index": [i, ...], "value": number}, with zero-based indices. Change as few weights as the task needs.
\end{lstlisting}

\textbf{Raw reply}
\begin{lstlisting}
[{"tensor": "net.0.weight", "index": [0, 30], "value": 0}, {"tensor": "net.0.weight", "index": [1, 30], "value": 0}, {"tensor": "net.0.weight", "index": [2, 30], "value": 0}, {"tensor": "net.0.weight", "index": [3, 30], "value": 0}, {"tensor": "net.0.weight", "index": [4, 30], "value": 0}, {"tensor": "net.0.weight", "index": [5, 30], "value": 0}, {"tensor": "net.0.weight", "index": [6, 30], "value [... 1478 more characters on this line omitted ...]
\end{lstlisting}

\textbf{The model's own account}
\begin{lstlisting}
I first scanned net.0.weight for columns with unusually large entries, which gave candidates 30, 31, 25 and 46 (column 30 had the largest sum of squares, about 0.63 against roughly 0.4 for a typical column). I then computed by hand the linearised downstream sensitivity of each first-layer unit, v = net.2.weight^T net.4.weight (ignoring ReLU gating), and projected candidate columns onto it: column 30 agrees in sign with v on nearly every one of the 32 hidden units (projection about +0.21), whereas column 31 largely cancels (about -0.005); columns 25 and 46 were not projected. I therefore zeroed all 32 entries of column 30 of net.0.weight, which removes any dependence of the output on input 30 and touches nothing else. Confidence that input 30 is the leak: about 88%.
\end{lstlisting}

\subsubsection{Editing a trained surrogate from its weights and a decile table}
{\raggedright Run folder: \texttt{experiments/\allowbreak e1\_\allowbreak surrogate\_\allowbreak head/\allowbreak runs/\allowbreak edit\_\allowbreak \_\allowbreak loop\_\allowbreak bulk\_\allowbreak modulus\_\allowbreak \_\allowbreak seed0} (Fable 5.2).\par}

\textbf{Prompt}
\begin{lstlisting}
You are given a small neural network as its raw weights. You cannot run it and you cannot train it.

# Task
This network is a surrogate inside a materials discovery loop. Inputs are the first 50 principal components of mean-pooled ORB-v3 node embeddings of a crystal. The target is the bulk modulus from a Birch-Murnaghan fit, computed with the eSEN-30M-OAM potential. Training rows are the loop's warm-start pool (500 known structures); test rows are the structures a generator (MatterGen, CrystalFlow or AD [... 241 more characters on this line omitted ...]

The network was trained on 500 rows whose target has mean 134.6 and standard deviation 72.14 GPa. The loop cares most about the candidates with the highest true values, and those are the ones this network gets wrong. Its errors on training rows it had not seen (five-fold cross-validation), grouped by decile of the true value:

decile of true value | true range (GPa) | mean true | mean predicted | mean error
1 | 7.512 to 46.47 | 33.75 | 45.08 | +11.33
2 | 46.47 to 64.85 | 56.33 | 60.69 | +4.364
3 | 64.85 to 87.27 | 74.83 | 80.4 | +5.573
4 | 87.27 to 106.7 | 96.37 | 97.57 | +1.197
5 | 106.7 to 124.1 | 115.1 | 121.1 | +6.027
6 | 124.1 to 145.8 | 135.2 | 136.3 | +1.04
7 | 145.8 to 178.3 | 160.1 | 164.2 | +4.053
8 | 178.3 to 200.8 | 187.9 | 181.6 | -6.209
9 | 200.8 to 233 | 215.8 | 203.3 | -12.48
10 | 233 to 385.2 | 271.1 | 244.8 | -26.24

Edit the weights so that the network stops underpredicting the highest values on new candidates, without losing its ranking of the rest.

# Architecture
Input x has 50 entries. h0 = x.
h1 = relu(net.0.weight @ h0 + net.0.bias)   (net.0.weight is 32 x 50, net.0.bias has 32 entries)
h2 = relu(net.2.weight @ h1 + net.2.bias)   (net.2.weight is 32 x 32, net.2.bias has 32 entries)
y = net.4.weight @ h2 + net.4.bias   (net.4.weight is 1 x 32, net.4.bias has 1 entries)

# Weights
## net.0.weight shape=(32, 50)
0.047 0.19 -0.081 -0.108 -0.201 0.092 0.094 0.186 -0.13 0.087 -0.124 -0.023 -0.083 -0.104 0.111 0.009 0.091 0.175 -0.14 -0.102 0.041 0.131 0.061 0.205 0.03 0.137 0.138 -0.079 -0.133 -0.013 0.01 0.173 -0.163 -0.022 -0.18 -0.252 -0.245 0.301 0.064 0.042 -0.068 -0.058 0.085 -0.251 0.003 -0.102 0.006 0.103 -0.014 0.035
0.175 0.166 0.091 0.106 0.071 -0.046 0.133 -0.12 -0.029 -0.061 -0.049 0.032 0.031 0.159 0.169 0.072 0.06 -0.005 0.156 0.315 -0.089 -0.137 0.089 0.07 0.094 0.101 0.064 -0.087 -0.044 -0.191 -0.005 0.026 0.11 -0.134 -0.023 -0.01 -0.032 0.004 0.021 -0.019 -0.113 0.123 0.102 0.111 -0.039 -0.052 0.073 0.111 0.038 -0.044
0.213 0.159 0.136 0.02 -0.144 0.078 -0.006 -0.167 -0.181 0.021 -0.118 -0.014 -0.077 0.202 0.107 -0.047 -0.101 -0.117 0.096 0.11 0.03 0.072 -0.078 0.07 -0.046 -0.044 0.023 -0.131 -0.064 -0.085 0.138 -0.043 -0.093 0.047 0 0.051 0.098 0.026 0.134 0.026 0.002 0.113 0.038 0.07 -0.168 -0.041 0.017 0.109 -0.111 -0.088
-0.178 -0.104 0.075 0.086 0.106 -0.014 0.084 -0.223 0.045 -0.044 0.071 -0.119 -0.202 -0.156 -0.063 0.038 -0.056 -0.06 -0.13 0.018 0.002 -0.164 0.207 0.105 -0.096 0.186 -0.11 -0.121 0.129 -0.12 0.17 -0.125 -0.014 0.014 0.158 -0.006 0.06 0.011 0.008 0.059 0.23 0.001 0.031 -0.043 -0.194 -0.174 0.09 0.107 0 -0.033
0.1 0.146 0.027 0.003 -0.139 0.158 0.038 0.052 -0.043 -0.056 -0.122 -0.098 0.09 -0.085 0.099 0.083 -0.212 -0.182 0.13 -0.215 0.063 0.111 0.035 -0.056 -0.081 0.081 -0.119 -0.155 -0.026 -0.05 0.149 -0.027 0.021 0.187 0.217 -0.002 0.086 -0.21 -0.111 0.116 -0.066 -0.042 0.025 -0.088 0.15 -0.142 0.116 0.029 -0.133 -0.033
[... 25 lines of numbers omitted here; the full file is in the repository ...]
0.221 -0.046 0.17 -0.022 0.051 0.006 0.031 0.085 -0.127 0.131 0.01 -0.06 -0.161 -0.11 -0.076 -0.002 0.036 0.06 0.068 -0.103 0.17 0.219 -0.089 -0.099 0.111 0.109 -0.075 0.089 -0.104 0.057 -0.024 -0.058 -0.006 -0.102 0.056 -0.046 -0.081 -0.018 0.149 0.043 0.136 -0.216 -0.022 -0.005 -0.015 0.138 0.118 -0.076 -0.012 0.026
0.022 0.121 0.122 0.009 -0.174 -0.06 -0.109 -0.177 -0.134 -0.189 -0.032 -0.049 0.009 -0.02 0.175 -0.031 -0.111 -0.158 -0.056 0.103 -0.047 -0.234 0.129 -0.103 -0.143 -0.017 0.082 -0.116 0.05 -0.052 -0.029 0.228 -0.043 0.188 -0.03 0.032 -0.031 0.108 -0.06 -0.056 0.062 0.062 -0.083 0.016 0.004 0.048 -0.107 0.084 -0.023 0.061
## net.0.bias shape=(32,)
0.009 0.157 0.141 0.027 0.023 0.086 0.061 0.03 -0.01 -0.008 0.063 0.075 -0.035 0.077 -0.078 0.096 0.085 -0.081 -0.009 -0.048 0.009 -0.093 -0.093 0.106 0.135 0.155 0.087 0.029 0.139 -0.031 0.092 -0.123
## net.2.weight shape=(32, 32)
0.206 0.091 -0.041 -0.318 0.167 0.119 0.042 -0.129 0.218 -0.073 -0.363 -0.299 -0.003 0.093 -0.127 0.05 0.016 0.006 0.143 0.086 -0.037 -0.162 0.178 0.201 0.097 0.02 0.153 0.016 -0.077 -0.012 0.11 0.196
-0.038 0.184 -0.207 0.231 0.098 0.035 0.139 0.148 -0.146 -0.031 0.135 0.014 0.22 -0.086 -0.092 0.038 -0.108 0.205 -0.084 -0.055 -0.061 -0.06 -0.135 0.002 -0.103 0.049 0.182 0.074 0.136 -0.072 0.08 0.148
0.134 0.235 -0.083 -0.122 0.11 -0.259 -0.008 -0.197 0.032 -0.057 -0.036 -0.218 -0.162 0.174 0.018 -0.072 -0.098 0.01 0.175 0.09 0.226 -0.042 0.232 -0.058 0.213 0 0.02 -0.048 -0.074 0.092 0.244 -0.031
0.04 0.058 -0.107 0.196 0.037 0.208 0.135 0.055 0.027 -0.127 0.205 0.139 -0.111 -0.004 0.05 0.076 0.02 -0.159 -0.07 0.018 -0.275 -0.01 -0.064 0.051 0.072 -0.112 0.035 0.01 -0.053 0.05 -0.154 0.057
-0.138 0.004 -0.161 -0.027 0.078 -0.101 -0.064 0.131 0.013 -0.142 -0.047 -0.019 -0.093 0.026 0.039 0.2 -0.145 -0.139 0.026 0.064 0.094 0.037 -0.032 -0.078 0.032 -0.16 -0.103 -0.077 -0.232 0.042 -0.04 0.062
[... 25 lines of numbers omitted here; the full file is in the repository ...]
-0.226 -0.144 -0.176 0.182 0.286 0.168 0.097 0.1 0.172 -0.182 -0.015 -0.097 0.011 -0.051 0.073 -0.056 -0.033 -0.104 -0.179 -0.044 -0.221 0.137 -0.209 -0.061 -0.093 -0.119 0.254 0.202 -0.023 -0.065 0.189 0.087
0.011 -0.256 0.077 -0.078 0.004 0.056 -0.006 -0.017 -0.258 -0.033 -0.042 0.216 -0.031 0.125 0.053 -0.012 -0.01 -0.035 -0.067 -0.042 -0.007 0.231 0.077 0.043 0.051 0.076 0.194 0.074 0.158 -0.218 0.076 -0.118
## net.2.bias shape=(32,)
0.039 0.024 0.099 0.241 -0.192 0.044 0.074 0.126 -0.036 0.204 0.09 0.01 0.029 0.049 0.058 0.044 0.11 0.097 -0.171 -0.025 0.09 0.063 0.039 0.208 -0.149 -0.024 0.011 0.235 -0.198 0.048 -0.082 0.039
## net.4.weight shape=(1, 32)
10.66 -7.465 18.747 -5.842 -2.627 10.081 13.585 -5.382 15.925 -13.07 13.271 13.874 13.562 7.08 11.554 -14.113 -12.891 -6.245 1.553 10.618 16.294 14.877 -6.863 -14.462 -11.671 11.63 -16.814 -18.426 -4.16 -10.873 -11.142 -10.513
## net.4.bias shape=(1,)
126.262

# Reply format
Reply with a JSON list of edits and no other text. Each edit is {"tensor": name, "index": [i, ...], "value": number}, with zero-based indices. Change as few weights as the task needs.
\end{lstlisting}

\textbf{Raw reply}
\begin{lstlisting}
[
{"tensor": "net.4.weight", "index": [0, 0], "value": 12.259},
{"tensor": "net.4.weight", "index": [0, 1], "value": -8.585},
{"tensor": "net.4.weight", "index": [0, 2], "value": 21.559},
{"tensor": "net.4.weight", "index": [0, 3], "value": -6.718},
{"tensor": "net.4.weight", "index": [0, 4], "value": -3.021},
{"tensor": "net.4.weight", "index": [0, 5], "value": 11.593},
{"tensor": "net.4.weight", "index": [0, 6], "value": 15.623},
{"tensor": "net.4.weight", "index": [0, 7], "value": -6.189},
{"tensor": "net.4.weight", "index": [0, 8], "value": 18.314},
{"tensor": "net.4.weight", "index": [0, 9], "value": -15.031},
{"tensor": "net.4.weight", "index": [0, 10], "value": 15.262},
{"tensor": "net.4.weight", "index": [0, 11], "value": 15.955},
{"tensor": "net.4.weight", "index": [0, 12], "value": 15.596},
{"tensor": "net.4.weight", "index": [0, 13], "value": 8.142},
{"tensor": "net.4.weight", "index": [0, 14], "value": 13.287},
{"tensor": "net.4.weight", "index": [0, 15], "value": -16.23},
{"tensor": "net.4.weight", "index": [0, 16], "value": -14.825},
{"tensor": "net.4.weight", "index": [0, 17], "value": -7.182},
{"tensor": "net.4.weight", "index": [0, 18], "value": 1.786},
{"tensor": "net.4.weight", "index": [0, 19], "value": 12.211},
{"tensor": "net.4.weight", "index": [0, 20], "value": 18.738},
{"tensor": "net.4.weight", "index": [0, 21], "value": 17.109},
{"tensor": "net.4.weight", "index": [0, 22], "value": -7.892},
{"tensor": "net.4.weight", "index": [0, 23], "value": -16.631},
{"tensor": "net.4.weight", "index": [0, 24], "value": -13.422},
{"tensor": "net.4.weight", "index": [0, 25], "value": 13.375},
{"tensor": "net.4.weight", "index": [0, 26], "value": -19.336},
{"tensor": "net.4.weight", "index": [0, 27], "value": -21.19},
{"tensor": "net.4.weight", "index": [0, 28], "value": -4.784},
{"tensor": "net.4.weight", "index": [0, 29], "value": -12.504},
{"tensor": "net.4.weight", "index": [0, 30], "value": -12.813},
{"tensor": "net.4.weight", "index": [0, 31], "value": -12.09},
{"tensor": "net.4.bias", "index": [0], "value": 126.451}
]
\end{lstlisting}

\textbf{The model's own account} (\texttt{model\_self\_report.md}; this run has no \texttt{self\_report.txt})
\begin{lstlisting}
# What the model said it did

Reported by the model after writing its reply; not verified beyond the scored reply itself.

> I multiplied all 32 output weights (net.4.weight) by 1.15 and set net.4.bias to 126.451, which makes the new output y' = 1.15*y - 18.75, a stretch of the predictions about a pivot of 125 GPa. The decile table shows a compressed prediction range, so the stretch lifts the top-decile mean from about 245 to about 263 GPa. Because the change is a positive affine map of the output, the ranking is unchanged; the hidden layers are untouched. The 1.15 factor was chosen by hand arithmetic from the decile means and nothing was run.
\end{lstlisting}

\subsubsection{One edit of a potential's last layer, with the permuted-values control (ORB-v3)}
{\raggedright Run folder: \texttt{supplementary/\allowbreak e2\_\allowbreak mlip\_\allowbreak readout\_\allowbreak edit/\allowbreak runs/\allowbreak orb\_\allowbreak edit\_\allowbreak \_\allowbreak opus\_\allowbreak \_\allowbreak ctrl1} (Opus 5.5).\par}

\textbf{Prompt}
\begin{lstlisting}
You are given the output layer of the energy head of a machine-learned interatomic potential (ORB-v3, orb_v3_conservative_inf_omat) as raw numbers. You cannot run it and you cannot train it.

# How the head works
The energy of a crystal with N atoms is
  E = 1.509531 * N * ( w2 . h + b2 ) + C,     h = silu(W1 x + b1)
x is the mean over the atoms of 256 features computed by the frozen message-passing layers; W1 (256 x 256) and b1 (256) are frozen and not shown; C does not depend on the head (reference energies of the elements, pair repulsion). You can change w2 (256 numbers) and b2 (1 number). Nothing else. b2 moves the energy per atom of every crystal by the same amount and cannot change a bulk modulus.

# The fault
On 200 crystals with DFT bulk moduli (Materials Project elastic tensors), this potential's bulk modulus, from a fit of energy against volume over +-3%, is wrong by -9.4 GPa on average (mean absolute error 12.7 GPa). By range of the DFT value:
DFT 10 to 70 GPa: mean error +0.4 GPa (+1.9%)
DFT 70 to 139 GPa: mean error -4.0 GPa (-3.6%)
DFT 139 to 201 GPa: mean error -11.8 GPa (-6.5%)
DFT 201 to 346 GPa: mean error -22.2 GPa (-8.9%)

# The task
Edit the numbers so that the bulk modulus error on NEW crystals is smaller, while the energy per atom of every crystal stays where it is: an average change above 5 meV per atom counts as a failure, because formation energies depend on it.

# What each weight of w2 does
For each weight j of w2: its value; the change of the bulk modulus in GPa per unit increase of the weight, mean over the 200 crystals and standard deviation across them; the change of the energy per atom in eV per unit increase, mean and standard deviation. The bulk-modulus numbers are measured by changing the weight by 0.002 and refitting energy against volume, so they include the shift of the eq [... 112 more characters on this line omitted ...]
j | w2[j] | dK mean | dK sd | dE mean | dE sd
0 | 1.6941e-06 | +0.005641 | 0.0633 | +2.749e-07 | 0.000265
1 | -1.005e-05 | +0.01236 | 0.0552 | -4.691e-05 | 0.000248
2 | -5.0398e-06 | +0.009736 | 0.0576 | +3.355e-05 | 0.000257
3 | -6.0875e-06 | +11.19 | 14.7 | -0.2216 | 0.142
4 | -9.8823e-06 | -0.003031 | 0.0787 | -4.419e-05 | 0.000272
[... 249 lines of numbers omitted here; the full file is in the repository ...]
254 | -2.4848e-06 | -110.4 | 102 | +0.7197 | 0.997
255 | 0.086948 | +0.00362 | 0.067 | -2.726e-06 | 0.000254

# The other number
## b2 shape=(1,)
0.023866

# Reply format
Reply with a JSON list of edits and no other text. Each edit is {"tensor": name, "index": [i], "value": number} with zero-based indices; `value` is the new value, not the change. Tensor names are w2 and b2. Change as few numbers as the task needs.
\end{lstlisting}

\textbf{Raw reply}
\begin{lstlisting}
[{"tensor": "w2", "index": [188], "value": 0.04516}, {"tensor": "b2", "index": [0], "value": 0.025359}]
\end{lstlisting}

\textbf{The model's own account}
\begin{lstlisting}
I ranked every w2 unit by mean bulk-modulus gain per unit of crystal-to-crystal energy spread (|dK mean| / dE sd), since b2 can cancel the mean energy shift but not the spread; unit 188 was best (about 116 GPa per eV of spread), and without knowing correlations between units a single best unit is the only choice whose energy spread is guaranteed, so I lowered w2[188] by 0.004 (spread about 3.6 meV per atom, expected mean dK about +0.4 GPa, scaling up for stiffer crystals) and raised b2 by 0.0014929 to cancel the -2.25 meV per atom mean shift. The 5 meV energy budget caps any w2 edit at well under 1 GPa of correction, so this fixes only a small part of the -9.4 GPa bias. The table looks inconsistent: the 44 large w2 values sit almost entirely on units whose activations are near zero, while the 44 units with large sensitivities mostly carry weights near 1e-5 (overlap of 8, about what a random permutation gives), which suggests the w2 column has been shuffled against the sensitivities; I still acted on the table as given. Confidence that this edit lowers the bulk-modulus error on new crystals without breaking the energy limit: about 40%.
\end{lstlisting}

\subsubsection{Fifth attempt of a feedback chain, with the four earlier attempts and their measured results in the prompt (ORB-v3)}
{\raggedright Run folder: \texttt{supplementary/\allowbreak e2\_\allowbreak mlip\_\allowbreak readout\_\allowbreak edit/\allowbreak runs/\allowbreak head\_\allowbreak orb\_\allowbreak \_\allowbreak fable\_\allowbreak \_\allowbreak c2\_\allowbreak \_\allowbreak r5} (Fable 5.2).\par}

\textbf{Prompt}
\begin{lstlisting}
You are given the last layer of the energy readout of a machine-learned interatomic potential (ORB-v3 (orb_v3_conservative_inf_omat), 25,510,582 weights in total) as raw numbers. You cannot run it and you cannot train it.

# How the last layer works
The energy of a crystal is
  E = h . w + (uniform term) + (terms that do not depend on these numbers)
h is a vector of 256 numbers computed for the crystal by the frozen layers before; it changes when the crystal is strained. You can change w (256 numbers) and b (1 number). b is a uniform term: increasing it by one unit raises the energy per atom of every crystal by 1.50953 eV and cannot change a bulk modulus.

# The fault
On 200 crystals with DFT bulk moduli (Materials Project elastic tensors), this potential's bulk modulus, from a fit of energy against volume over +-3%, is wrong by -9.4 GPa on average (mean absolute error 12.7 GPa). By range of the DFT value:
DFT 10 to 70 GPa: mean error +0.4 GPa (+1.9%)
DFT 70 to 139 GPa: mean error -4.0 GPa (-3.6%)
DFT 139 to 201 GPa: mean error -11.8 GPa (-6.5%)
DFT 201 to 346 GPa: mean error -22.2 GPa (-8.9%)

# The task
Edit the numbers so that the bulk modulus error on NEW crystals is smaller, while the energy per atom of every crystal stays where it is: a mean absolute change above 5 meV per atom counts as a failure, because formation energies depend on it.

# What each weight does
For each weight j of w: its value; the change of the bulk modulus in GPa per unit increase of the weight, mean over the 200 crystals and standard deviation across them; the change of the energy per atom in eV per unit increase, mean and standard deviation. The bulk-modulus numbers are measured by changing the weight by 0.002 and refitting energy against volume, so they include the shift of the equ [... 110 more characters on this line omitted ...]
j | w[j] | dK mean | dK sd | dE mean | dE sd
0 | -5.082e-06 | +0.005641 | 0.0633 | +2.749e-07 | 0.000265
1 | -1.321e-05 | +0.01236 | 0.0552 | -4.691e-05 | 0.000248
2 | -1.107e-05 | +0.009736 | 0.0576 | +3.355e-05 | 0.000257
3 | 0.014675 | +11.19 | 14.7 | -0.2216 | 0.142
4 | 3.7407e-07 | -0.003031 | 0.0787 | -4.419e-05 | 0.000272
[... 249 lines of numbers omitted here; the full file is in the repository ...]
254 | -0.022598 | -110.4 | 102 | +0.7197 | 0.997
255 | 8.0737e-06 | +0.00362 | 0.067 | -2.726e-06 | 0.000254

# The other number
## b shape=(1,)
0.0238665

# Your earlier attempts
This is attempt 5 of 5. Each earlier reply was applied to the original numbers and measured. Edits do not accumulate: every reply is applied to the original numbers. This is your last attempt. The reply that counts is the one with the lowest mean absolute error among those within the energy limit.

## Attempt 1
Your reply:
[{"tensor": "w", "index": [188], "value": -0.025107}, {"tensor": "w", "index": [180], "value": 0.0046987}, {"tensor": "w", "index": [3], "value": 0.029675}, {"tensor": "b", "index": [0], "value": 0.0270995}]
Measured on the same 200 crystals: bulk modulus mean error -8.53 GPa, mean absolute error 11.98 GPa; mean absolute change of energy per atom 3.81 meV (within the limit).
DFT 10 to 70 GPa: mean error +0.5 GPa (+2.1%)
DFT 70 to 139 GPa: mean error -3.4 GPa (-3.0%)
DFT 139 to 201 GPa: mean error -10.8 GPa (-5.9%)
DFT 201 to 346 GPa: mean error -20.4 GPa (-8.2%)

## Attempt 2
Your reply:
[{"tensor": "w", "index": [188], "value": -0.027107}, {"tensor": "w", "index": [161], "value": -0.054161}, {"tensor": "w", "index": [180], "value": 0.0186987}, {"tensor": "w", "index": [135], "value": 0.0011387}, {"tensor": "b", "index": [0], "value": 0.0245655}]
Measured on the same 200 crystals: bulk modulus mean error -8.04 GPa, mean absolute error 11.44 GPa; mean absolute change of energy per atom 4.79 meV (within the limit).
DFT 10 to 70 GPa: mean error +0.5 GPa (+2.0%)
DFT 70 to 139 GPa: mean error -3.1 GPa (-2.7%)
DFT 139 to 201 GPa: mean error -10.4 GPa (-5.7%)
DFT 201 to 346 GPa: mean error -19.2 GPa (-7.7%)

## Attempt 3
Your reply:
[{"tensor": "w", "index": [38], "value": -0.075672}, {"tensor": "w", "index": [90], "value": -0.039503}, {"tensor": "w", "index": [77], "value": -0.102609}, {"tensor": "w", "index": [49], "value": -0.040377}, {"tensor": "b", "index": [0], "value": 0.0241308}]
Measured on the same 200 crystals: bulk modulus mean error -7.61 GPa, mean absolute error 11.30 GPa; mean absolute change of energy per atom 4.15 meV (within the limit).
DFT 10 to 70 GPa: mean error +0.5 GPa (+1.9%)
DFT 70 to 139 GPa: mean error -3.1 GPa (-2.7%)
DFT 139 to 201 GPa: mean error -9.6 GPa (-5.2%)
DFT 201 to 346 GPa: mean error -18.3 GPa (-7.4%)

## Attempt 4
Your reply:
[{"tensor": "w", "index": [38], "value": -0.065672}, {"tensor": "w", "index": [77], "value": -0.095609}, {"tensor": "w", "index": [151], "value": 0.078565}, {"tensor": "w", "index": [186], "value": 0.066448}, {"tensor": "b", "index": [0], "value": 0.040857}]
Measured on the same 200 crystals: bulk modulus mean error -7.16 GPa, mean absolute error 11.01 GPa; mean absolute change of energy per atom 23.33 meV (ABOVE the limit: a failure).
DFT 10 to 70 GPa: mean error -1.0 GPa (-1.4%)
DFT 70 to 139 GPa: mean error -4.0 GPa (-3.6%)
DFT 139 to 201 GPa: mean error -8.3 GPa (-4.4%)
DFT 201 to 346 GPa: mean error -15.4 GPa (-6.2%)

# Reply format
Reply with a JSON list of edits and no other text. Each edit is {"tensor": name, "index": [i], "value": number} with zero-based indices; `value` is the new value, not the change. Tensor names: w and b.
\end{lstlisting}

\textbf{Raw reply}
\begin{lstlisting}
[{"tensor": "w", "index": [38], "value": -0.074672}, {"tensor": "w", "index": [77], "value": -0.102109}, {"tensor": "w", "index": [254], "value": -0.030598}, {"tensor": "w", "index": [161], "value": -0.052161}, {"tensor": "w", "index": [49], "value": -0.040013}, {"tensor": "w", "index": [90], "value": -0.040995}, {"tensor": "w", "index": [65], "value": -0.03543}, {"tensor": "w", "index": [84], "va [... 68 more characters on this line omitted ...]
\end{lstlisting}

\textbf{The model's own account}
\begin{lstlisting}
I built on attempt 3, whose measured bulk-modulus shift (+1.79 GPa) matched the linear prediction from the table: make weights with a large |dK| per unit of energy more negative (38, 77, 254, 161) and raise weights with a small |dK| per unit of energy (49, 90, 65, 84), choosing the amounts by hand arithmetic so that both the mean energy change and the summed energy standard deviation cancel, with b absorbing the remaining 0.5 meV mean. The edit is spread over four pairs instead of two, on the guess that the per-crystal energy residuals of different pairs are partly independent, so the same energy budget buys more stiffness. The linear prediction is a bulk-modulus shift of about +2.2 GPa (mean error near -7.2 GPa); I am about 55% confident it stays under the 5 meV limit and improves on attempt 3, since I have no per-crystal data to check the residual correlation.
\end{lstlisting}

\subsubsection{Writing an error-risk predictor for a potential}
{\raggedright Run folder: \texttt{supplementary/\allowbreak e2\_\allowbreak mlip\_\allowbreak readout\_\allowbreak edit/\allowbreak runs/\allowbreak risk\_\allowbreak \_\allowbreak opus\_\allowbreak \_\allowbreak rep1} (Opus 5.5).\par}

\textbf{Prompt}
\begin{lstlisting}
You are given a small neural network as its raw weights. You cannot run it and you cannot train it.

# Task
Write the weights of this network from what you know about materials and about machine-learned interatomic potentials. No training data is given. A universal potential trained on PBE calculations (one of MACE-MPA-0, ORB-v3, eSEN-30M-OAM, GRACE-2L-OAM) has relaxed an inorganic crystal from the Materials Project and predicted its bulk modulus from a fit of energy against volume. The output y must be [... 239 more characters on this line omitted ...]

Inputs, not standardised:
x[0] = volume_per_atom_A3
x[1] = mean_atomic_mass_u
x[2] = mean_pauling_electronegativity
x[3] = mean_elemental_melting_point_K
x[4] = frac_alkali
x[5] = frac_alkaline_earth
x[6] = frac_3d_transition
x[7] = frac_4d_5d_transition
x[8] = frac_lanthanide_actinide
x[9] = frac_p_block_metal
x[10] = frac_metalloid
x[11] = frac_C_N_O
x[12] = frac_S_Se_P
x[13] = frac_halogen
x[14] = frac_H
x[15] = bulk_modulus_predicted_by_the_potential_GPa

volume_per_atom_A3 is the DFT-relaxed cell volume divided by the number of atoms. The means are over the atoms of the cell. The fractions are:
frac_alkali: atomic fraction of Li Na K Rb Cs
frac_alkaline_earth: atomic fraction of Be Mg Ca Sr Ba
frac_3d_transition: atomic fraction of Sc Ti V Cr Mn Fe Co Ni Cu Zn
frac_4d_5d_transition: atomic fraction of Y Zr Nb Mo Tc Ru Rh Pd Ag Cd Hf Ta W Re Os Ir Pt Au Hg
frac_lanthanide_actinide: atomic fraction of La Ce Pr Nd Pm Sm Eu Gd Tb Dy Ho Er Tm Yb Lu Ac Th Pa U Np Pu
frac_p_block_metal: atomic fraction of Al Ga In Sn Tl Pb Bi
frac_metalloid: atomic fraction of B Si Ge As Sb Te
frac_C_N_O: atomic fraction of C N O
frac_S_Se_P: atomic fraction of S Se P
frac_halogen: atomic fraction of F Cl Br I
frac_H: atomic fraction of H
Typical ranges: volume per atom 8 to 60, mean atomic mass 7 to 200, mean electronegativity 0.8 to 3.5, mean melting point 300 to 3700, predicted bulk modulus 10 to 350.

# Architecture
Input x has 16 entries. h0 = x.
h1 = relu(net.0.weight @ h0 + net.0.bias)   (net.0.weight is 16 x 16, net.0.bias has 16 entries)
y = net.2.weight @ h1 + net.2.bias   (net.2.weight is 1 x 16, net.2.bias has 1 entries)

# Weights
## net.0.weight shape=(16, 16)
0 0 0 0 0 0 0 0 0 0 0 0 0 0 0 0
0 0 0 0 0 0 0 0 0 0 0 0 0 0 0 0
0 0 0 0 0 0 0 0 0 0 0 0 0 0 0 0
0 0 0 0 0 0 0 0 0 0 0 0 0 0 0 0
0 0 0 0 0 0 0 0 0 0 0 0 0 0 0 0
[... 9 lines of numbers omitted here; the full file is in the repository ...]
0 0 0 0 0 0 0 0 0 0 0 0 0 0 0 0
0 0 0 0 0 0 0 0 0 0 0 0 0 0 0 0
## net.0.bias shape=(16,)
0 0 0 0 0 0 0 0 0 0 0 0 0 0 0 0
## net.2.weight shape=(1, 16)
0 0 0 0 0 0 0 0 0 0 0 0 0 0 0 0
## net.2.bias shape=(1,)
0

# Reply format
Reply with the complete weights in exactly the format shown above: one '## name shape=(...)' header per tensor, then its rows of numbers. No other text.
\end{lstlisting}

\textbf{Raw reply}
\begin{lstlisting}
## net.0.weight shape=(16, 16)
0 0 0 0 0 0 0 0 0 0 0 0 0 0 0 1
0 0 0 0 0 0 0 0 0 0 0 0 0 0 0 1
0 0 0 0 0 0 0 0 0 0 0 0 0 0 0 1
0 0 0 0 0 0 150 0 0 0 0 0 0 0 0 1
0 0 0 0 0 0 0 150 0 0 0 0 0 0 0 1
[... 9 lines of numbers omitted here; the full file is in the repository ...]
0 0 0 0 0 0 0 0 0 0 0 0 0 0 0 0
0 0 0 0 0 0 0 0 0 0 0 0 0 0 0 0
## net.0.bias shape=(16,)
0 -120 -250 -150 -150 -150 0 14 120 -500 5 0 0 0 0 0
## net.2.weight shape=(1, 16)
0.06 0.04 0.04 0.06 0.03 0.08 1 0.5 0.4 0.3 1.5 0 0 0 0 0
## net.2.bias shape=(1,)
1.5
\end{lstlisting}

\textbf{The model's own account}
\begin{lstlisting}
I built the network by hand so the predicted absolute error grows with the potential's own bulk modulus: a base slope of about 6% of B, with extra slope above 120 GPa and again above 250 GPa, because softening and absolute errors are largest for stiff crystals. Gated units (relu(B + 150*frac - 150)) make the error grow faster with B for 3d transition metals (magnetism, +U), 4d/5d metals and f-elements. An additive chemistry unit adds a few GPa for those classes and for metalloids and S/Se/P. Further units flag small volume per atom, a predicted B that is implausibly low for a dense non-hydride, non-alkali, non-halide cell (layered, framework or failed fits), a predicted B that is implausibly high for a large volume per atom, and a near-zero or negative predicted B. I did not have to find a specific fact; I am about 55% confident that this ranking clearly beats using the predicted B alone.
\end{lstlisting}

\end{document}